\documentclass[12pt]{article}
\pdfoutput=1  
\usepackage{hyperref}
\usepackage[pdftex]{graphicx}
\usepackage{amsmath,amssymb}
\usepackage{cancel}
\usepackage{appendix}
\usepackage{mathtools}
\usepackage{xcolor}
\usepackage[mathscr]{eucal}
\newcommand{\gs}{\ensuremath{g_s}} 
\newcommand{\ls}{\ensuremath{l_s}} 

\def\p{\partial}

\newcommand{\tr}{\mathop{\rm Tr}}

\newcommand{\cN}{{\mathcal{N}}}
\newcommand{\cO}{{\mathcal{O}}}

\newcommand{\bZ}{{\mathbf{Z}}}

\newcommand{\vev}[1]{{\left< {#1} \right>}}
\newcommand{\Pm}{\ensuremath{P'_{-}}}

\begin{document}

\begin{titlepage}


\begin{center}
\Large \bf \textbf{On the Underlying Nonrelativistic 
\\
Nature of Relativistic Holography}
\end{center}

\begin{center}
Alberto G\"uijosa

\vspace{0.2cm}
Departamento de F\'{\i}sica de Altas Energ\'{\i}as, Instituto de Ciencias Nucleares, \\
Universidad Nacional Aut\'onoma de M\'exico,
\\ Apartado Postal 70-543, CdMx 04510, M\'exico\\
 \vspace{0.2cm}
\vspace{0.2cm}
{\tt alberto@nucleares.unam.mx}
\vspace{0.2cm}
\end{center}

\begin{center}
{\bf Abstract}
\end{center}
\noindent
Over the past quarter century, considerable effort has been invested in the study of nonrelativistic (NR) string theory, its U-dual NR brane theories, and their geometric foundations in (generalized) Newton-Cartan geometry. Many interesting results have been obtained, both for their intrinsic value and in the hope that they hold useful lessons for relativistic string/M theory. 
By synthesizing two strands of recent developments (especially, arXiv:2312.13243 and arXiv:2410.03591), we argue that this hope has already come to fruition, because standard, \emph{relativistic} holography can now be recognized as a statement within a corresponding \emph{nonrelativistic} brane theory. Our main conclusions are general, but in the familiar example of D3-brane based holography, they read as follows: 
(i) $\cN=4$ SYM is exactly the worldvolume theory of D3-branes within `NR D3-brane theory';
(ii) AdS$_5\times$S$^5$ is exactly the corresponding RR black 3-brane,  and  includes  an  asymptotically flat-Newton-Cartan region; \linebreak[3]
(iii) AdS/CFT duality is precisely synonymous with black-brane/D-brane (i.e., closed-string/open-string) duality 
within NR D3-brane theory; 
(iv) Newton-Cartan geometry is the underlying structure upon which entanglement of the D3-brane degrees of freedom builds relativistic spacetime.

 \vspace{0.4cm}
\smallskip
\end{titlepage}

\tableofcontents

\section{Introduction and Summary}\label{introsec}

 \subsection{Antecedents}
 \label{antecedentssubsec}
 
Nonrelativistic (NR) string theory \cite{Danielsson:2000gi,Gomis:2000bd,Danielsson:2000mu}
is an interesting simplified version of string theory, discovered on the heels of the formulation of the Matrix theory \cite{Banks:1996vh,Susskind:1997cw} and AdS/CFT \cite{Maldacena:1997re,Gubser:1998bc,Witten:1998qj} conjectures. It can be studied either as a special limit (given below in Eq.~(\ref{nrf1limit})) of any one of the standard string theories \cite{Danielsson:2000gi,Danielsson:2000mu}, or starting from an intrinsic worldsheet definition that makes no reference to a limit \cite{Gomis:2000bd,Danielsson:2000mu}.\footnote{Makes no reference, that is, after having been itself obtained by applying the same limit once and for all at the level of the Polyakov action.}  Its most prominent features are: a spacetime foliation that distinguishes time and one other `longitudinal' direction from the remaining `transverse' directions; the nonrelativistic character of the transverse dynamics, complemented by relativistic dynamics within the worldvolume of longitudinal D-branes;  
and the slimming down of the spectrum, such that the theory almost exclusively 
contains strings and other branes that carry \emph{positive} string winding along the longitudinal direction.\footnote{This feature motivated the alternative name `wound string theory', proposed in \cite{Danielsson:2000gi,Danielsson:2000mu} but subsequently disused.} This last property originates from the fact that the NR limit scales up the longitudinal components of the metric and Kalb-Ramond fields in such a way as to produce infinite tensional and electric energy costs, that cancel against one another only when the string is oriented along the positive direction of the longitudinal axis. So, just like the usual nonrelativistic particle limit suppresses antiparticles, the nonrelativistic string limit suppresses antistrings. 
For a review of the wealth of research carried out on NR string theory over a score of years, see \cite{Oling:2022fft}.  

The present paper will highlight some inferences that follow from bringing together two of the strands of recent developments related to NR string theory, worked out respectively in \cite{Danielsson:2000mu,Guijosa:2023qym,Avila:2023aey} and \cite{Blair:2020ops,Blair:2023noj,Gomis:2023eav,Blair:2024aqz}. The first strand refers to the nature of gravity in this theory, a topic where credences have  evolved in a curious manner over the years. NR string theory was uncovered as a 10-dimensional  framework unifying the $p$-dimensional noncommutative \emph{open} string (NCOS) theories \cite{Seiberg:2000ms,Gopakumar:2000na}, originally thought to be surprising examples of string theories that \emph{did not} contain gravity.  
In an important paper \cite{Klebanov:2000pp}, it was shown that, upon longitudinal compactification, \emph{closed} strings were also present, but only if they carried positive longitudinal winding, $w>0$, which still seemed to leave gravity out of the picture. Nonetheless, wound string scattering amplitudes revealed an instantaneous gravitational force \cite{Gomis:2000bd}, which was promptly understood to be due to off-shell `Newtonian gravitons' (and their superpartners) \cite{Danielsson:2000mu}. These are massless modes that, on shell, can only evade the $w>0$ requirement if they have vanishing transverse momentum, rendering them of measure zero as asymptotic states. Despite this, being the only massless states in the theory, their off-shell contributions mediate long-distance interactions, and open the door to considering curved backgrounds.  

NR string theory was thus recognized as a UV-complete theory of nonrelativistic quantum gravity.
Given that a geometrization of Newtonian gravity analogous to that of general relativity is provided by the Newton-Cartan (NC) formalism \cite{Cartan:1923zea,Cartan:1924yea}  (which in itself has been the subject of vigorous contemporary research \cite{Hartong:2022lsy,Bergshoeff:2022eog}), it was only natural for the formulation of NR string theory on curved backgrounds
 to be based on a stringy generalization of the NC framework, known as string Newton-Cartan (SNC) \cite{Andringa:2012uz,Harmark:2017rpg,Bergshoeff:2018yvt,Harmark:2018cdl,Gomis:2019zyu,Gallegos:2019icg,Harmark:2019upf,Bergshoeff:2021bmc,Bidussi:2021ujm}. This has led in recent years to a very fruitful body of work, reviewed in \cite{Oling:2022fft}.  

Yet another unexpected enrichment of this first strand, foreshadowed in \cite{Danielsson:2000mu} and substantiated in \cite{Guijosa:2023qym,Avila:2023aey}, was the reemergence of \emph{relativistic} gravity in NR string theory when a large grouping of objects is considered. This has been demonstrated in various settings, from alternate viewpoints. One clear-cut example is a stack of the basic objects of the theory, positively-wound strings. If we view it first in the parent string theory from which NR string theory descends, we know that it gravitates, giving rise to the familiar, relativistic black string \cite{Dabholkar:1989jt,Dabholkar:1990yf}. The natural expectation is that in the NR limit this should collapse down to a SNC geometry, but the actual outcome is more interesting: the strings create a large tubular bubble within which gravity is in fact \emph{relativistic}, and it is only far away that the background asymptotes to SNC form \cite{Avila:2023aey}.\footnote{As emphasized in \cite{Avila:2023aey}, the only way to prevent the relativistic bubble from being generated is if we completely deplete the strength of the source, formally scaling the number of strings down to zero.} Within the intrinsic definition of NR string theory \cite{Gomis:2000bd}, it is well-understood \cite{Gomis:2019zyu,Gallegos:2019icg,Yan:2019xsf,Blair:2020ops,Gomis:2020fui,Yan:2021lbe,Oling:2022fft} that the most general worldsheet $\sigma$-model indeed includes a marginal term, denoted originally $\beta\bar{\beta}$ and 
lately $\lambda\bar{\lambda}$, that has been shown \cite{Blair:2020ops} to be an instance of the notable $T\bar{T}$ deformation \cite{Smirnov:2016lqw,Cavaglia:2016oda,Jiang:2019epa}, and has precisely the effect of deforming back towards relativistic string theory. In recent years there seems to be a preference to reserve the term `NR string theory' to the subset of states where the coefficient of this term is fine-tuned to zero,\footnote{For quantum-mechanical consistency, this requires the enforcement of special constraints or symmetry principles \cite{Gomis:2019zyu,Yan:2021lbe}.} so that the backgrounds under scrutiny are purely of SNC form. But semantics aside, the physical lesson of this first strand of recent developments is that {\bf collections of objects in the theory discovered in \cite{Danielsson:2000gi,Gomis:2000bd} do gravitate \emph{relativistically}, distorting the geometry such that it is only \emph{{asymptotically} SNC}}. Perhaps the most compelling illustration is that the mere act of adding a boundary to the worldsheet, as one ought to do to describe longitudinal D-branes in the theory, is enough to source\footnote{As we will emphasize in Section~\ref{blackpbranesubsec}, the sourcing effect is really due the strings that are dissolved within the D-branes.} the $\lambda\bar{\lambda}$ operator \cite{Guijosa:2023qym}. 

The second strand has to do with some of the analogous theories that are obtained by acting on NR string theory with the familiar set of dualities \cite{Danielsson:2000gi,Gomis:2000bd}. Each such theory is what results after
 applying to a parent string/M theory the corresponding dual image of the NR string limit. This again leads to the cancellation of tensional energy against the electric energy associated with the dual of the Kalb-Ramond field, thereby selecting as lightest excitation the dual of the fundamental string. 

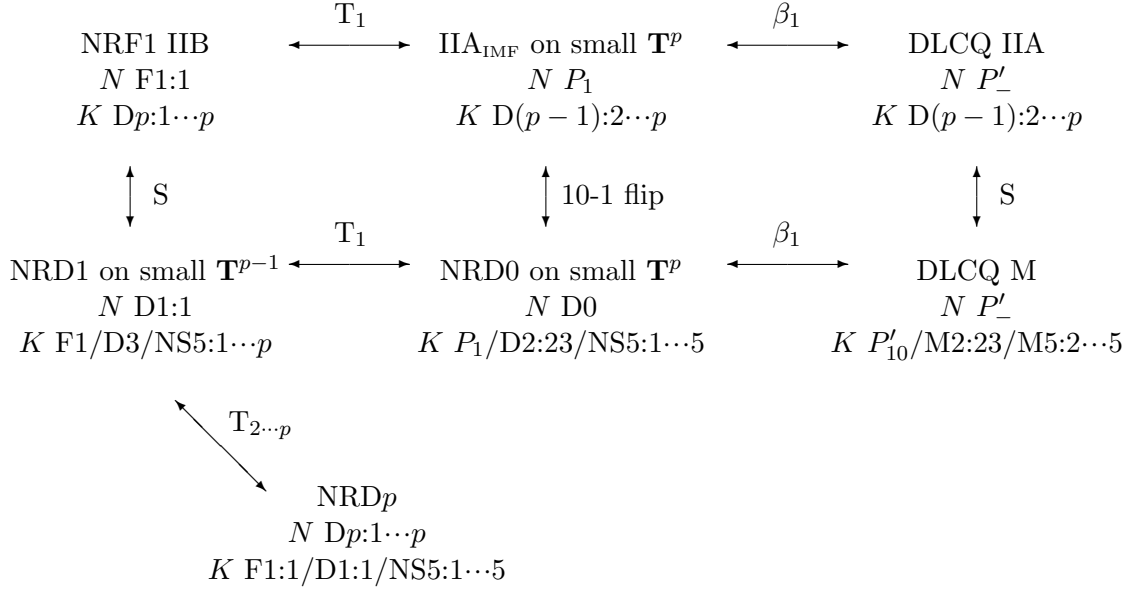
\begin{figure}[htb]
\begin{center}
\begin{picture}(150,45) {\small
\put(0,40){\begin{minipage}[t]{4cm}
\begin{center}
NRF1 IIB \\
$N$ F1:1 \\
$K$ D$p$:1$\cdot\!\cdot\!\cdot p$
\end{center}
\end{minipage}}
\put(0,10){\begin{minipage}[t]{4cm}
\begin{center}
NRD1
on small $\mathbf{T}^{p-1}$\\
$N$ D1:1 \\
$K$ F1/D3/NS5:1$\cdot\!\cdot\!\cdot p$
\end{center}
\end{minipage}}
\put(28,-20){\begin{minipage}[t]{4cm}
\begin{center}
NRD$p$ \\
$N$ D$p$:1$\cdot\!\cdot\!\cdot p$ \\
$K$ F1:1/D1:1/NS5:1$\cdot\!\cdot\!\cdot$5
\end{center}
\end{minipage}}
\put(55,40){\begin{minipage}[t]{4cm} 
\begin{center}
IIA$_{\mbox{\tiny IMF}}$ on small $\mathbf{T}^{p}$\\
$N$ $P_{1}$ \\
$K$ D($p-1$):2$\cdot\!\cdot\!\cdot p$
\end{center}
\end{minipage}}
\put(55,10){\begin{minipage}[t]{4cm}
\begin{center}
NRD$0$ on small $\mathbf{T}^{p}$\\
$N$ D$0$\\
$K$ $P_{1}$/D2:23/NS5:1$\cdot\!\cdot\!\cdot$5
\end{center}
\end{minipage}}
\put(110,40){\begin{minipage}[t]{4cm}
\begin{center}
DLCQ IIA \\
$N$ \Pm \\
$K$ D($p-1$):2$\cdot\!\cdot\!\cdot p$
\end{center}
\end{minipage}}
\put(110,10){\begin{minipage}[t]{4cm}
\begin{center}
DLCQ M \\
$N$ \Pm \\
$K$ $P'_{10}$/M2:23/M5:2$\cdot\!\cdot\!\cdot$5
\end{center}
\end{minipage}}
\put(30,-12){\vector(-1,1){6}}
\put(30,-12){\vector(1,-1){6}}
\put(31,-10){T$_{2\cdot\cdot\cdot p}$}
\put(47,41){\vector(-1,0){8}}
\put(47,41){\vector(+1,0){8}}
\put(45,44){T$_{1}$}
\put(105,41){\vector(-1,0){8}}
\put(105,41){\vector(+1,0){8}}
\put(103,44){$\beta_{1}$}
\put(47,12){\vector(-1,0){8}}
\put(47,12){\vector(+1,0){8}}
\put(45,15){T$_{1}$}
\put(105,12){\vector(-1,0){8}}
\put(105,12){\vector(+1,0){8}}
\put(103,15){$\beta_{1}$}
\put(130,21){\vector(0,-1){4}}
\put(130,21){\vector(0,+1){4}}
\put(133,20){S}
\put(73,21){\vector(0,-1){4}}
\put(73,21){\vector(0,+1){4}}
\put(75,20){10-1 flip}
\put(18,21){\vector(0,-1){4}}
\put(18,21){\vector(0,+1){4}}
\put(21,20){S} }
\end{picture}
\end{center}
\par
\vspace*{2.5cm}
\caption{{\protect\small A portion of the duality web for Type II NR/DLCQ
theories on a transverse $\mathbf{T}^{p-1}$ for $p=1,3,5$, including the
images of $N$ fundamental strings and $K$ longitudinal D$p$-branes, in the
various descriptions. $N$ must be
strictly positive, but $K$ is arbitrary. Non-vanishing-size compactifications are \emph{not}
mentioned. As explained in footnote~\ref{Bfoot}, this figure can be alternatively presented with the second column omitted, implying a direct connection between the first and third columns via longitudinal T duality. See main text for full
 description.}}
\label{fig:dualities}
\end{figure}

Figure~\ref{fig:dualities}, reproduced directly from \cite{Danielsson:2000gi} save for some name relabelings, shows a portion of the duality web in the particular case of NR Type IIB string theory, hereafter NRF1 IIB theory, 
or simply NRF1, for short. The `F1' component of this acronym is of course the standard abbreviation for fundamental string.

The top line of each entry in the figure indicates the name 
of the theory in question,\footnote{References \cite{Danielsson:2000gi} and \cite{Gomis:2000bd} adopted different viewpoints with respect to the content of NR string theory. The former reference followed the convention that is standard for relativistic string theory, considering superselection sectors with different types of solitonic charge as part of a single underlying theory. In that view, NRF1 IIB encompasses \emph{all} of the finite-energy states that result from applying the limit to the \emph{entire} relativistic Type IIB string theory (and likewise for other types of string theories). The authors of \cite{Gomis:2000bd} came at this from a different perspective: given that the superselection sector with longitudinal D$p$-branes was already being explored under the NCOS moniker, and was known to display a \emph{relativistic} spectrum for open strings, they focused almost exclusively on the sector without D-branes, regarded as a theory of nonrelativistic \emph{closed} strings (believed at that point to be non-gravitational). 
As a result of this difference in perspective, the dual theories denoted in \cite{Gomis:2000bd} as `Galilean brane theories' (e.g., GD$p$) referred to a specific subsector of the dual theories that in \cite{Danielsson:2000gi} were denominated `Wrapped brane theories' (e.g., WD$p$). In the present paper we adopt the same perspective as in \cite{Danielsson:2000gi,Danielsson:2000mu}, but use for the dual theories the NR name proposed in \cite{Gomis:2000bd}, which over the years has been the one in common usage (e.g., NRD$p$ here refers to the theory that was named WD$p$ in \cite{Danielsson:2000gi}). Incidentally, these differences in physical perspective might be related to the distinction in  semantic preferences mentioned three paragraphs above. One must bear in mind, however, that the two issues are \emph{not} identical: the more stringent terminology of \cite{Gomis:2000bd} disallows worldsheet boundaries and consequently avoids their associated $\lambda\bar{\lambda}$ deformation \cite{Guijosa:2023qym}, but it certainly allows longitudinally-wound closed strings, which deform the $\sigma$-model in the same manner, taking one away from the pure SNC formalism \cite{Avila:2023aey}.\label{wrappedfoot}}  with the phrase `small $\mathbf{T}^{p}$' referring to a $p$-dimensional torus that is vanishingly small in the limit that defines said theory. The second line identifies the lightest excitation. 
Naturally, when the F1s are transmuted into branes with larger or smaller dimensionality, the number of directions that count as longitudinal increases  or decreases accordingly. The integer $N>0$ refers to the corresponding number of units of longitudinal string winding or brane wrapping. $P_{i}$ stands for Kaluza-Klein (KK) units of momentum along $x^{i}$. 

The third line of the first entry alludes to longitudinal D-branes as examples of other objects that are allowed within NRF1, as long as they carry some number $0<n\leq N$ of units of the total F1 charge. In contrast with $N$ and $n$, the D$p$ charge $K$ can be positive, negative or zero. In other words, even though NRF1 does not contain anti-F1s, it does contain anti-D$p$s. The third line in the remaining entries then shows the corresponding objects in the dual theories. 
 The notation 
X$q$:$i_{1}\cdot\cdot\cdot i_{q-1}$ indicates an X$q$-brane
wrapping the $i_{1},\ldots,i_{q-1}$ cycles of the torus. Triple entries on the third line
apply respectively to the cases $p=1,3,5$. 

\begin{figure}[tb]
\begin{center}
\vspace{0.5cm} 
\begin{picture}(150,75) {\small
\put(0,70){\begin{minipage}[t]{4cm} 
\begin{center}
NRF1 IIA \\
$N$ F1:1 \\
$K$ D$p$:1$\cdot\!\cdot\!\cdot p$
\end{center}
\end{minipage}}
\put(42,70){\begin{minipage}[t]{4cm}
\begin{center}
IIB$_{\mbox{\tiny IMF}}$
on small $\mathbf{T}^{p}$\\
$N$ $P_{1}$ \\
$K$ D($p-1$):2$\cdot\!\cdot\!\cdot p$
\end{center}
\end{minipage}}
\put(83,70){\begin{minipage}[t]{4cm}
\begin{center}
DLCQ IIB \\
$N$ \Pm \\
$K$ D($p-1$):2$\cdot\!\cdot\!\cdot p$
\end{center}
\end{minipage}}
\put(0,40){\begin{minipage}[t]{4cm}
\begin{center}
NRM2 on small $\mathbf{T}^{p-1}$\\
$N$ M2:1(10) \\
$K$ M2:12/M5:1234(10)
\end{center}
\end{minipage}}
\put(35,40){\begin{minipage}[t]{4cm}
\begin{center}
NRF1 IIB \\
$N$ F1:1 \\
$K$ D1:1
\end{center}
\end{minipage}}
\put(77,40){\begin{minipage}[t]{4cm}
\begin{center}
IIA$_{\mbox{\tiny IMF}}$ on small $\mathbf{T}^{p}$ \\
$N$ $P_{1}$ \\
$K$ D0
\end{center}
\end{minipage}}
\put(117,40){\begin{minipage}[t]{4cm}
\begin{center}
DLCQ IIA \\
$N$ \Pm \\
$K$ D0
\end{center}
\end{minipage}}
\put(-2,10){\begin{minipage}[t]{4cm}
\begin{center}
NRD$p$\\
$N$ D$p$:1$\cdot\!\cdot\!\cdot p$ \\
$K$ F1:1
\end{center}
\end{minipage}}
\put(37,10){\begin{minipage}[t]{4cm}
\begin{center}
NRD1 on small $\mathbf{T}^{p-1}$\\
$N$ D1:1\\
$K$ F1:1
\end{center}
\end{minipage}}
\put(80,10){\begin{minipage}[t]{4cm}
\begin{center}
NRD0 on small $\mathbf{T}^{p}$ \\
$N$ D0 \\
$K$ $P_{1}$
\end{center}
\end{minipage}}
\put(118,10){\begin{minipage}[t]{4cm}
\begin{center}
DLCQ M \\
$N$ \Pm \\
$K$ $P_{10}$
\end{center}
\end{minipage}}
\put(38,72){\vector(-1,0){6}}
\put(38,72){\vector(+1,0){6}}
\put(36,75){T$_{1}$}
\put(86,72){\vector(-1,0){6}}
\put(86,72){\vector(+1,0){6}}
\put(84,75){$\beta_{1}$}
\put(18,51){\vector(0,-1){4}}
\put(18,51){\vector(0,+1){4}}
\put(21,50){S}
\put(39,51){\vector(-1,1){6}}
\put(39,51){\vector(1,-1){6}}
\put(40,53){T$_{2\cdot\!\cdot\!\cdot p}$}
\put(77,51){\vector(-1,1){6}}
\put(77,51){\vector(1,-1){6}}
\put(78,53){T$_{2\cdot\!\cdot\!\cdot p}$}
\put(122,51){\vector(-1,1){6}}
\put(122,51){\vector(1,-1){6}}
\put(123,53){T$_{2\cdot\!\cdot\!\cdot p}$}
\put(72,41){\vector(-1,0){6}}
\put(72,41){\vector(+1,0){6}}
\put(70,36){T$_{1}$}
\put(120,41){\vector(-1,0){6}}
\put(120,41){\vector(+1,0){6}}
\put(118,36){$\beta_{1}$}
\put(18,19){\line(0,1){1}}
\put(18,21){\line(0,1){1}}
\put(18,18){\vector(0,-1){2}}
\put(18,23){\vector(0,+1){2}}
\put(21,20){S}
\put(57,21){\vector(0,-1){4}}
\put(57,21){\vector(0,+1){4}}
\put(60,20){S}
\put(98,21){\vector(0,-1){4}}
\put(98,21){\vector(0,+1){4}}
\put(101,20){1-10 flip}
\put(137,21){\vector(0,-1){4}}
\put(137,21){\vector(0,+1){4}}
\put(140,20){S}
\put(32,11){\vector(-1,0){5}}
\put(32,11){\vector(+1,0){5}}
\put(28,14){T$_{2\cdot\cdot\cdot p}$}
\put(79,11){\vector(-1,0){4}}
\put(79,11){\vector(+1,0){4}}
\put(77,14){T$_{1}$}
\put(123,11){\vector(-1,0){6}}
\put(123,11){\vector(+1,0){6}}
\put(121,14){$\beta_{1}$}
}
\end{picture}
\end{center}
\par
\vspace*{-0.5cm}
\caption{{\protect\small Same as Fig.~\ref{fig:dualities}, for $p=2,4$. 
The two figures are related through transverse
T-duality. Again, compact directions are not mentioned unless
they have vanishing size in the relevant metric.
Double entries for the object with multiplicity $K$ apply
respectively to the $p=2,4$ cases. The dotted arrow connecting NRM2 to 
NRD$p$ applies only in the $p=2$ case.} \label{fig:dualities2} }
\end{figure}
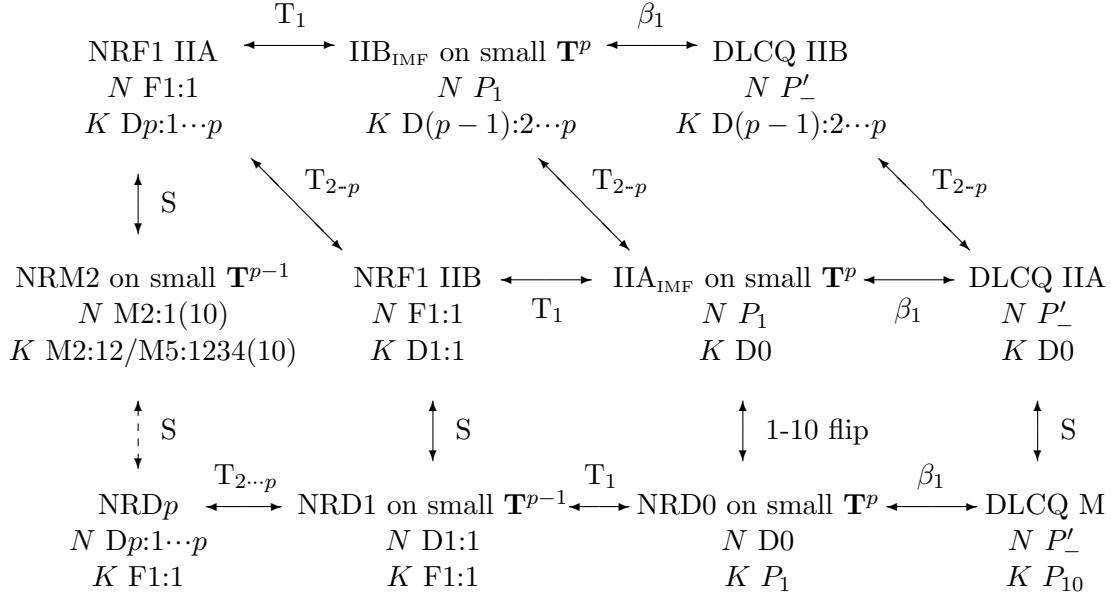

The arrows of the figure labeled  T$_1$ and T$_{2\cdot\cdot\cdot p}$ refer to T duality along the directions indicated by the subindices, those labeled S  refer to S duality, and the ones with 
$\protect\beta_{1}$ are large boosts along $x^{1}$.\footnote{As in \cite{Danielsson:2000gi}, where the near-critical $B$ field was shown to be optional (see footnote \ref{nobfoot} below), Fig.~\ref{fig:dualities} takes NRF1 IIB to have been defined with $B=0$. The theory that is longitudinally-T-dual to it, labeled IIA$_{\mbox{\tiny IMF}}$ in Fig.~\ref{fig:dualities}, is then IIA in a version of the infinite momentum frame (IMF) with a shrinking longitudinal circle. (The longitudinal momentum here is infinite precisely because the size of the circle vanishes while the number $N$ of KK units remains fixed.) This IMF description becomes DLCQ IIA after applying a boost with parameter $\beta_1\to 1$.  This boost is equivalent \cite{Danielsson:2000gi} to switching on the near-critical $B$ field in NRF1 IIB, so that alternative presentation of the NR string theory is \emph{directly} T$_{1}$-dual to DLCQ IIA (this would amount to removing the central column of  Fig.~\ref{fig:dualities}).  Ditto for NRF1 IIA vs.~DLCQ IIB in Fig.~\ref{fig:dualities2}. Having explained what the IMF subindex refers to, we note that the theory labeled NRD0 could alternatively be denominated M$_{\mbox{\tiny IMF}}$, because it is precisely M theory in the infinite momentum frame (on a vanishingly small spatial $p$-torus).\label{Bfoot}} 

Figure~\ref{fig:dualities2}, related to Fig.~\ref{fig:dualities} through transverse T duality and also taken from \cite{Danielsson:2000gi}, displays the corresponding portion of the web for the cases $p=2,4$. 

A particularly noteworthy feature emphasized in \cite{Danielsson:2000gi} is that the arguments \cite{Seiberg:1997ad,Sen:1997we} that established that $U(N)$ maximally supersymmetric Yang-Mills (MSYM) in $p+1$ dimensions does provide a non-perturbative formulation \cite{Banks:1996vh,Susskind:1997cw} for the discrete light-cone quantization (DLCQ)\footnote{Compactification on a lightlike circle can be problematic for a field theory \cite{Hellerman:1997yu}, unlike what happens in string theory on account of the basic excitations being extended objects \cite{Guijosa:1998rq,Bilal:1998ys}. The connection with NR string theory established in Figs.~\ref{fig:dualities} and \ref{fig:dualities2} ensures that the DLCQ of string/M theory is well defined.} of M theory on a transverse $\mathbf{T}^{p}$, and that likewise Matrix string theory \cite{Dijkgraaf:1997vv} fully describes DLCQ IIA string theory on a transverse $\mathbf{T}^{p-1}$, started from DLCQ M (respectively DLCQ IIA), followed the horizontal
arrows in the middle (top) line
of Fig.~\ref{fig:dualities}, and then proceeded down and
diagonally to arrive at the NRD$p$ theory. For $p=2,4$, the argument starts at the lower-right corner of Fig.~\ref{fig:dualities2} and proceeds to the left along the bottom line.
These connections (noted also in \cite{Hyun:2000cw}) entail that the action for D$p$-branes within NRD$p$ theory is precisely the corresponding Matrix theory action, i.e.,  MSYM in $p+1$ dimensions \cite{Witten:1995im,Banks:1996vh,Susskind:1997cw,Seiberg:1997ad}.\footnote{On the other hand, as emphasized in \cite{Danielsson:2000gi}, NRD$p$ theory \emph{as a whole} is not synonymous with $(p+1)$-dimensional MSYM.}  

Various other theories of interest are included within or beyond the portions of the duality web displayed in these two figures. For instance, open membrane (OM) theory \cite{Gopakumar:2000ep} is the worldvolume theory of M5 branes in NRM2 theory \cite{Danielsson:2000gi,Garcia:2002fa}, 
little string theory (LST) \cite{Seiberg:1997zk,Losev:1997hx,Aharony:2019zsx} is the worldvolume theory of NS5-branes in NRNS5 theory \cite{Danielsson:2000gi}, and $(p+1)$-dimensional noncommutative Yang-Mills (NCYM) theory \cite{Sheikh-Jabbari:1997qke,Sheikh-Jabbari:1998aur,Seiberg:1999vs} is the worldvolume theory for D$p$-branes in NRD($p-2$) theory \cite{Danielsson:2000gi,Danielsson:2000mu}.
It was in seeking a space-time generalization of the space-space noncommutativity displayed by NCYM that the NCOS theories were discovered \cite{Seiberg:2000ms,Gopakumar:2000na}, with the surprise that they were stringy. 
Over the years, the web of dualities has been further explored and extended \cite{Bergshoeff:2018yvt,Blair:2020ops,Bergshoeff:2022iss,Blair:2023noj,Gomis:2023eav,Fontanella:2024rvn,Lambert:2024uue,Lambert:2024yjk,Fontanella:2024kyl,Blair:2024aqz,Lambert:2024ncn},
leading to useful connections to yet 
 more theories, such as Carroll strings \cite{Cardona:2016ytk,Bagchi:2023cfp,Harksen:2024bnh,Bagchi:2024rje}, spin matrix theory \cite{Harmark:2014mpa} and tensionless strings \cite{Lindstrom:1990qb,Lindstrom:1990ar,Isberg:1992ia,Bagchi:2013bga,Bagchi:2015nca,Bagchi:2016yyf}, and to decoupled subsectors of the AdS/CFT correspondence \cite{Gomis:2005pg,Sakaguchi:2007ba,Harmark:2017rpg,Harmark:2018cdl,Harmark:2020vll}. 

Crucial for us is the very recent, remarkable paper \cite{Blair:2024aqz}, where among many interesting results obtained by following up on \cite{Blair:2020ops,Blair:2023noj,Gomis:2023eav}, it was  discovered that {\bf the NRD$p$, NRM2 and NRM5 theories of \cite{Danielsson:2000gi,Gomis:2000bd} 
are precisely the setting of the most familiar examples of the holographic correspondence \cite{Maldacena:1997re,Itzhaki:1998dd}.} To be more precise, let us refer in turn to the two sides of the AdS/CFT correspondence. In the field theory side, due to the above-reviewed relation to Matrix theory  
(on the basis of which the authors of \cite{Blair:2024aqz} used for the NRD$p$ theories the alternative name {\bf M$p$T}), the additional connection with holography was known since the very inception of NR brane theories in \cite{Danielsson:2000gi,Gomis:2000bd}. And indeed, the fact that the \emph{same} MSYM theories played a certain role in Matrix theory and a somewhat similar role in holography gave rise to various attempts to clarify the relation between the Matrix theory and AdS/CFT conjectures--- see, e.g., \cite{Balasubramanian:1997kd,Hyun:1997zt,Hyun:1998bi,Hyun:1998iq,deAlwis:1998ki,Polchinski:1999br}. The surprising discovery made in \cite{Blair:2024aqz} is that the NR brane theories also contain \emph{the gravity side of holographic duality}, because {\bf the NR brane limit is exactly the same as Maldacena's near-horizon limit}. Oddly enough, this highly consequential fact had been overlooked for nearly a quarter century. A number of valuable implications have already been worked out in \cite{Blair:2024aqz}.

\subsection{Outline and main results}
\label{outlinesubsec}

In this paper, 
we will combine the first and second strands we just reviewed, to show that the  results of \cite{Avila:2023aey,Guijosa:2023qym} throw useful light on the more recent discovery in \cite{Blair:2024aqz}, allowing us to gain some further insight into the underlying \emph{nonrelativistic} nature of relativistic holography. The outline and summary of the paper are as follows. 

In Section~\ref{stringsec}, we provide a short overview of nonrelativistic string theory, starting with the flat-space formulation of \cite{Danielsson:2000gi,Gomis:2000bd,Danielsson:2000mu} in Section~\ref{flatsubsec}, where we focus on the closed-string sector, displaying the resulting nonrelativistic spectrum, Eq.~(\ref{closednrf1}). Section~\ref{dbranesubsec} discusses D-branes of the two varieties that are present in NRF1 theory, longitudinal and transverse, as well as their associated open-string sectors. Significant for our story is the long-known fact that open strings on longitudinal D-branes have a \emph{relativistic} spectrum, seen in (\ref{openlong}). 

Section~\ref{sncsubsec} summarizes the basic elements of the string Newton-Cartan formalism. 

The last two subsections of Section~\ref{stringsec}
recall the main results of the recent works \cite{Guijosa:2023qym,Avila:2023aey}, which are vital for our reasoning in this paper. The central lesson is that, generally, black branes in NR string theory are \emph{not} SNC geometries. Instead, they are \emph{asymptotically} SNC, and interpolate between an inner curved Lorentzian region and an outer flat SNC region. 
This is exemplified in Section~\ref{blackstringsubsec} by the black string (\ref{blackstring}), and in Section~\ref{blackpbranesubsec} by the longitudinal RR black $p$-brane (\ref{longdp}). 

In both cases, the physical interpretation, 
schematized in Figs.~\ref{fig:blackstring} and \ref{fig:blackbraneinnrf1},
 is that the interactions of the corresponding stack of strings or of (F1-carrying) D-branes have an alternative, geometric description in terms of a tubular bubble within which physics is \emph{relativistic}. In other words, the F1s source the previously mentioned $\lambda\bar{\lambda}$ deformation, with a position-dependent coefficient that switches off as one moves away from the source. The resulting relativistic bubble has a radial width set by the size of the string or D-brane bound state,\footnote{As we indicate below, this radial size is much larger than the string length, and therefore geometrically reliable, only when the total F1 charge $N$ or charge density $\nu$ is sufficiently large.} and is translationally invariant along the dimensions spanned by said stack. As we emphasize, the relativistic character of \emph{closed} strings within the bubble in the longitudinal RR black brane is the direct counterpart of the known relativistic character of the \emph{open} strings associated with the longitudinal D-brane stack. In both descriptions, the strings are excitations of the bubble/stack, not of the surrounding flat SNC space, and it is for this reason that they cannot exist without their host bubble/stack. 
 
 We must stress that our description of the background as a relativistic bubble embedded in flat nonrelativistic spacetime is only a useful schematic. For a more accurate account, one must bear in mind that there is an intermediate throat region, such that the passage from relativistic to nonrelativistic physics is completely smooth. The question of where the inner relativistic bubble ends and the outer Newton-Cartan spacetime begins is then a matter of degree: we can choose the transition point depending on how good an approximation we wish the flat SNC description to be, i.e., how small we wish to keep the coefficient of the  $\lambda\bar{\lambda}$ deformation. This schematic naturally accompanies any form of asymptotic structure. For instance, asymptotically flat spacetimes are curved everywhere, yet it is useful to conceptualize them as having an outer flat region. Similarly, asymptotically SNC geometries remain Lorentzian throughout, but they are conveniently described as possessing an outer string Newton-Cartan region.
 
 A corollary of these findings is that purely SNC backgrounds, which in themselves are coherent states of off-shell \emph{unwound} strings (i.e., Newtonian gravitons \cite{Danielsson:2000mu}), describe the allowed geometries in NR string theory \emph{in the absence of wound strings}. More precisely, when we place on a SNC background a grouping of objects carrying positive longitudinal string winding, we can describe the system either in terms of the objects' intrinsic dynamics, or alternatively (when the total F1 charge is large enough), as an asymptotically SNC background, which will have at its core an emergent relativistic bubble of the type discovered in \cite{Guijosa:2023qym,Avila:2023aey}. 
 
 In this second description, the emergence of the bubble implies that the full background is a solution of the equations of motion of \emph{relativistic} supergravity, but this does \emph{not} mean that we are suddenly dealing with a state of the original parent relativistic string theory. The crucial point is that a particular theory is defined not just by its action or equations of motion, but also by its \emph{boundary conditions}.\footnote{We know very well that the marked differences in the asymptopia of, e.g., Minkowski, de Sitter and anti-de Sitter spacetimes are of great consequence for the nature of the theories formulated on them.} 
 In the black string (\ref{blackstring}) and the RR black brane (\ref{longdp}), it is the external flat SNC region that determines the asymptotic states of the theory. {}From a nonlinear-$\sigma$-model perspective, the fact that the coefficient of the $\lambda\bar{\lambda}$ term switches off asymptotically is physically significant: it implies that the most useful description of the physics of the outer region is as a small deformation of the 10-dimensional flat SNC geometry. 
 
 In Section~\ref{nrbranesubsec} we collect the main ideas and formulas relevant after the passage, via U-duality as in Figs.~\ref{fig:dualities} and \ref{fig:dualities2}, from NR string theory to NR \emph{brane} theories. This includes the further generalization of Newton-Cartan to $p$-brane Newton-Cartan ($p$NC) \cite{Kluson:2017abm,Pereniguez:2019eoq,Blair:2021waq,Ebert:2021mfu,Novosad:2021tlq,Bergshoeff:2023rkk,Ebert:2023hba,Bergshoeff:2024ipq}. Motivated by the connections discussed earlier in this Introduction, we concentrate specifically on nonrelativistic D$p$-brane (NRD$p$) theories. As indicated in Figs.~\ref{fig:dualities} and \ref{fig:dualities2}, the most salient features of NRD$p$ theory are the role of longitudinally-positively-wrapped D$p$-branes as the lightest excitations, the absence of anti-D$p$s, and the nonrelativistic character of physics along the $9-p$ transverse spatial dimensions.  
 
 Section~\ref{holographysubsec} then describes the second key insight for this paper: the very surprising discovery of \cite{Blair:2024aqz}\footnote{In the present paper we will only employ this particular finding of \cite{Blair:2024aqz}, but we wish to stress again that that paper contains many other valuable results, and is therefore a highly recommendable read.} that the limit (\ref{nrdplimit}) that defines NR D$p$-brane theory is exactly the same as the near-horizon limit that leads to the gravitational description in D$p$-based \emph{relativistic} holography. By virtue of having absorbed in Sections~\ref{blackstringsubsec} and \ref{blackpbranesubsec} the lessons of \cite{Guijosa:2023qym,Avila:2023aey}, we are able to present this crucial finding of \cite{Blair:2024aqz} from that other, complementary
 perspective. We are led to conclude that, for each value of $p$, \emph{the} AdS$_{p+2}\times$S$^{8-p}$ \emph{near-horizon geometry} (\ref{nrblackbrane}) \emph{is nothing more and nothing less than the RR black $p$-brane in NRD$p$ theory}.   

Armed with these elements, in Section~\ref{nrinrsec} we develop and further substantiate this new perspective. As discussed in and around Eqs.~(\ref{polchinski})-(\ref{polchinskip}), our central result is that {\bf relativistic D$p$-based holographic duality is exactly synonymous with open-string/closed-string duality \emph{within} nonrelativistic D$p$-brane theory}, i.e., with the statement (\ref{polchinskip}) that a stack of $N$ D$p$-branes embedded in 10-dimensional flat $p$NC spacetime has an alternative description in terms of the 10-dimensional, asymptotically flat $p$NC RR black brane.\footnote{The corresponding statements hold as well for relativistic M2-, M5- or NS5-based holographic duality \cite{Maldacena:1997re,Itzhaki:1998dd,Aharony:1998ub}, and for the multi-near-horizon limits examined recently in \cite{Blair:2024aqz}, an example of which is discussed in Section~\ref{ads3subsec} below.}

In Section~\ref{awfulsubsec} we elaborate on a particular aspect of this statement, exemplifying with the case $p=3$ from that point on: the fact that the near-horizon/NRD3 limit does \emph{not} in fact get rid of the 10-dimensional (asymptotically) flat region, but merely transmutes its character from Lorentzian to Newton-Cartan. As indicated in (\ref{polchinskip}) and (\ref{polchinski3}), the theory remains 10-dimensional on \emph{both} sides of the duality. See Fig.~\ref{fig:ads}. This is further discussed for the open-string or field-theory side (i.e., (3+1)-dimensional MSYM) in Section~\ref{trulysubsec}, where the 10-dimensional 3NC structure is seen to provide a natural language to jointly consider, within a seemingly $(3+1)$-dimensional theory, the spacetime metric and the internal space metric. 
Additional verification of the overall picture is provided by a D3-brane probe in Section~\ref{d3subsec} (see Fig.~\ref{fig:d3probe}), and a D3-F1 or D3-D1 probe in Section~\ref{f1d1subsec}.

 The following two subsections describe a couple of generalizations. Section~\ref{ads3subsec} explains that the similarity (pointed out already in \cite{Danielsson:2000mu}) with the characteristics of short versus long strings in AdS$_3$ is not a mere analogy, but a complete identity. Section~\ref{nonflatsubsec} emphasizes that one can equally well consider asymptotically \emph{non}-flat $p$NC geometries, and AdS$_5\times X^5$ backgrounds with $X^5\neq\mathrm{S}^5$, such as \cite{Klebanov:1998hh}, provide a class of examples. 

Section~\ref{indeedsubsec} emphasizes the fact, already mentioned above for the black string in NRF1, that a theory is defined by its equations of motion in combination with its boundary conditions. For the NR D3-brane system, the relativistic equations of motion in  AdS$_5\times$S$^5$ are supplemented by the \emph{nonrelativistic} physics enforced by the asymptopia. 
 
 Section~\ref{cmsubsec} makes the simple point that, from the new perspective developed in this paper, it is more natural to consider the MSYM gauge group to be $U(N)$ rather than $SU(N)$. Knowing that the surrounding flat space region is in fact \emph{not} lost in the near-horizon/NRD3 limit, it becomes desirable to retain the $U(1)$ factor that describes the center-of-mass motion of the D3-brane stack within the ambient spacetime. Up to now, this $U(1)$ has been normally omitted from the AdS/CFT statement of equivalence, as in (\ref{maldacena}).   

In Section~\ref{disassemblingsubsec}, we discuss the well-known option of separating the D3-branes into more than one stack. In the open-string description, this is of course the  Coulomb branch of vacua of $3+1$ MSYM. In the closed-string description, it amounts to considering a correspondingly multi-centered black 3-brane. Around Eq.~(\ref{midpointbkg}), we find that, just as expected, each one of the centers in the black brane is surrounded by a relativistic bubble, dual to the corresponding D3 stack, and that upon moving these stacks/bubbles far from one another, we are left precisely with the underlying 10-dimensional flat 3NC geometry, now at intermediate rather than asymptotic locations in the spacetime. See Fig.~\ref{fig:2bubbles}. In other words, the AdS$_5\times$S$^5$ spacetime can be disassembled into chunks of Lorentzian geometry,  moving about on the backdrop of Newton-Cartan geometry. 

This picture leads naturally to the idea, presented in Section~\ref{d3scatsubsec}, of scattering these chunks of relativistic spacetime off each other, i.e., scattering the D3 stacks off one another. As recalled a few pages back in this Introduction, this is precisely the type of calculation that was considered in the heyday of Matrix theory \cite{Bachas:1995kx,Lifschytz:1996iq,Douglas:1996yp,Becker:1997wh,Becker:1997xw,Dine:1997sz,Becker:1997cp,Buchbinder:2001ui,Taylor:2001vb}, because of its dual interpretation in terms of graviton scattering in DLCQ M theory. 
Up to now, D-brane scattering amplitudes have not been emphasized as fundamental observables in AdS/CFT. They provide a direct point of contact between AdS/CFT and Matrix theory, along the lines anticipated in \cite{Polchinski:1999br}.  

We are then prompted to wonder about the meaning, within the new NRD3 perspective, of the familiar GKPW recipe \cite{Gubser:1998bc,Witten:1998qj} for MSYM correlation functions. In Section~\ref{gravitonscatsubsec}, we see that it computes scattering of the Newtonian gravitons of the NRD3 theory off the D3-branes. 

Finally, in Section~\ref{entanglesubsec} we motivate the notion that the usual Ryu-Takayanagi holographic entanglement entropy formula \cite{Ryu:2006bv,Ryu:2006ef} can be used for computations directly carried out on the pure 10-dimensional flat 3NC geometry, or in other words, in $U(N)$ MSYM with $N=0$. All such entropies correctly vanish, as expected from the complete lack of degrees of freedom. This shows that the non-Lorentzian geometry is not in itself emergent. The contrast between RT surfaces in AdS$_\times$S$^5$ and in purely flat 3NC spacetime is illustrated in Fig.~\ref{fig:entanglement}. {}From this we conclude that the {\bf flat 3-brane Newton-Cartan geometry is the 10-dimensional underlying structure on top of which entanglement of the D3-brane degrees of freedom builds the \emph{relativistic} 10-dimensional AdS$_5\times$S$^5$ spacetime}. As follows from the relation to NRF1 theory,  
the backreaction of the D3s turns flat 3NC into AdS$_5\times$S$^5$ through a particular deformation of the MSYM action, involving a term that is the T$_{23}\,\circ$\,S-dual image of the above-mentioned $\lambda\bar{\lambda}$ deformation of NRF1 theory. Because of the connection discovered in \cite{Blair:2020ops}, in all NR brane theories this effect can be identified with the $T\bar{T}$ deformation. The specifics have been explored in some detail in \cite{Blair:2024aqz}.

Altogether, then, we find that the results of \cite{Danielsson:2000mu,Guijosa:2023qym,Avila:2023aey}
and \cite{Blair:2020ops,Blair:2023noj,Blair:2024aqz} converge nicely, in a quite unexpected manner. The ensuing approach brings together many seemingly disparate ideas, to provide a fresh and insightful perspective on familiar, relativistic holography, with many possible extensions to novel relativistic and nonrelativistic examples (a task that was started already in \cite{Blair:2024aqz}). More work is surely called for to better appreciate and profit from this new perspective.


\section{Brief Review of Nonrelativistic String Theory} \label{stringsec}

We will now review some of the specifics of NRF1, using mostly the notation in \cite{Blair:2024aqz}, for ease of comparison. 

\subsection{Flat-space NRF1}
\label{flatsubsec}
Consider a relativistic string theory on 10-dimensional Minkowski spacetime. A NR version can be obtained for any one of the familiar types of string theories, but in this paper we will concentrate specifically on Type IIA and IIB. 
Following recent usage, we will employ hatted and unhatted  variables to refer respectively to quantities in the parent relativistic and the limiting nonrelativistic string theory.\footnote{In  \cite{Blair:2024aqz} this distinction is made with uppercase and lowercase variables. However, we prefer to reserve case changes for more standard uses, such as distinguishing between target space coordinates $x$ and the corresponding embedding fields $X$, between an induced metric $g$ and the corresponding ambient metric $G$, etc.}
The limit will involve a $2+8$ foliation of the spacetime, that distinguishes between `longitudinal' and `transverse' directions 
\begin{equation}
\hat{x}^{A}\equiv(\hat{t},\hat{x}_1)~,
\qquad
\hat{x}^{A'}\equiv(\hat{x}_2,\ldots,\hat{x}_9)~.
\label{f1foliation}
\end{equation} 
Along the longitudinal spatial direction, for convenience we turn on a constant critical Kalb-Ramond field $\hat{B}_{01}$, and compactify on a circle of radius $\hat{R}$.\footnote{The physical content of the theory in the limit (\ref{nrf1limit}) is most clearly understood with this compactification. Once we understand this content, we can certainly decompactify by scaling to infinity the radius $R$ defined in (\ref{nrf1limit}).} Fundamental strings are then able to wind around the longitudinal circle any number of times, $w\in\mathbf{Z}$. Knowing that string winding is conserved, we can describe the full theory as the union of superselection sectors with different values of the \emph{total} longitudinal winding $\hat{N}$.  Altogether, our starting background is 
\begin{equation}
\hat{G}_{\mu\nu}=\eta_{\mu\nu}~,\quad
\hat{B}=d\hat{t}\wedge d\hat{x}_1~,\quad
\hat{x}_1\simeq \hat{x}_1+2\pi\hat{R}~,
\label{prelimitbkg}
\end{equation}
with string coupling $\hat{g}_s$ and string length $\hat{l}_s$. 
We then apply the NR string limit \cite{Danielsson:2000gi,Gomis:2000bd}
\begin{equation}
\begin{aligned}
\omega \to \infty, & \quad \mbox{with}
\\
\hat{x}^A &= \omega\, x^A 
\quad (A=0,1)~,
\\
\hat{x}^{A'}&=x^{A'}
\quad (A'=2,\ldots,9)~,
\\
\hat{g}_s&=\omega\, g_s~,
\quad
\hat{l}_s=l_s~,
\\
\hat{N}&=N~, 
\quad
\hat{R}=\omega R~.
\end{aligned}
\label{nrf1limit}
\end{equation}
The dimensionless parameter $\omega$ that implements the limit in (\ref{nrf1limit}) was first employed in \cite{Gomis:2004pw}, and is related to the dimensionless parameter $\delta\to 0$ of the original references \cite{Danielsson:2000gi,Danielsson:2000mu} through $\omega\equiv 1/\sqrt{\delta}$. The coordinate scaling in (\ref{nrf1limit}) turns (\ref{prelimitbkg}) into\footnote{As explained in \cite{Danielsson:2000gi}, even though introducing a critical $B$ field is a very convenient way to subtract the divergent tensional energy proportional to longitudinal string winding $w$, it is not strictly necessary, because the same decoupling of the $w>0$ sector occurs even if $B=0$. Turning on that $\cO(\omega^2)$ piece in $B$ is in fact T$_1$-dual to performing the boost $\beta_1$ in the DLCQ story \cite{Seiberg:1997ad}, depicted in Figs.~\ref{fig:dualities} and \ref{fig:dualities2}. Through U duality, this optional nature of the near-critical $B$ field in NRF1 naturally extends to each of the dual NR brane theories in the figures, and beyond.\label{nobfoot}}  
\begin{equation}
\begin{aligned}
\widehat{ds}^2&= \omega^{2}\left(-dt^2 + dx_1^2\right) + \left(dx_2^2+\ldots+dx_9^2\right)~,
\\
\hat{B}&=\omega^2\, dt\wedge dx_1~,
\\
x_1&\simeq x_1+2\pi R~,
\end{aligned}
\label{limitbkg}
\end{equation}
so we could equally well consider this scaling of the background part of the definition of the NR limit.  
As is customary in analyses of recent years, the choice of units here is such that the string length is held fixed, in contrast with the $l_s\propto\omega^{-1/4}\to 0$ convention in \cite{Danielsson:2000gi,Gomis:2000bd,Danielsson:2000mu}. But of course, there is agreement between both presentations with respect to the dimensionless ratios that control the physics of the limit, such as $\widehat{ds}^2/l_s^2$.
Beyond the $\cO(\omega^2)$ term in $\hat{B}_{01}$, one can include a constant finite piece $B_{01}$ that is freely adjustable \cite{Danielsson:2000mu}.

As seen in (\ref{limitbkg}), the physical effect of the limit is to pull the spacetime metric apart  into separate longitudinal and transverse pieces $\eta_{AB},\delta_{A'B'}$. 
Together, these structures and the $(x^A,x^{A'})$ coordinates describe the geometry inhabited by the flat NRF1 theory.  {}From the perspective of the parent string theory,  $\omega\to\infty$ opens up and completely flattens the lightcone in the eight transverse directions, while maintaining it at 45 degrees in the longitudinal direction. This is the origin of the NR character of the transverse dynamics. 

\vspace*{-0.4cm}
\begin{figure}[thb]
\begin{picture}(150,50) 
  \centering
  \includegraphics[width=7.0cm]{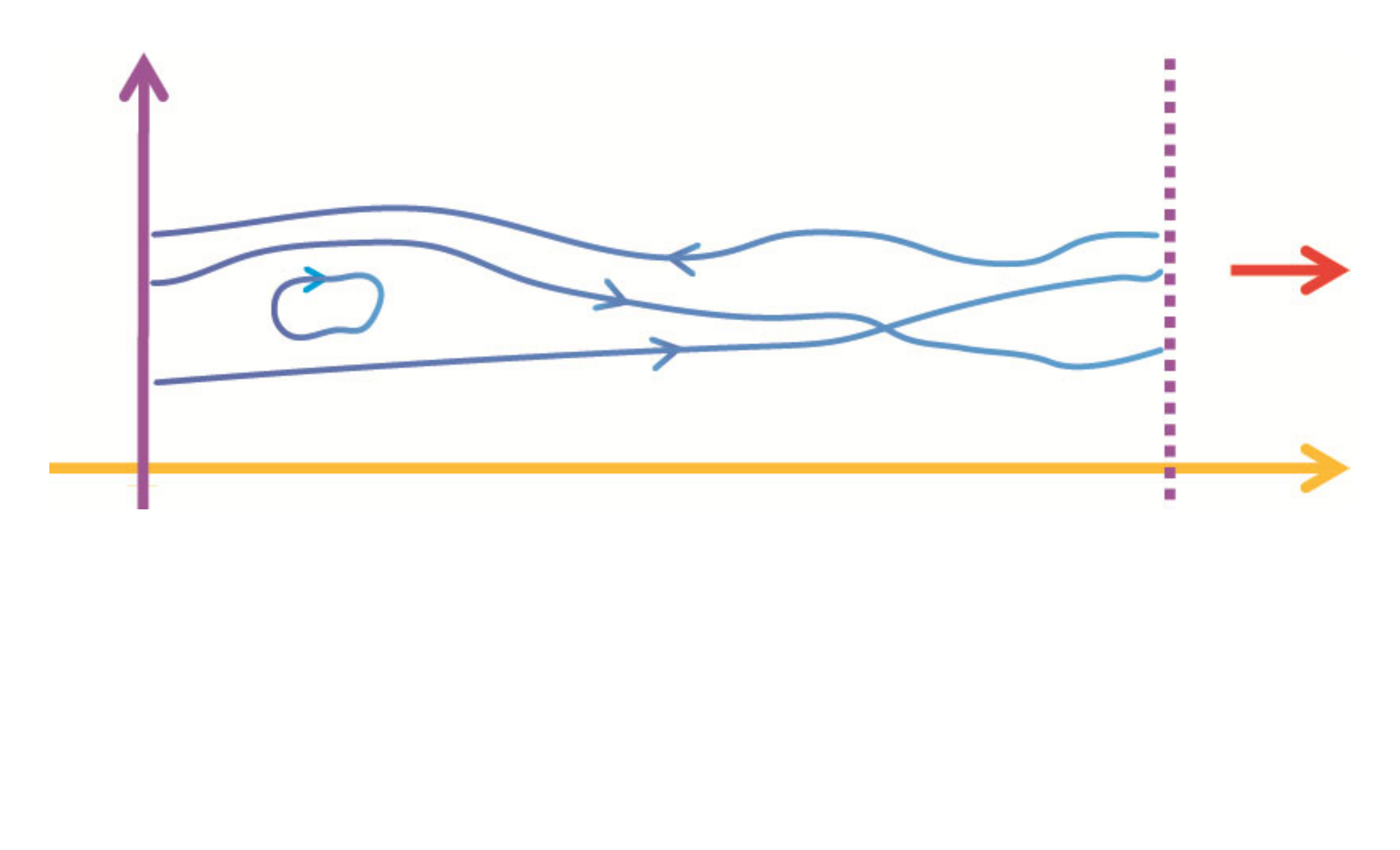}
  \hspace*{0.5cm}
  \includegraphics[width=7.0cm]{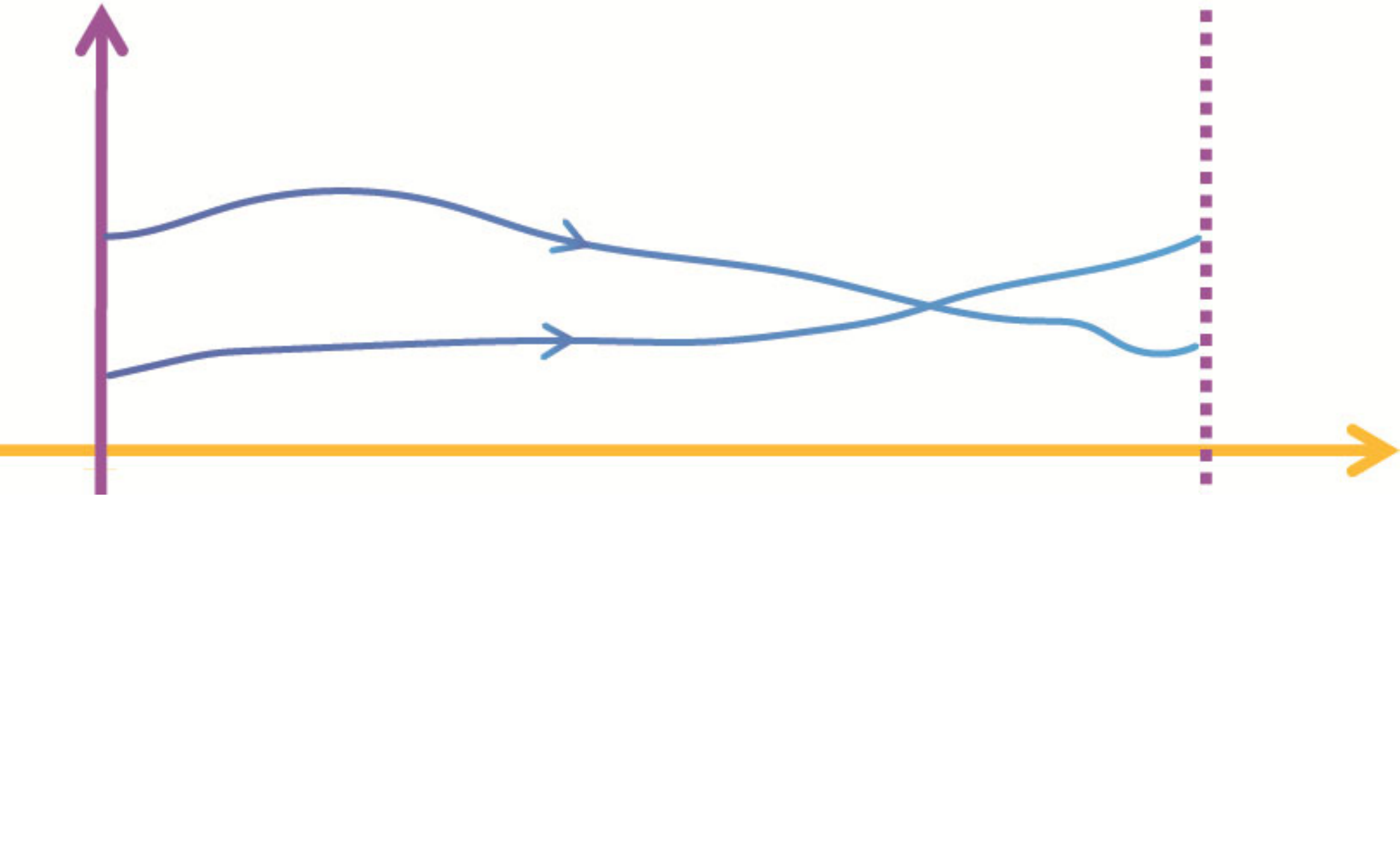}
  \put(-80,18){$\hat{x}_1$}
  \put(1,19){$x_1$}
  \put(-147,42){$\hat{x}_2,\ldots,\hat{x}_9$}
  \put(-71,44){$x_2,\ldots,x_9$}
  \put(-87,32){\color{red}$\hat{B}_{01}$}
  \put(-93,14){$2\pi\hat{R}$}
  \put(-13,15){$2\pi R$}
 \end{picture}
\vspace*{-1.5cm}
  \caption{Truncation of the closed string spectrum induced by the nonrelativistic string limit (\ref{nrf1limit}). Left: In the parent, relativistic string theory, fundamental strings can have arbitrary winding number along the compact longitudinal direction (the image illustrates the cases $w=-1,2,0$). Right: in the nonrelativistic string theory that results from taking the limit, strings must have positive longitudinal winding (such as $w=2$ in the example of the image).}
  \label{fig:nrf1} 
  \end{figure}
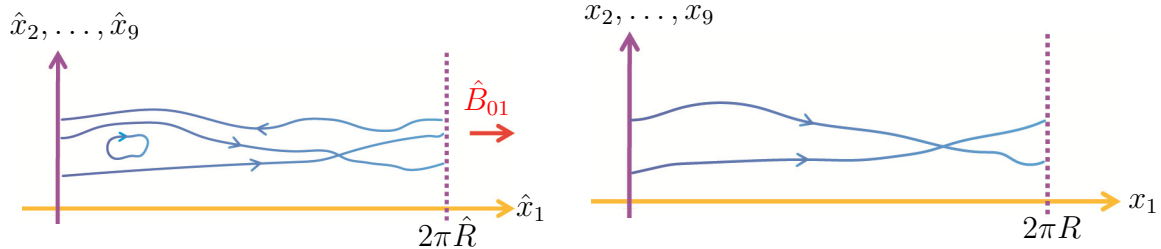

As mentioned in the Introduction and schematized in Fig.~\ref{fig:nrf1}, the NR limit (\ref{nrf1bkg}) scales up the longitudinal components of the metric and Kalb-Ramond fields in such a way as to produce infinite tensional and electric energy costs for fundamental strings, which cancel against one another only when the strings are oriented along the positive direction of the longitudinal axis. For this reason, closed string states generically survive the limit only if they carry positive longitudinal winding \cite{Klebanov:2000pp,Danielsson:2000gi,Gomis:2000bd}.
The energy of a closed string with transverse momentum $\vec{p}_{\perp}$, 
longitudinal winding $w>0$, $m\in\mathbf{Z}$ units of longitudinal momentum, and left- and right-excitation levels N$_{\mbox{\tiny{L}}}$ and N$_{\mbox{\tiny{R}}}$ is given by \cite{Klebanov:2000pp,Danielsson:2000gi,Gomis:2000bd}
\begin{equation}
p_{0}=\frac{l_s^2\, \vec{p}_{\perp}^{\,2}} {2wR}+\frac{\mbox{N}_{\mbox{\tiny{L}}}+\mbox{N}_{\mbox{\tiny{R}}}}{wR}~, \qquad \mbox{N}_{\mbox{\tiny{L}}}-\mbox{N}_{\mbox{\tiny{R}}}=mw~.
\label{closednrf1}
\end{equation}
We have chosen to omit here the (finite) tensional contribution to the string energy, proportional to $\mu\equiv wR/\ls^2$. As explained in \cite{Danielsson:2000mu}, the coefficient of that term is controlled by the optional $B_{01}$ piece mentioned below (\ref{limitbkg}), and is freely adjustable due to the conservation of  total string winding. The kinetic energy in (\ref{closednrf1}) identifies $\mu$ as the NR mass of the string.  

The spectrum (\ref{closednrf1}) follows from applying the limit (\ref{nrf1limit}) to the standard relativistic closed string spectrum \cite{Klebanov:2000pp,Danielsson:2000gi}, or alternatively, from the worldsheet action that gives an intrinsic definition of NRF1 \cite{Gomis:2000bd},
\begin{equation}
S_{\mbox{\tiny GO}}=\frac{1}{4\pi\alpha'}\int d^{2}\sigma\left(\lambda\bar{\p}X^{+}+\bar{\lambda}\p X^{-}+\p \vec{X}_{\perp}\cdot \bar{\p}\vec{X}_{\perp}\right)~.
\label{go}
\end{equation}
Here, $\vec{X}_{\perp}\equiv X^{A'}$ denotes the eight transverse embedding fields, and the dot product refers to contraction with the transverse metric $\delta_{A'B'}$. 
The novelty in the action is purely in the longitudinal sector, which involves the lightcone embedding fields $X^+\equiv X^0+X^1$ and $X^-\equiv -X^0+X^1$, as well as the Lagrange-multiplier fields $\lambda,\bar{\lambda}$ mentioned in the Introduction. 
If we add a $\lambda\bar{\lambda}$ term to (\ref{closednrf1}), the longitudinal sector can be recognized as a rewriting of the usual relativistic term $\p X^A\bar{\p}X^B\eta_{AB}$. The effect of the limit (\ref{nrf1limit}) is to remove the $\lambda\bar{\lambda}$ term \cite{Gomis:2000bd}. One can optionally include in (\ref{go}) a $\p X^+\bar{\p} X^-$ term, whose effect is to shift (\ref{closednrf1}) by the tensional energy contribution mentioned in the previous paragraph. Such a term can be omitted when describing closed strings, but is indispensable \cite{Danielsson:2000mu} if we consider longitudinal D-branes (see Section~\ref{dbranesubsec}). 
The action (\ref{go}) is invariant under the string Galilean group \cite{Gomis:2000bd,Danielsson:2000mu,Batlle:2016iel}, which includes translations, longitudinal boosts, transverse rotations,
and string Galilean boosts. 

The form of the spectrum (\ref{closednrf1}) suggests that unwound strings, $w=0$, have divergent energy, except if $\vec{p}_{\perp}=0$, N$_{\mbox{\tiny L}}=0$, N$_{\mbox{\tiny R}}=0$. Such massless closed string modes would then have energy $p_0=|p_1|=|m|/R$. This conclusion turns out to be correct, even if it cannot  be properly derived from (\ref{closednrf1}) or (\ref{go}), because the case $w=0$ requires exceptional treatment, and in particular, it forces one to retain the $\lambda\bar{\lambda}$ term \cite{Danielsson:2000mu}.
On shell, these `Newtonian gravitons' are of measure zero as asymptotic states, due to the $\vec{p}_{\perp}=0$ requirement \cite{Danielsson:2000mu}. But off shell, they give rise to long-distance instantaneous Newtonian interactions \cite{Gomis:2000bd,Danielsson:2000mu}, and to the possibility of considering NRF1 theory on a curved background, as we will discuss in Section~\ref{sncsubsec}. 

\subsection{D-branes in NRF1}
\label{dbranesubsec}

D$p$-branes are present in the nonperturbative spectrum of the Type II nonrelativistic string theories, 
with $p$ even in NRF1 IIA and odd in NRF1 IIB.
Longitudinally-extended D-branes must necessarily have positive F1 winding $n>0$ along $x^1$.  In other words, they must be D$p$-F1 bound states, which they achieve by carrying a constant worldvolume electric field $E\equiv F_{01}$  \cite{Witten:1995im}. 
Compactifying the transverse directions spanned by such a D$p$-brane on a torus with radii $R_2,\ldots,R_p$, we can define the F1 charge density \cite{Danielsson:2000gi}
\begin{equation}
\nu\equiv\frac{n\,l_s^{\,p-1}}{R_2\ldots R_p}~,
\label{nu}
\end{equation}
which is related to the electric field through $E=1-(2\nu^2 g_s^2)^{-1}$, and determines the D-brane's dynamical tension\footnote{Viewed from the parent string theory, this is the finite tension of the bound state that remains after taking the limit (\ref{nrf1limit}), with the leading divergent contribution due to the F1 tension having been subtracted by the Kalb-Ramond field.} \cite{Danielsson:2000gi}, 
\begin{equation}
T^{\,\scriptscriptstyle ||}_{\mbox{\tiny D$p$},\nu}=\frac{1}{2(2\pi)^p \nu g_s^2 l_s^{p+1}}~.
\label{longtension}
\end{equation}
At the level of the worldsheet action (\ref{go}), the presence of this electric field or F1 density is reflected through the appearance of a $\p X^+\bar{\p} X^-$ term with the specific coefficient $1/4\nu^2 g_s^2$
\cite{Danielsson:2000mu}.

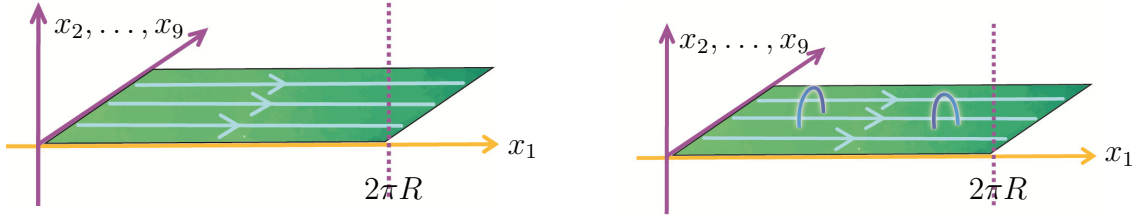
\begin{figure}[thb]
\begin{picture}(150,50)
  \centering
  \includegraphics[width=7.0cm]{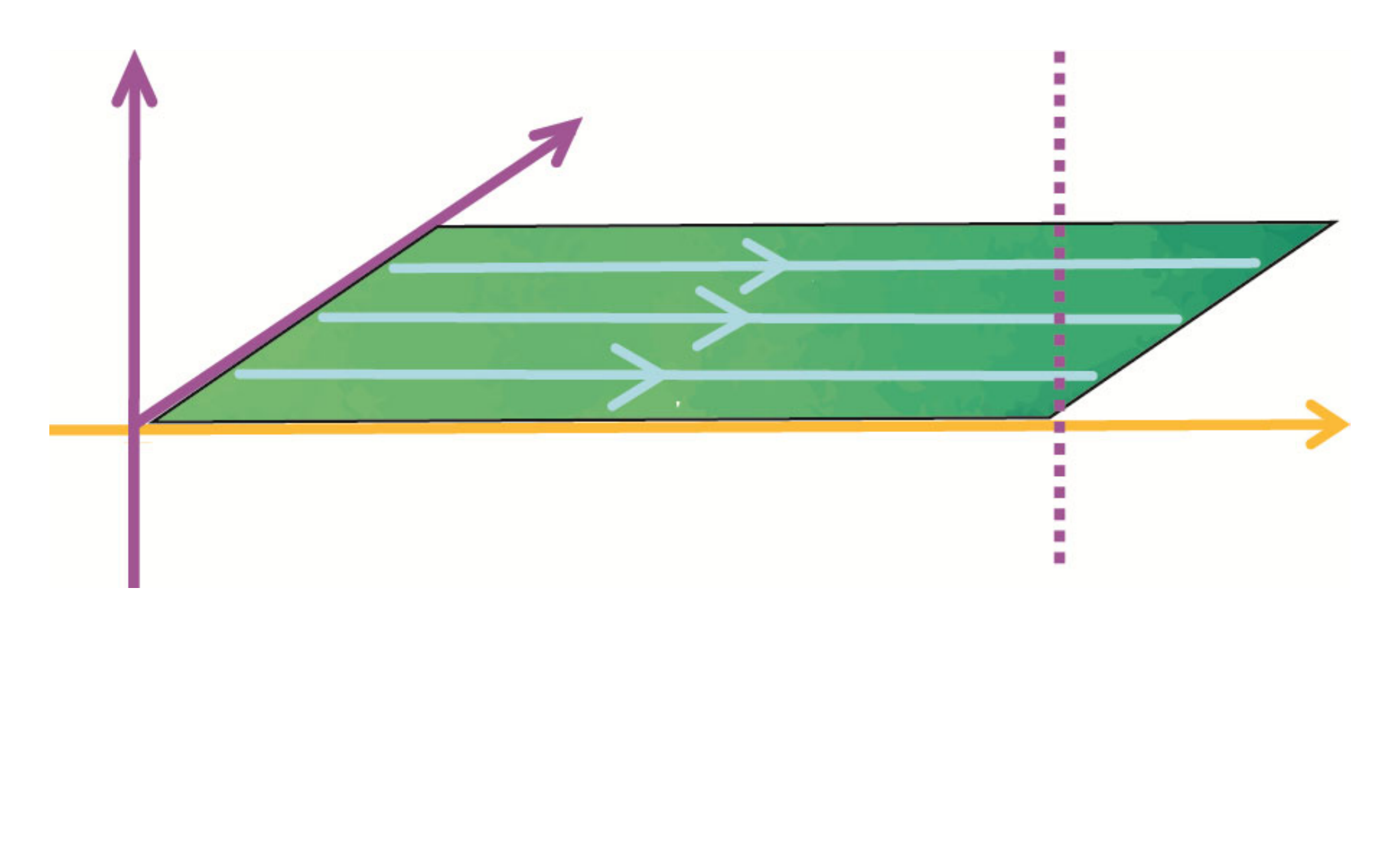}
  \hspace*{0.8cm}
  \includegraphics[width=7.0cm]{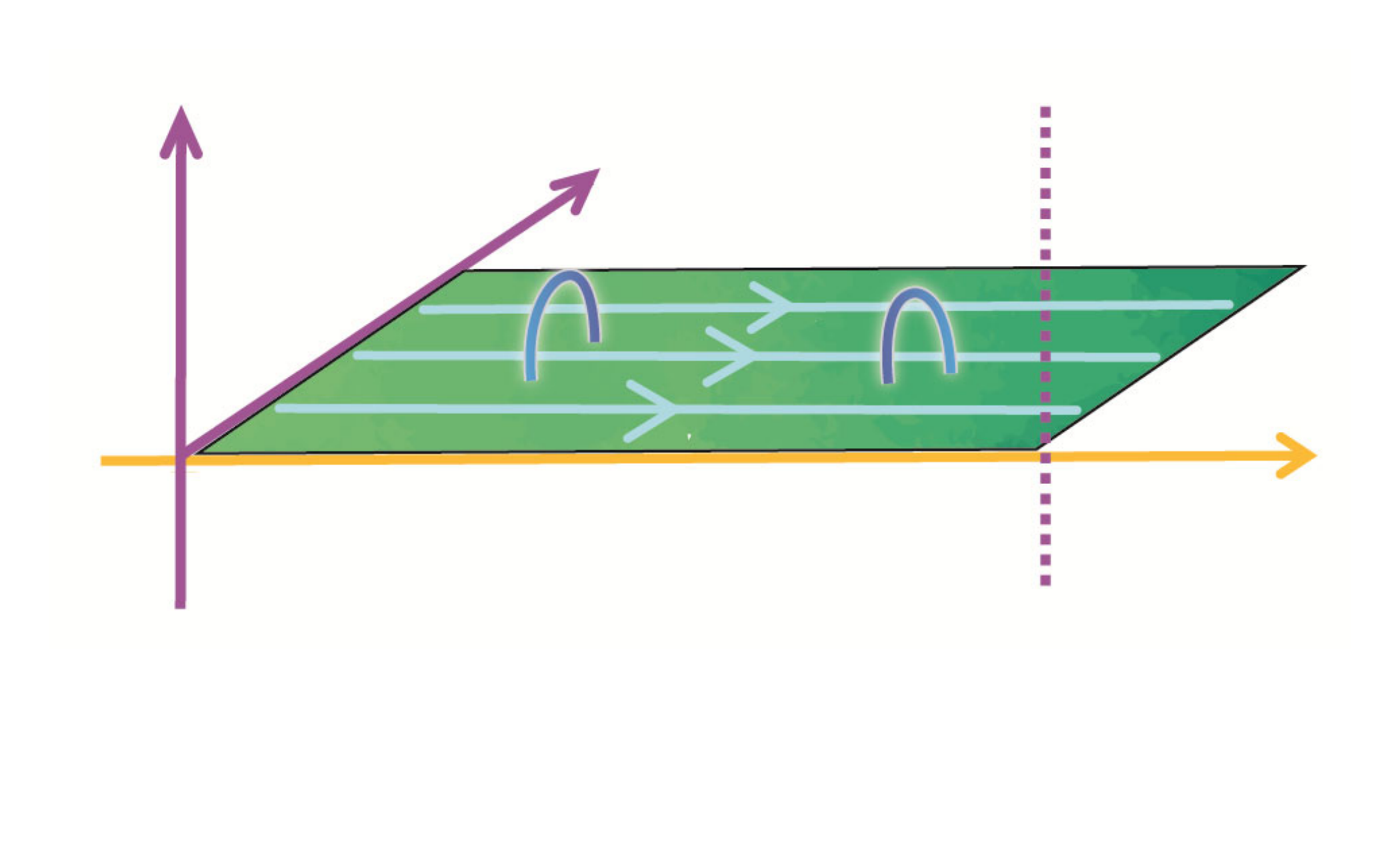}
  \put(-82,20){$x_1$}
  \put(-3,19){$x_1$}
  \put(-142,36){$x_2,\ldots,x_9$}
  \put(-60,34){$\,x_2,\ldots,x_9$}
  \put(-101,14){$2\pi R$}
  \put(-21,14){$2\pi R$}
 \end{picture}
\vspace*{-1.5cm}
  \caption{Longitudinal D-brane in NR string theory. Left: The D-brane must carry positive string winding along the longitudinal direction, i.e., it must have fundamental strings dissolved inside its worldvolume. Right: Small excitations of the D-brane are described by open strings, with the \emph{relativistic} spectrum (\ref{openlong}).}
  \label{fig:longdbrane} 
  \end{figure}
  
A notable aspect discovered in the NCOS papers \cite{Seiberg:2000ms,Gopakumar:2000na} is that the open strings that describe the excitations of longitudinal D-branes in NRF1 theory have a \emph{relativistic} spectrum, albeit with a rescaled metric  \cite{Seiberg:1999vs,Danielsson:2000mu}
\begin{equation}
\nu^2 g_s^2\left[(p_0)^2-(p_1)^2\right]-\vec{p}_{\perp}^{\,2}=\frac{\mbox{N}}{l_s^2}~.
\label{openlong}
\end{equation}
With hindsight, the appearance a quarter century ago of this relativistic spectrum within NR string theory can now be understood as a presage of the recent results of \cite{Guijosa:2023qym,Avila:2023aey}, which are central for the present paper, and will be reviewed in Sections~\ref{blackstringsubsec} and \ref{blackpbranesubsec}.

Compared to their counterparts in standard relativistic string theory, the scattering amplitudes of these open strings display three noteworthy features: an open string coupling set by the F1 density ($g_{\mbox{\tiny open}}^2=1/\nu$)
\cite{Seiberg:1999xz,Seiberg:2000ms,Gopakumar:2000na}, Moyal phases that encode noncommutativity between space and time \cite{Seiberg:2000ms,Gopakumar:2000na,Danielsson:2000gi}, and the vanishing of amplitudes that include any number of massless open string vertices with purely longitudinal momenta, reflecting the existence of a free $U(1)$ sector \cite{Klebanov:2000pp,Gomis:2000bd}.

Transverse D$p$-branes can only carry longitudinal F1 charge through the open strings that describe their excitations, which are subject to the $w>0$ requirement, and have a spectrum identical to that of closed strings (with $p_1=m/R=0$) \cite{Danielsson:2000gi,Danielsson:2000mu},
\begin{equation}
p_{0}=\frac{l_s^2\, \vec{p}_{\perp}^{\,2}} {2wR}+\frac{\mbox{N}}{2wR}~.
\label{opentransnrf1}
\end{equation}
Interestingly, even when unexcited and therefore devoid of F1 charge, the tension of these branes is finite all by itself
\cite{Danielsson:2000gi}, 
\begin{equation}
T^{\,\scriptstyle \perp}_{\mbox{\tiny D$p$}}=\frac{1}{(2\pi)^p g_s l_s^{p+1}}~.
\label{transtension}
\end{equation}
This is the same formula as in the relativistic case.

As indicated in \cite{Danielsson:2000gi}, NRF1 theory also contains branes other than D-branes, including NS5-branes. These can easily be seen to have finite tension if they are extended longitudinally (with the same formula for the tension as in the relativistic case), and vanishing tension if they are purely transverse \cite{Guijosa:2001notes}. 

Overall, a pattern becomes clear here. Branes in the spectrum of NRF1 theory would normally be subject to the requirement of carrying positive longitudinal F1 charge. However, exceptional cases exist in which the manner of carrying this charge allows it to be taken to infinity without incurring any energy cost, implying that the corresponding `bare' brane must by itself have a non-divergent energy. For instance, any wound open strings attached to a transverse D$p$-brane can be displaced arbitrarily far way along the transverse directions occupied the brane. Similarly, starting with a longitudinal NS5-brane carrying positive string winding, the NS5$\scriptstyle ||$-F1 bound state is at threshold (unlike in the D$p\scriptstyle ||$-F1 bound state),  meaning that the F1s can be taken out of the NS5 and moved away to infinity, entailing that bare longitudinal NS5-branes (unlike bare longitudinal D$p$-branes) have a finite tension.

\subsection{String Newton-Cartan (SNC) backgrounds}
\label{sncsubsec}

As mentioned in the Introduction, the extension of NRF1 to curved backgrounds leads naturally to the concept of string Newton-Cartan (SNC) geometries \cite{Andringa:2012uz,Harmark:2017rpg,Bergshoeff:2018yvt,Harmark:2018cdl,Gomis:2019zyu,Gallegos:2019icg,Harmark:2019upf,Bergshoeff:2021bmc,Bidussi:2021ujm}. We will now highlight a few of the salient features, mostly following the summary of \cite{Bergshoeff:2018yvt,Gomis:2019zyu} provided in \cite{Avila:2023aey}. 
The essential idea is to start with the 10-dimensional manifold and equip it with a distinguished $2+8$ foliation by introducing separate longitudinal and transverse vielbeine
$\tau_{\mu}^{A}(x)$ ($A=0,1$) and $E_{\mu}^{A'}(x)$ ($A'=2,\ldots,9$), which allow us to generalize (\ref{limitbkg}) by replacing $dx^A\to\tau_{\mu}^{A}dx^{\mu}$ and $dx^{A'}\to E_{\mu}^{A'}dx^{\mu}$. 
The inverses $\tau^{\mu}_{A}$ and $E^{\mu}_{A'}$ satisfy
\begin{equation}
    \tau_{\mu}^{A}\tau^{\mu}_{B}=\delta^{A}_{B}~,
    \quad
    E_{\mu}^{A'}E^{\mu}_{B'}=\delta^{A'}_{B'}~,
    \quad
    \tau^{\mu}_{A}E_{\mu}^{A'}=\tau_{\mu}^{A}E^{\mu}_{A'}=0,
    \quad
    \tau_{\mu}^{A}\tau^{\nu}_{A}+E_{\mu}^{A'}E^{\nu}_{A'}=\delta_{\mu}^{\nu}~.
\label{sncvielbeine}
\end{equation}
The associated longitudinal metric and transverse co-metric read 
\begin{equation}
    \tau_{\mu\nu}\equiv \tau_{\mu}^{A}\tau_{\nu}^{B}\eta_{AB}~,
    \qquad
    E^{\mu\nu}\equiv E^{\mu}_{A'}E^{\nu}_ {B'}\delta^{A'B'}~,
\label{sncmetrics}
\end{equation}
and satisfy $\tau_{\mu\nu}E^{\nu\lambda}=0$.
A SNC geometry is characterized by these vielbeine together with two longitudinal gauge fields $m^A_{\mu}$ (associated with a non-central extension of the string Galilei algebra). 
Alternatively, the latter information can be repackaged \cite{Bidussi:2021ujm} into a single 2-form field $m_{\mu\nu}$.

This SNC structure incorporates the string Galilean boosts $\delta_{\scriptstyle\Sigma}\tau^A_{\mu}=0,\delta_{\scriptstyle\Sigma} E^{A'}_{\mu}=\Sigma^{A'}{}_{\!\!\! A}\,\tau^A_{\mu}\,$, or in terms of the inverses (\ref{sncvielbeine}),
$\delta_{\scriptstyle\Sigma}\tau_A^{\mu}=-\Sigma^{A'}{}_{\!\!\! A}E_{A'}^{\mu}~,$ $\delta_{\scriptstyle\Sigma}E_{A'}^{\mu}=0$.
The combinations 
\begin{equation}
\tau^{\mu\nu}\equiv \tau^{\mu}_{A}\tau^{\nu}_{B}\eta^{AB}~, 
\qquad 
E_{\mu\nu}\equiv E_{\mu}^{A'}E_{\nu}^{B'}\delta_{A'B'}
\label{emunulower}
\end{equation}
are not boost-invariant,
but taking into account that $\delta_{\scriptstyle\Sigma}m_{\mu}^{A}=\Sigma^{A}{}_{\!\!\! A'}\,E^{A'}_{\mu}$, one can form the invariant combinations
\begin{equation}
    \begin{aligned}
    H_{\mu\nu}&\equiv E_{\mu\nu}+\left(\tau_{\mu}^{A}m_{\nu}^{B}+\tau_{\nu}^{A}m_{\mu}^{B}\right)\eta_{AB}~,
    \\
    \mathcal{T}^{\mu\nu}&\equiv\tau^{\mu\nu}-\left(E^{\mu}_{A'}\tau^{\nu}_{A}+E^{\nu}_{A'}\tau^{\mu}_{A}\right)m_{\lambda}^{A}E^{\lambda}_{A'}~.
    \end{aligned}
\label{sncmetrics2}
\end{equation}
With these elements, Christoffel symbols and curvature tensors can be defined, and various flavors of SNC geometry are obtained upon imposing different torsion constraints \cite{Oling:2022fft}. 

In \cite{Bergshoeff:2019pij} it was shown that, \emph{if} a parameter $\omega$ can be found with respect to which the metric, Kalb-Ramond and dilaton fields of a relativistic 10-dimensional Type II supergravity background happen to scale in an appropriate manner, then a proper SNC background can be obtained by taking $\omega\to\infty$. 
The required scaling is 
\begin{equation}
\begin{aligned}
\hat{g}_{\mu\nu}&=\omega^{2}\tau_{\mu\nu}+
\left[
H_{\mu\nu}-\left(\tau_{\mu}^{A}c_{\nu}^{B}+\tau_{\nu}^{A}c_{\mu}^{B}\right)\eta_{AB}
\right]
+\frac{1}{\omega^{2}}\left(m_{\mu}^{A}-c_{\mu}^{A}\right)\left(m_{\nu}^{B}-c_{\nu}^{B}\right)\eta_{AB}~,
\\
\hat{B}_{\mu\nu}&=\omega^{2}\tau_{\mu}^{A}\tau_{\nu}^{B}\epsilon_{AB}
+\left[B_{\mu\nu}+\left(\tau_{\mu}^{A}c_{\nu}^{B}-\tau_{\nu}^{A}c_{\mu}^{B}\right)\epsilon_{AB}\right]-\frac{1}{\omega^{2}}c_{\mu}^{A}c_{\nu}^{B}\epsilon_{AB}~,
\\ 
\exp(\hat{\Phi})&=\omega\exp(\Phi)~.
\end{aligned}
\label{nrf1bkg}
\end{equation}
The $c_{\mu}^{A}$ here are arbitrary functions that do not appear in the equations of motion, but are necessary to take the NR limit. The reformulation \cite{Bidussi:2021ujm} in terms of $m_{\mu\nu}$ mentioned above  dispenses with this redundancy. 
The behavior of the dilaton in (\ref{nrf1bkg}) is a generalization of the scaling of the string coupling constant in (\ref{nrf1limit}).

To describe a NR string propagating on a nontrivial SNC background characterized by $\tau^A_{\mu}$, $H_{\mu\nu}$, 
$B_{\mu\nu}$ and $\Phi$, one must as usual add to (\ref{go}) the corresponding vertex operators. The result is the $\sigma$-model action
\cite{Gomis:2019zyu,Bergshoeff:2018yvt,Bergshoeff:2019pij}
\begin{equation}
\begin{aligned}
S_{\sigma}=&\frac{1}{4\pi\alpha'}\int d^{2}\sigma\sqrt{h}\left[\lambda\bar{\mathcal{D}}X^{\mu}\tau_{\mu}+\bar{\lambda}\mathcal{D}X^{\mu}\bar{\tau}_{\mu}+\mathcal{D}X^{\mu}\bar{\mathcal{D}}X^{\nu}\left(H_{\mu\nu}+B_{\mu\nu}\right)\right]\\
&+\frac{1}{4\pi}\int d^{2}\sigma\sqrt{h} R^{(2)}\left(\Phi-\frac{1}{4}\ln g\right)~.
\end{aligned}
\label{nrf1sigma}
\end{equation}
where $h_{\alpha\beta}=e_{\alpha}^{a}e_{\beta}^{b}\delta_{ab}$ is the intrinsic worldsheet metric, $h\equiv\text{det}(h_{\alpha\beta})$, $R^{(2)}$ is the Ricci scalar of $h_{\alpha\beta}\,$,
\begin{equation}
\tau_{\mu}\equiv \tau_{\mu}^{0}+\tau_{\mu}^{1}~, \qquad \bar{\tau}_{\mu}\equiv \tau_{\mu}^{0}-\tau_{\mu}^{1}~,
\end{equation}
\begin{equation}
\mathcal{D}\equiv \frac{i}{\sqrt{h}}\epsilon^{\alpha\beta}\bar{e}_{\alpha}\nabla_{\beta}~, \qquad \bar{\mathcal{D}}\equiv \frac{i}{\sqrt{h}}\epsilon^{\alpha\beta}e_{\alpha}\nabla_{\beta}~,
\end{equation}
\begin{equation}
e_{\alpha}\equiv e_{\alpha}^{1}+ie_{\alpha}^{2}, 
\qquad 
\bar{e}_{\alpha}\equiv-e_{\alpha}^{1}+ie_{\alpha}^{2}~,
\end{equation}
 $\nabla_{\alpha}$ is the covariant derivative compatible with $h_{\alpha\beta}$, $\epsilon^{\alpha\beta}$ is the two-dimensional Levi-Civita symbol,
and
\begin{equation}
g\equiv \text{det}^{(8)}\left(H_{\mu\nu}\right)\text{det}^{(2)}\left(\tau_{\rho}^{A}H^{\rho\sigma}\tau_{\sigma}^{B}\right)~.
\label{detg}
\end{equation}
As usual, the background must be constrained by requiring that the path integral be Weyl invariant, which leads to the equations of motion of the SNC target space fields \cite{Gomis:2019zyu}.

In preparation for the following subsections, we recall the emphasis made in Section~\ref{antecedentssubsec} on the fact that (\ref{nrf1sigma}) is \emph{not} the most general renormalizable nonlinear $\sigma$-model associated with the Gomis-Ooguri action (\ref{go}). For full generality, we must also allow a $\lambda\bar{\lambda}$ term.

\subsection{Black string in NRF1}
\label{blackstringsubsec}

Knowing that the elementary excitations of nonrelativistic string theory  are positively-wound closed strings, the simplest collection of objects that we can consider is a static stack of $N$ singly-wound coincident F1s extending purely along $x^1$. Such a stack can be studied still with $x^1$ compact, or in the decompactification limit $R\to\infty$.  
As shown in \cite{Avila:2023aey}, for 
$N g_s^2 \gg 1$
the stack distorts the geometry appreciably, and gives rise to the following background:\footnote{In the remainder of the paper, we will consider various black brane backgrounds, each characterized by some curvature radius. For simplicity, we will label all of these different length scales with the same letter $L$, providing explicitly in each case the context-dependent definition of $L$. \label{Lfoot}}
\begin{equation}
\begin{aligned}
ds^{2}&=\frac{r^{6}}{L^{6}}\left(-dt^{2}+dx^{2}\right)+dx_2^{2}+\ldots+dx_9^2~,
\\
B&=\frac{r^{6}}{L^{6}}\,dt\wedge dx~,\\
\exp\left(2\Phi\right)&=\frac{g_{s}^{2}r^{6}}{L^{6}}~,
\\
L^6&\equiv 32\pi^{2}N g_{s}^{2}l^{\,6}_{s}~. 
\end{aligned}
\label{blackstring}
\end{equation}
By construction, this background describes a black string in NR string theory: it is what results from starting in the parent string theory and applying the limit (\ref{nrf1limit}) to the extremal relativistic Dabholkar-Harvey black string \cite{Dabholkar:1989jt,Dabholkar:1990yf}. 

Remarkably, the black string (\ref{blackstring}) is not a SNC geometry; it is only \emph{asymptotically} SNC \cite{Avila:2023aey}. This assertion entails two separate but complementary properties. 
The first is that, if we follow \cite{Danielsson:2000mu} in identifying\footnote{Any qualms about identifying the constant parameter $\omega$ with a function of the radial coordinate can be assuaged by defining 
$r\equiv r_0+\mathrm{r}$, with $r_0$ an arbitrarily large constant, and then setting $\omega^2=r_0^6/L^6$. Below we will see that, from the perspective of the $\sigma$-model (\ref{nrf1sigma}),  the black string background (\ref{blackstring}) includes a $\lambda\bar{\lambda}$ term with a coefficient whose  $r$-dependence is essential for the correct physical interpretation.} $\omega^2= r^6 /L^6$ in the asymptotic region  $r\gg L$, then  the background (\ref{blackstring}) is seen to take precisely the form (\ref{nrf1limit}).
As $\omega\to\infty$, we are left with the flat SNC background
\begin{equation}
\tau_{\mu}^{A}=\delta_{\mu}^{A}~, \quad E_{\mu}^{A'}=\delta_{\mu}^{A'}~, \quad m_{\mu}^{A}=0~, \quad 
B_{\mu\nu}=0~, \quad \Phi=\ln g_{s}~.
\label{flatsnc}
\end{equation} 
 This implies that, on the black string background (\ref{blackstring}), only positively-wound strings can escape to $r\to\infty$, and such strings in fact sense a flat SNC geometry. This is true due to the interplay between the metric and the Kalb-Ramond fields in (\ref{blackstring}), in spite of the fact that (\ref{blackstring}) no longer includes the additive 1 originally present in the harmonic function 
 $1+\hat{L}^6/\hat{r}^{\,6}$ that characterizes the
 parent relativistic background.
 
 The second crucial property is that, at $r\, {\scriptstyle\lesssim}\, L$ ($\omega\, {\scriptstyle\lesssim}\, 1$), the warping in (\ref{blackstring}) gives rise to significantly different physics. One notable sign of this is the fact that the black string (\ref{blackstring}) is a solution not of the SNC equations of motion worked out in \cite{Gomis:2019zyu,Gallegos:2019icg,Bergshoeff:2019pij,Yan:2019xsf,Bergshoeff:2021bmc,Bergshoeff:2021tfn}, but of
 the \emph{relativistic} supergravity equations of motion \cite{Avila:2023aey}. This implies that the region $r\, {\scriptstyle\lesssim}\, L$ can be safely inhabited by the complete cast of branes of standard string theory, including F1s with any winding whatsoever, and moreover, the worldvolume actions describing the dynamics of such branes take the usual \emph{relativistic} form.  But importantly, we should not lose sight of the fact that  a particular theory is defined not just by its action or equations of motion, but also by its boundary conditions.
 In the previous paragraph we have verified that the asymptotics of (\ref{blackstring}) enforce the expected  NRF1 scaling. It is this external region that determines the spectrum of asymptotic states of the theory. Particularly telling is the fact that, if we employ as a probe of (\ref{blackstring}) a positively-wound F1  described by the standard relativistic $\sigma$-model action, this action reduces to (\ref{nrf1sigma}) as $r/L\to\infty$ ($\omega\to\infty$), but \emph{acquires a $\lambda\bar{\lambda}$ correction as we move inward} \cite{Avila:2023aey}. This term has clearly been sourced by the stack of $N$ strings, and gives rise to relativistic features that naturally become more pronounced as we approach said stack. 
 Our F1 probe is thus described by \emph{the most general form of the $\sigma$-model in NRF1}, with a $\lambda\bar{\lambda}$ term that switches off asymptotically but contributes appreciably in the inner region. 

One possible source of
confusion is the comparison between (\ref{nrf1limit}) and (\ref{nrf1bkg}). Why did we deduce the black string in NRF1 by applying the limit that defines NRF1 \emph{in flat space}, Eq.~(\ref{nrf1limit}), instead of the curved space version of the limit, Eq.~(\ref{nrf1bkg})? It might seem like this amounts to only enforcing the limit asymptotically, which could then be the reason why the resulting background only asymptotes to SNC form at infinity. The answer is that (\ref{nrf1bkg}) is not in fact a general prescription: it is one thing to hypothesize that relativistic backgrounds exist which happen to scale in the manner indicated there, and quite another \emph{to find, within a given family of backgrounds, a parameter that in actual fact implements that scaling}. As shown in \cite{Avila:2023aey}, in the particular case of the Dabholkar-Harvey black string, the only way to achieve (\ref{nrf1bkg}) is to deviate from (\ref{nrf1limit}) and demand instead that $\hat{N}=\omega^{-4}N\to 0$. Since the number of F1s is discrete, this scaling is clearly unphysical. This failed attempt reveals that the unachievability of (\ref{nrf1bkg}) is due to the very existence of the source: we can only obtain a fully SNC background if we completely remove the stack of F1s. 

Relatedly, it is important to note that, in detail, the black string (\ref{blackstring}) differs from the Dabholkar-Harvey solution everywhere, not just at infinity, so it is \emph{not} true that it is obtained by applying the limit (\ref{nrf1limit}) only in the asymptotic region. This point is even clearer in the string-stack side, where the limit has the effect of collapsing the surrounding 10-dimensional Minkowski space to 10-dimensional flat SNC, \emph{and additionally}, of simplifying the intrinsic dynamics of the stack of strings. 
Given the definition of NRF1 as the result of applying the limit (\ref{nrf1limit}) \emph{to all objects} of the parent relativistic theory \cite{Danielsson:2000gi,Danielsson:2000mu}, to arrive at (\ref{blackstring}) we have simply done that for one of the better-known objects, the Dabholkar-Harvey black string. The essential point is that, when we say we are defining a \emph{gravitational} theory on flat spacetime, we really mean \emph{asymptotically} flat, because spacetime can be deformed dynamically. For our case, the correct statement then is that \emph{the limit (\ref{nrf1limit}) defines NRF1 theory on asymptotically flat SNC spacetime}.

\begin{figure}[thb]
\begin{center}
\begin{picture}(150,50)
  \includegraphics[width=7.0cm]{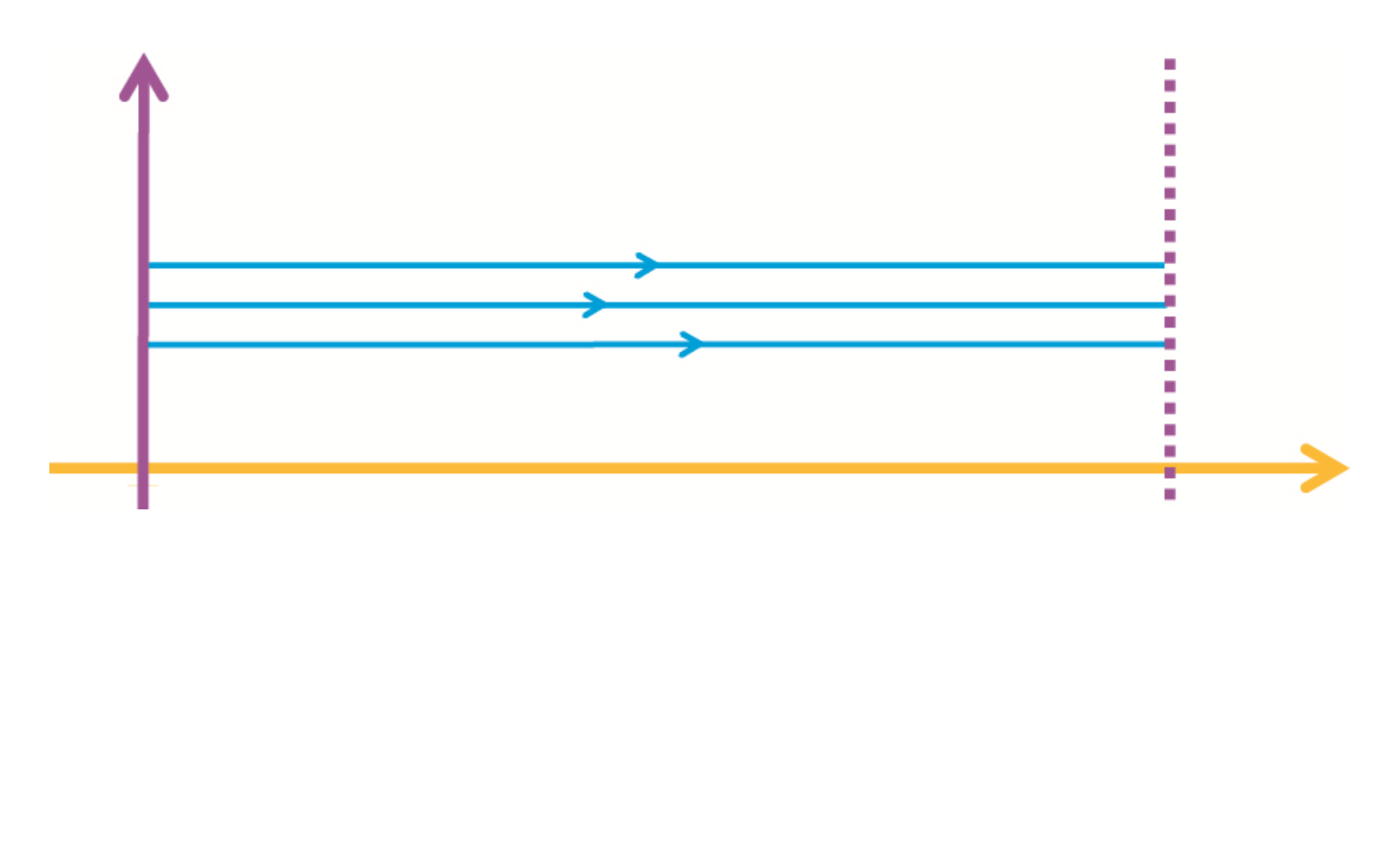}
  \hspace*{0.8cm}
  \includegraphics[width=7.0cm]{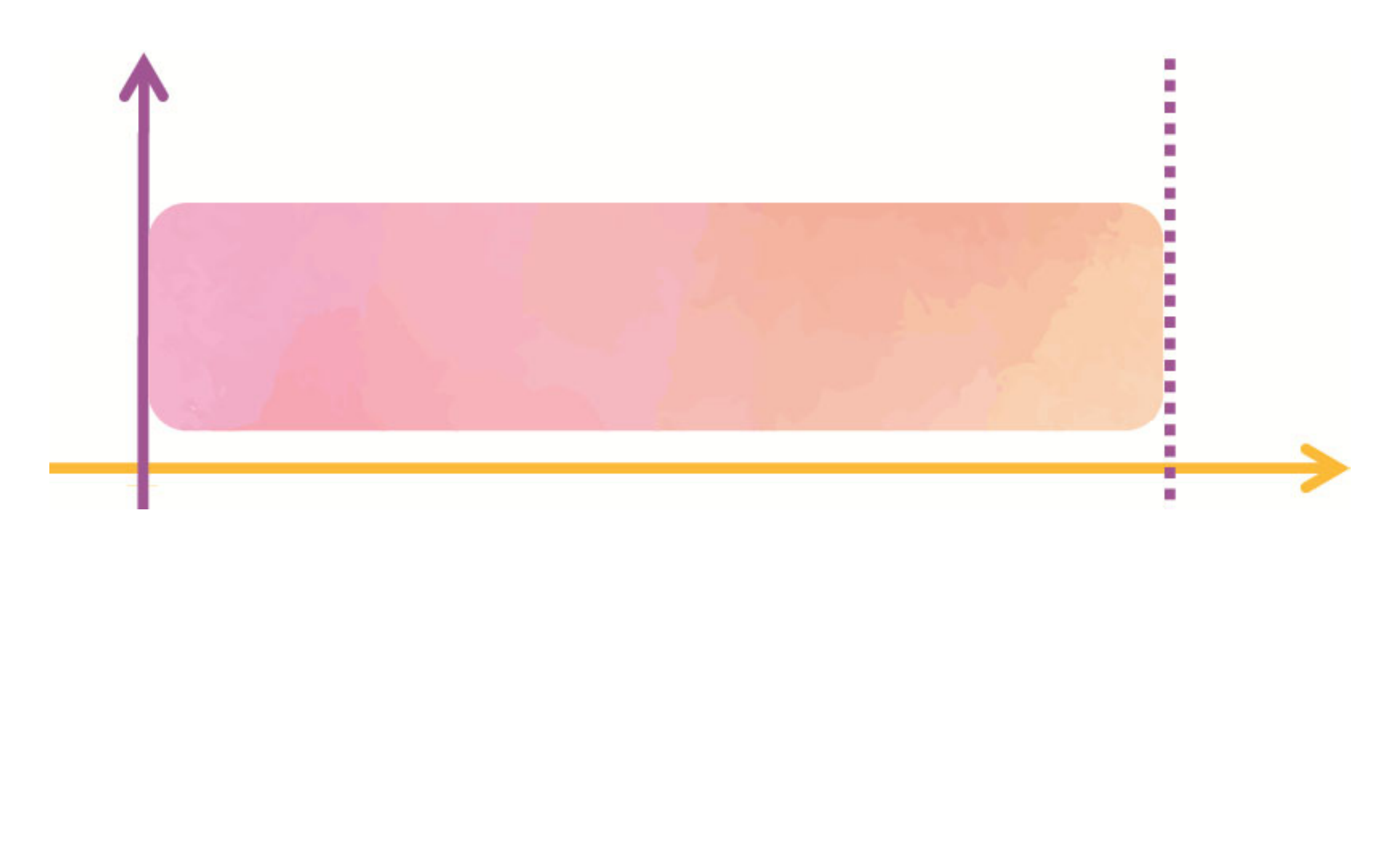}
  \put(-82,18){$x_1$}
  \put(-1,18){$x_1$}
  \put(-148,42){$x_2,\ldots,x_9$}
  \put(-67,42){$x_2,\ldots,x_9$}
  \put(-96,14){$2\pi R$}
  \put(-15,14){$2\pi R$}
 \end{picture}
 \vspace*{-1.5cm}
 \begin{picture}(75,35) 
   \includegraphics[width=7.0cm]{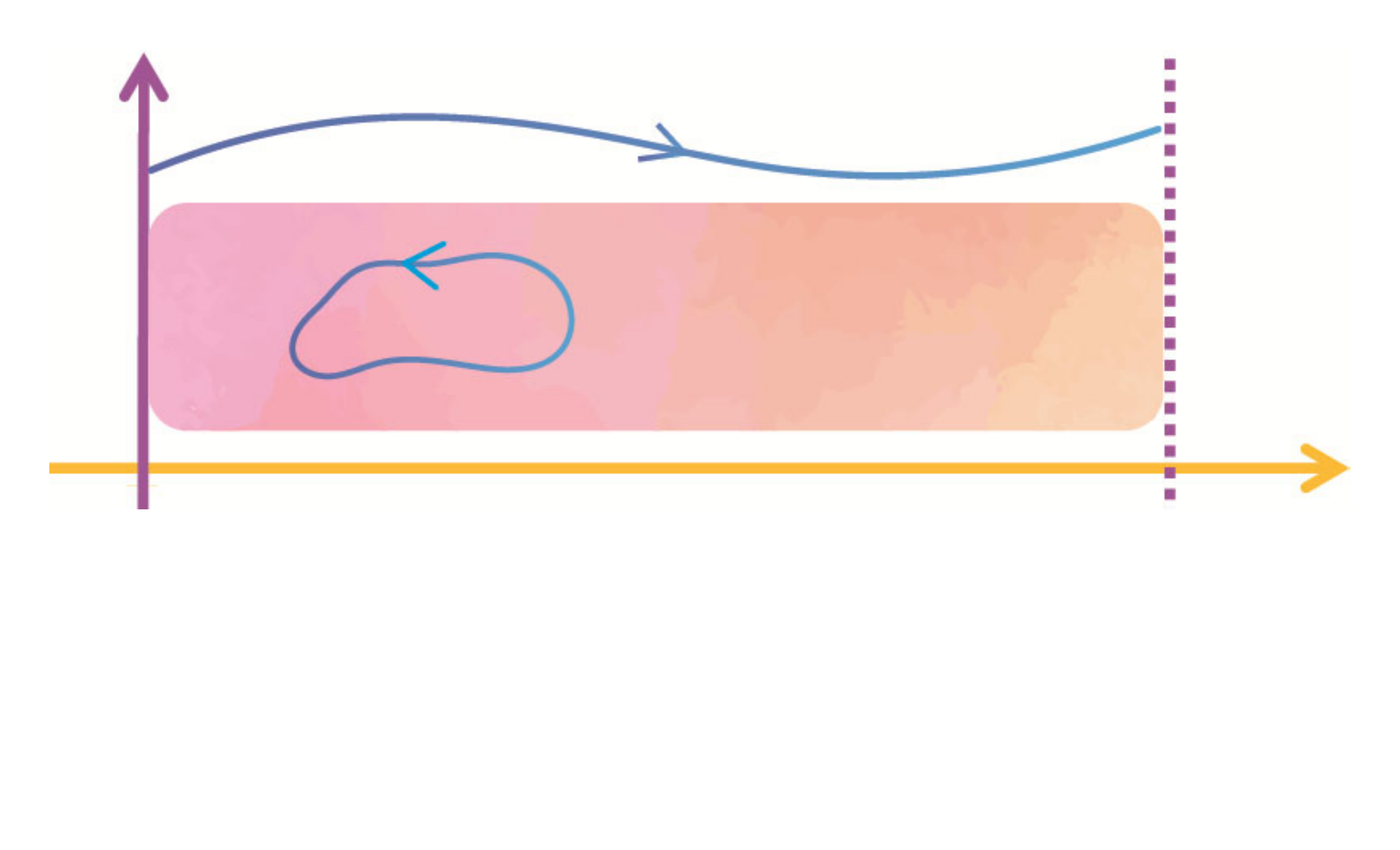}
   \put(-1,18){$x_1$}
   \put(-67,42){$x_2,\ldots,x_9$}
  \put(-15,14){$2\pi R$}
 \end{picture}
 \end{center}
\vspace*{-0.5cm}
  \caption{Stack of $N$ coincident positively-wound fundamental strings.
  Top left: The stack embedded in the 10-dimensional flat string Newton-Cartan (SNC) background. 
  Top right: When $Ng_s^2\gg 1$, an alternative description emerges in terms of the black string geometry (\ref{blackstring}), which can be described schematically as consisting of an inner tubular region $r\, {\scriptstyle\lesssim}\, L$ that corresponds to the stack, and an outer region that corresponds to the surrounding flat SNC geometry. 
  Bottom: The two schematic regions display significantly different properties. The inner region is a \emph{relativistic} bubble that can host the whole repertoire of perturbative and nonperturbative excitations of standard string theory, including unwound strings. The outer region, on the other hand, is asymptotically flat SNC, and is a welcoming environment only for excitations in the spectrum of \emph{nonrelativistic} string theory, such as positively-wound closed strings.}
  \label{fig:blackstring} 
  \end{figure}
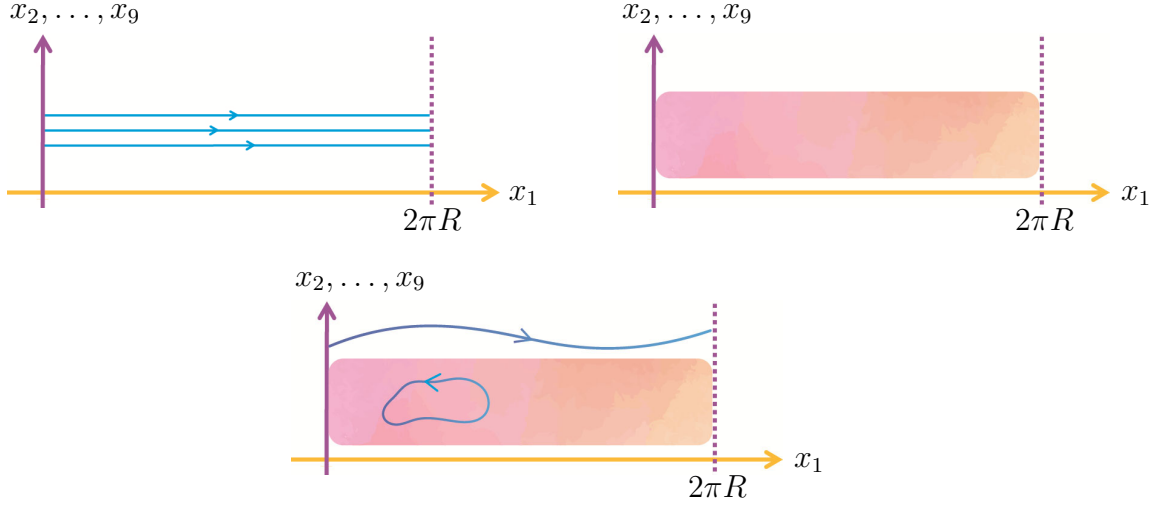

 To summarize, {\bf the black string of NRF1, Eq.~(\ref{blackstring}), consists schematically of two regions: an inner tubular bubble $r\, {\scriptstyle\lesssim}\, L$, extended uniformly along $x^1$, within which physics is locally that of ordinary relativistic string theory, and an outer region where the background asymptotes to flat SNC, and physics is consequently nonrelativistic.} See Fig.~\ref{fig:blackstring}. We should emphasize that the notion of a bubble is only schematic, not sharp, because there is an intermediate throat region, such that the transition from interior to exterior is smooth. The bubble is generated by the stack of $N$ F1s placed on the flat SNC background (\ref{flatsnc}), or better said, it is an alternative, geometric description of the effects produced by said stack. Just like in the relativistic black string, the existence of the asymptotic flat region implies that it is possible to exit the bubble, which is equivalent to exiting the stack of F1s.  
 {\bf One important effect of the NRF1 limit is precisely to convert
 the asymptotically flat \emph{Lorentzian} region that featured in the Dabholkar-Harvey black string to an asymptotically flat  \emph{string Newton-Cartan} region.} 
 
\subsection{RR black branes in NRF1}
\label{blackpbranesubsec}

Other objects of NRF1 that can naturally be subjected to the same type of analysis are  RR-charged black branes. These were also analyzed in \cite{Avila:2023aey}, and we briefly recall the results here.  

Consider first a stack of \emph{longitudinal} D-branes. Due to the requirement of positive fundamental string winding, we are interested in the extremal black $p$-brane sourced by a bound state of $\hat{K}=K$ D$p$-branes and $\hat{n}=n$ longitudinal F1s, with the latter corresponding through (\ref{nu}) to an F1 density $\hat{\nu}=\nu$.
Starting from such background in relativistic IIA/B string theory \cite{Schwarz:1995dk,Russo:1996if,Costa:1996zd,Lu:1999uca},  the NRF1 limit 
$\omega\to\infty$ with scaling (\ref{nrf1limit}) was implemented in \cite{Gopakumar:2000na,Harmark:2000wv}, yielding\footnote{See footnote~\ref{Lfoot}.}
\begin{equation}
\begin{aligned}
\widehat{ds}^2 &= 
\frac{r^{7-p}}{L^{7-p}}\mathsf{H}^{\frac{1}{2}}
\left( -dt^2+dx^2  \right) +\mathsf{H}^{-\frac{1}{2}}\left( dx_2^2+\dots +dx_p^2  \right)
\\ 
&   \qquad+\mathsf{H}^{\frac{1}{2}}\left( dx_{p+1}^2+\dots + dx_9^2  \right)~,
\\
\hat{B}&=\frac{r^{7-p}}{L^{7-p}}dt\wedge dx~, 
\\
\exp(2\hat{\Phi})&=\frac{r^{7-p}}{L^{7-p}}g_s^2\mathsf{H}^{\frac{5-p}{2}}
~,
\end{aligned}
\label{longdp}
\end{equation}
where
\begin{equation}
\begin{aligned}
\mathsf{H}&=1+\frac{K^2 L^{7-p}}{g_s^2\nu^2 r^{7-p}}~,
\\
r^2&\equiv x_{p+1}^2+\dots +x_{9}^2~,
\\
L^{7-p}&\equiv \frac{(2\pi)^{7-p}\nu g_s^2 l_s^{\,7-p}}{(7-p)\Omega_{8-p}}~,
\end{aligned}
\label{hlong}
\end{equation}
with 
\begin{equation}
    \Omega_{8-p}=\frac{2\pi^{(9-p)/2}}{\Gamma\left(\frac{9-p}{2}\right)}
    \label{Omega}
\end{equation}
the area of the unit $(8-p)$-dimensional sphere. Together with the NS-NS fields (\ref{longdp}), the background includes two RR fields: the ($p+2)$-form field strength $F_{(p+2)}$ sourced by the $K$ D$p$-branes \cite{Horowitz:1991cd,Polchinski:1995mt}, and the $p$-form field strength $F_{(p)}$ due to the D$(p-2)$ charge induced by the electric worldvolume field \cite{Douglas:1995bn}.  

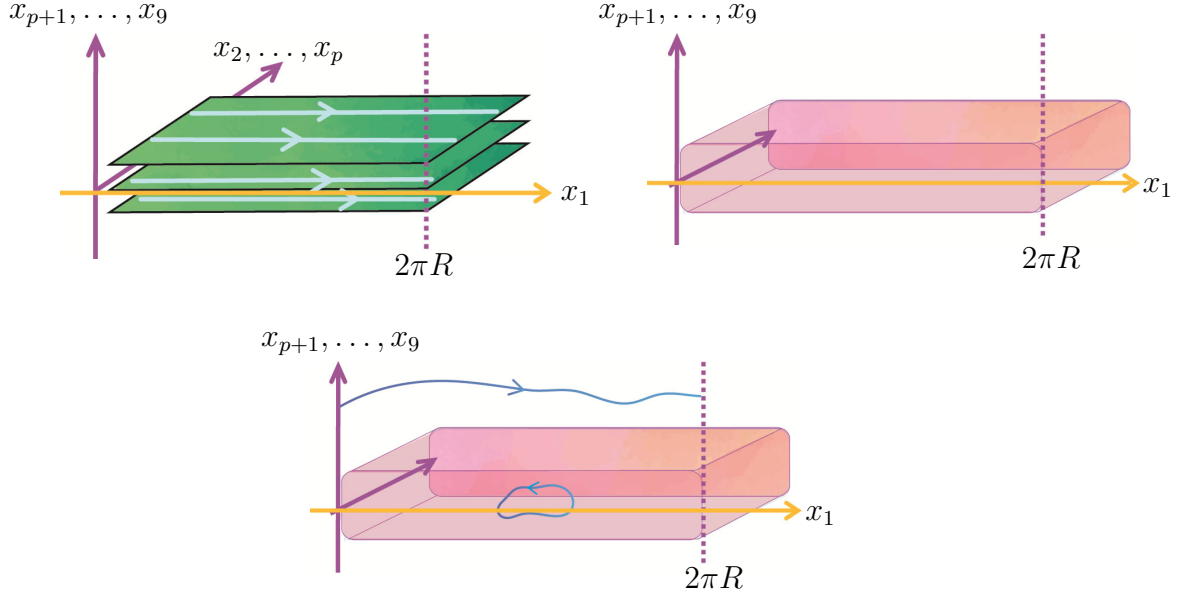
\begin{figure}[thb]
\begin{center}
\begin{picture}(150,50)
   \hspace*{0.3cm}
   \includegraphics[width=7.0cm]{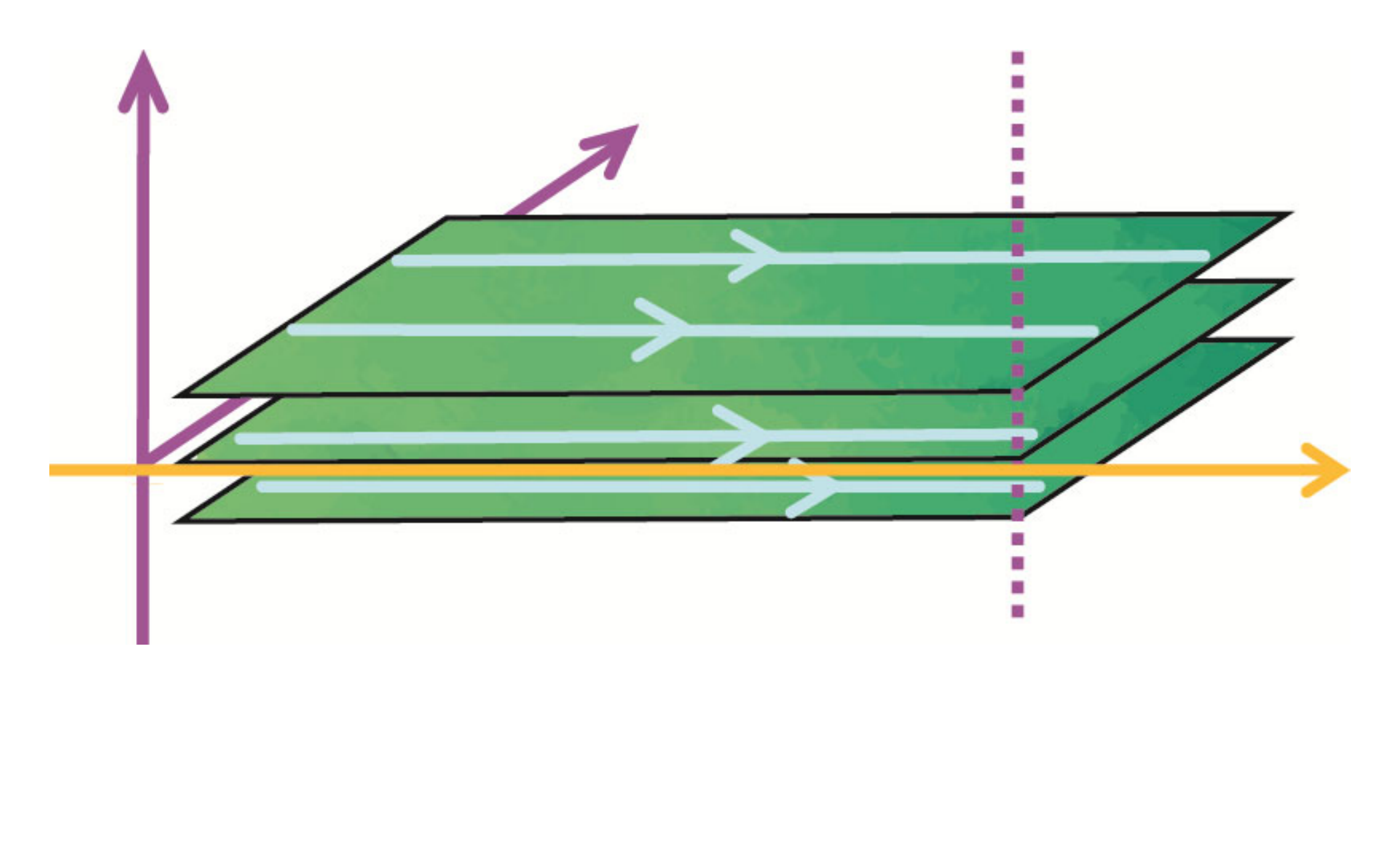}
  \hspace*{0.5cm}
  \includegraphics[width=7.0cm]{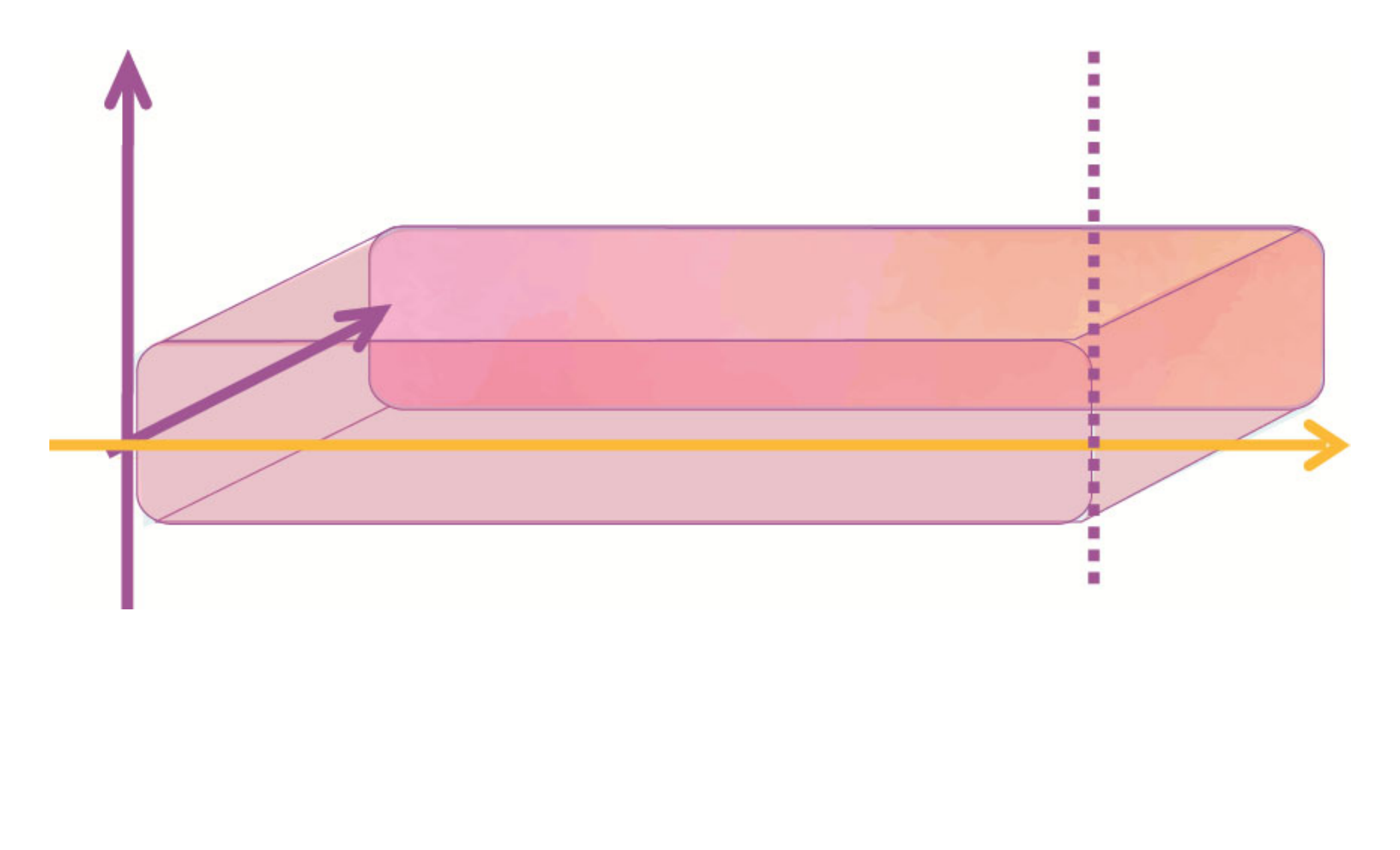}
  \put(-79,18){$x_1$}
  \put(-2,19){$x_1$}
  \put(-152,42){$x_{p+1},\ldots,x_9$}
  \put(-125,37){$x_2,\ldots,x_p$}
  \put(-74,42){$x_{p+1},\ldots,x_9$}
  \put(-101,8){$2\pi R$}
  \put(-18,9){$2\pi R$}
 \end{picture}
 \begin{picture}(75,43) 
   \includegraphics[width=7.0cm]{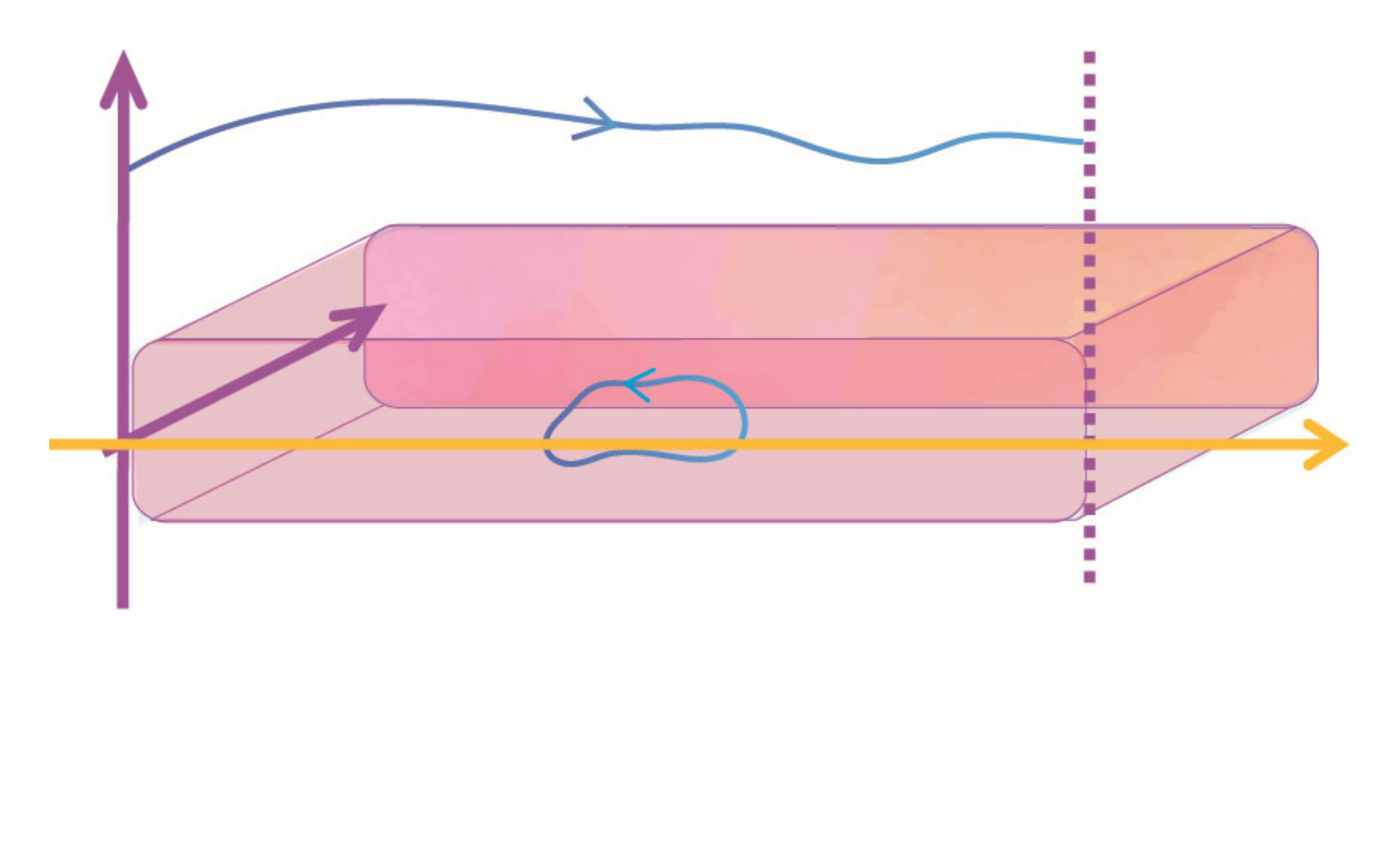}
   \put(-2,19){$x_1$}
   \put(-74,42){$x_{p+1},\ldots,x_9$}
  \put(-18,10){$2\pi R$}
 \end{picture}
 \end{center}
\vspace*{-1.5cm}
  \caption{Stack of $K$ coincident longitudinal D$p$-branes carrying a density $\nu>0$ of longitudinal fundamental string charge.
  Top left: The stack embedded in the 10-dimensional flat string Newton-Cartan (SNC) background. 
  Top right: When $\nu g_s^2\gg 1$, an alternative description emerges in terms of the RR black brane geometry (\ref{longdp}), which can be described schematically as consisting of an inner tubular region $r\, {\scriptstyle\lesssim}\, L$ that corresponds to the stack, and an outer region that corresponds to the surrounding flat SNC geometry.
  Bottom: The two schematic regions display significantly different properties. The inner region is a \emph{relativistic} bubble that can host the whole repertoire of perturbative and nonperturbative excitations of standard string theory, including unwound strings. The outer region, on the other hand, is asymptotically flat SNC, and is a welcoming environment only for excitations in the spectrum of \emph{nonrelativistic} string theory, such as positively-wound closed strings.}
  \label{fig:blackbraneinnrf1} 
  \end{figure}

By construction, the background (\ref{longdp}) is the RR black $p$-brane in NRF1. Just like it happened for the black string in the preceding subsection, (\ref{longdp}) is a solution of the  relativistic supergravity equations of motion, and becomes SNC only asymptotically \cite{Avila:2023aey}, with the identification\footnote{Naturally, the radial dependence of $\omega$ is identical to that of the NSNS black string (\ref{blackstring}) only when $p=1$. For $p>1$, the dependence is modified due to the uniform smearing of the positive F1 charge within the transverse directions occupied by the D-branes.} 
$\omega^2\equiv r^{7-p}/L^{7-p}$. The leading correction away from this flat SNC asymptopia was successfully reproduced in \cite{Guijosa:2023qym} via a calculation based on the worldsheet action (\ref{go}), which showed very explicitly that longitudinal D-branes induce a $\lambda\bar\lambda$ deformation. For this reason, we again find that the effect of the D$p$ stack can be described geometrically by the existence of an inner tubular bubble (now extended along $x^1,x^2,\ldots,x^p$), and that this bubble is surrounded by an outer region that asymptotes to flat SNC. See Fig.~\ref{fig:blackbraneinnrf1}.

As before, physics within the bubble is relativistic, and objects in that region can exit to the outer flat SNC region only if they are properly part of the spectrum of NRF1 theory. E.g., positively-wound F1s can escape, but unwound strings generally cannot. It is interesting to realize that this is in complete parallel with the overview of longitudinal D-branes given in Section~\ref{dbranesubsec}. The tubular bubble, extended along $p+1$ directions, corresponds to the D-brane (+F1) stack, and its transverse width is set by $L$. It is much larger than the string scale when $\nu g_s^2\gg 1$, which agrees with the analogous condition for the black string (\ref{blackstring}). The relativistic unwound closed strings that can only slide within the bubble are dual to the relativistic open strings that can only slide along the D-brane stack. And in both descriptions there is an outer 10-dimensional flat SNC region, which can only be explored as such by the certified inhabitants of NRF1 theory.

Early evidence for this picture was obtained already in \cite{Danielsson:2000mu}. It was shown there that, when considering a static F1 probe of (\ref{longdp}),  the energy cost for moving the probe from $r=0$ to $r\to\infty$ precisely matches the energy required for an F1 to become unbound from the D$p$-F1 stack. This was argued  in \cite{Danielsson:2000mu} to support the perspective (further developed later in \cite{Guijosa:2023qym}\cite{Avila:2023aey}) that the longitudinal RR brane in NRF1 includes a flat region external to the stack, and that, for this reason, the equivalence between the black brane and the corresponding D$p$ stack is analogous not to AdS/CFT-type duality, as initially believed, but to the open-string/closed-string duality between D-branes \cite{Dai:1989ua,Horava:1989ga} and RR black branes \cite{Horowitz:1991cd}, uncovered by \cite{Polchinski:1995mt,Strominger:1996sh,Gubser:1996wt,Callan:1996dv,Gubser:1996de,Maldacena:1996ix,Klebanov:1997kc,Gubser:1997yh,Gubser:1997se,Kruczenski:2007jg}.

The remaining alternative is a stack of $K$ \emph{transverse} D$p$-branes, which gives rise to the transverse RR black $p$-brane  \cite{Guijosa:2023qym,Avila:2023aey}. Due to the absence of F1 charge, this turns out to be a proper SNC background everywhere,\footnote{See footnote \ref{Lfoot}.}
\begin{equation}
\begin{aligned}
& \tau_{\mu}^{0}=\mathsf{H}^{-1/4}\delta^0_{\mu}~, \quad
\tau_{\mu}^{1}=\mathsf{H}^{1/4}\delta^1_{\mu}~, 
\\
E_{\mu\scriptscriptstyle ||}^{A'}&=\mathsf{H}^{-1/4}\delta_{\mu\scriptstyle ||}^{A'}~, \quad 
E_{\mu\scriptstyle \perp}^{A'}=\mathsf{H}^{1/4}\delta_{\mu\scriptstyle \perp}^{A'}~,
\\
m_{\mu}^{A}&=0~,
\\
\Phi&=\ln\left(g_{s}\,\mathsf{H}^{\frac{3-p}{4}}\right)~,
\\
B&=-\mu\, dt\wedge dx~,
\\
\mathsf{H} &= 1+\frac{L^{7-p}}{r_\perp^{6-p}}
\delta(x^1)~,
\\
L^{7-p}&\equiv 2^{5-p}\pi^{\frac{6-p}{2}}\,\Gamma\!\left(\frac{6-p}{2}\right)K g_s l_s^{\,7-p}~,
\end{aligned}
\label{dptrans}
\end{equation}
where the subindices $\mu\scriptstyle ||$ and
$\mu\scriptstyle\perp$
denote the transverse directions that are respectively parallel ($2,\ldots,p+1$) and perpendicular ($p+2,\ldots,9$) to the brane. 
Due to the form of $\mathsf{H}$, (\ref{dptrans}) is flat almost everywhere, with the exception of the 8-dimensional plane $x^1=0$. Within that plane, it becomes flat as $r_{\scriptstyle\perp}\to\infty$.
So, curiously, contrary to the black string and longitudinal RR brane, which were sources so strong that they distort the geometry from SNC form, the transverse RR brane is so weak that it produces a negligible effect except right on the plane that sits at the longitudinal position of the D$p$ stack. An interesting observation made recently in \cite{Lambert:2024uue,Fontanella:2024kyl,Fontanella:2024rvn,Blair:2024aqz} 
is that one can dial up the strength of the source by sending $\hat{K}\to\infty$ in such a way 
as to obtain a nontrivial SNC geometry everywhere. Notice that, unlike the unphysical switch-off scaling $\hat{N}\to 0$ that we briefly alluded to in the third-to-last paragraph of Section~\ref{blackstringsubsec}, the discrete nature of $\hat{K}$ is not fundamentally at odds with the strengthening that is of special interest in the case of the transverse RR brane.  

Comparing the outcomes for the black string and RR longitudinal black brane against the one for the transverse black brane, we see that, in all cases, the $\lambda\bar{\lambda}$ deformation that takes us out of the SNC formalism by generating an inner relativistic bubble is induced by the positively-wound F1s. This entails that even excited transverse D-branes can be treated within the pure SNC realm, but only if the number of open F1s ending on them (characterized by the NR spectrum (\ref{opentransnrf1})) is kept small. A large collection of such coincident open strings would again have a geometric description in terms of a black brane background that is only \emph{asymptotically} SNC \cite{Avila:2023aey}.

Other brane backgrounds of NRF1 theory, such as those generated by stacks of NS5-branes, can also be worked out by starting with the known solution in relativistic string theory and taking the limit \cite{Guijosa:2001notes}. We will refrain from writing down the explicit expressions here, but the result for   
coincident transverse NS5-branes is a background that is delta-function localized at the position of the branes, and for longitudinal NS5-branes, it is a background that is purely SNC.\footnote{We are reporting this in modern parlance here, not available at the time of \cite{Guijosa:2001notes}.}
Longitudinal NS5-branes can carry F1 charge by forming a threshold bound state. As expected, in that case the NRF1 limit does yield an \emph{asymptotically} SNC background, again due to the substantial backreaction of the fundamental strings. 

In all cases, we come to understand SNC geometries as the sector of solutions of NRF1 theory where no significant sources with positive longitudinal F1 winding are present. This sector is important because it naturally plays the role of asymptopia for the most general solutions.

\section{NR Brane Theories and Relativistic Holography}
\label{branesec}

In this section we review the most salient properties of the NR brane theories of \cite{Danielsson:2000gi,Gomis:2000bd}, as well as their very significant overlap with standard, relativistic holography. The latter was discovered recently in \cite{Blair:2024aqz}. To keep advancing towards the main goal of the present work, we will review this discovery of \cite{Blair:2024aqz} not within the approach of that paper, but from the perspective of \cite{Guijosa:2023qym,Avila:2023aey}, which we have just reviewed in Sections~\ref{blackstringsubsec} and \ref{blackpbranesubsec}.

\subsection{Nonrelativistic Brane Theories}
\label{nrbranesubsec}

As explained in \cite{Danielsson:2000gi,Gomis:2000bd}, and recalled in and around Figs.~\ref{fig:dualities} and \ref{fig:dualities2} in the Introduction, the NRF1 limit (\ref{nrf1limit}) can be U-dualized to obtain analogous limits where branes of string or M theory other than the fundamental string become the lightest excitations,
 thereby defining nonrelativistic \emph{brane} theories. For these we will use as in Figs.~\ref{fig:dualities},\ref{fig:dualities2} the acronym 
NRX$p$ (see footnote~\ref{wrappedfoot}), where X refers to the type of brane that is fundamental, and $p$ declares the brane's spatial dimensionality, e.g., NRD1 or NRM2. These are theories defined in $d=10$ or $d=11$ dimensions with a distinguished $(p+1)+(d-p-1)$ foliation. They contain a variety of objects, most of which are required to carry positive X$p$ wrapping number $w$ (equivalently, positive X$p$ charge) along the $p+1$ longitudinal directions $x^A$ ($A=0,1,\ldots,p$). Due to the conservation of X$p$ charge, just like in NRF1 we can restrict attention to a superselection sector with a fixed total longitudinal wrapping number $N$. Physics along the $d-p-1$ transverse directions $x^{A'}$ ($A'=p+1,p+2,\ldots,d-1$) is nonrelativistic, and there are exceptional `Newtonian gravitons' \cite{Danielsson:2000mu}, massless modes of unwrapped ($w=0$) X$p\,$s 
that are required to have vanishing transverse momentum when on shell, and that mediate instantaneous long-range interactions when off shell. 

For our purposes here, we will focus on the setup that defines the (asymptotically) flat-space NRD$p$ theory, which as indicated in Figs.~\ref{fig:dualities} and \ref{fig:dualities2}, is obtained from NRF1 IIB via S duality followed by T$_{2\ldots,p}$ duality. As in (\ref{prelimitbkg}), we start with Type IIA/B string theory on 10-dimensional Minkowski space, with a foliation that distinguishes between $p+1$ longitudinal and $9-p$ transverse directions, turning on a constant critical value for the RR $(p+1)$-form field, and compactifying the spatial longitudinal directions on a $p$-torus with size specified by a common radius $\hat{R}$:
\begin{equation}
\hat{G}_{\mu\nu}=\eta_{\mu\nu}~,
\quad
\hat{C}_{(p+1)}=d\hat{t}\wedge d\hat{x}_1\wedge\cdots\wedge d\hat{x}_p~,
\quad
\hat{x}_{1,\ldots,p}\simeq \hat{x}_{1,\ldots,p}+2\pi\hat{R}~.
\label{prelimitpbranebkg}
\end{equation}
Then we apply the limit that is the T$_{2\ldots p}\,\circ$\,S -dual image of (\ref{nrf1limit})
\cite{Danielsson:2000gi,Gomis:2000bd,Blair:2024aqz},
\begin{equation}
\begin{aligned}
\omega \to \infty, & \quad \mbox{with}
\\
\hat{x}^A&=\omega^{1/2} x^A\quad (A=0,\ldots,p),
\\
\hat{x}^{A'}&=\omega^{-1/2} x^{A'}\quad (A'=p+1,\ldots,9)~.
\\
\hat{g}_s&=\omega^{\frac{p-3}{2}}g_s~,
\quad
\hat{l}_s =l_s~,
\\
\hat{N}&=N~, 
\quad
\hat{R}=\omega^{1/2} R~.
\\
\end{aligned}
\label{nrdplimit}
\end{equation}
This converts (\ref{prelimitpbranebkg}) to
\begin{equation}
\begin{aligned}
\widehat{ds}^2&=\omega\, dx^A dx^B \eta_{AB}
+\omega^{-1}dx^{A'} dx^{B'} {\delta}_{A'B'}~,
\\
\hat{C}_{(p+1)}&=
\omega^{\frac{p+1}{2}}\,dt\wedge dx_1\wedge\ldots\wedge dx^p~,
\\
x_{1,\ldots,p}&\simeq
{x}_{1,\ldots,p}+2\pi R~.
\end{aligned}
\label{nrdpbkglimit}
\end{equation}
In Section~\ref{nrinrsec} we will be especially interested in the NRD3 limit,
\begin{equation}
\begin{aligned}
\omega \to \infty, & \quad \mbox{with}
\\
\hat{x}^A&=\omega^{1/2} x^A\quad (A=0,\ldots,3),
\\
\hat{x}^{A'}&=\omega^{-1/2} x^{A'}\quad (A'=4,\ldots,9)~.
\\
\hat{g}_s&=g_s~,
\quad
\hat{l}_s =l_s~,
\\
\hat{N}&=N~, 
\quad
\hat{R}=\omega^{1/2} R~.
\end{aligned}
\label{nrd3scaling}
\end{equation}
which goes along with
\begin{equation}
\begin{aligned}
\omega \to \infty, & \quad \mbox{with}
\\
\widehat{ds}^2&=\omega\, dx^A dx^B \eta_{AB}
+\omega^{-1}dx^{A'} dx^{B'} {\delta}_{A'B'}~,\\
\hat{C}_{(4)}&=
\omega^2\, dt\wedge dx_1\wedge dx_2\wedge dx_3 + \mbox{self-dual}~,
\end{aligned}
\label{nrd3bkgscaling}
\end{equation}
where `self-dual' alludes to the additional components of $\hat{C}_{(4)}$ that ensure the self-Hodge-duality of $F_{(5)}\equiv dC_{(4)}$.

The natural setting to generalize the flat-space NRX$p$ theories of \cite{Danielsson:2000gi,Gomis:2000bd} to curved space is a $p$-brane generalization of Newton-Cartan geometry \cite{Kluson:2017abm,Pereniguez:2019eoq,Blair:2021waq,Ebert:2021mfu,Novosad:2021tlq,Bergshoeff:2023rkk,Ebert:2023hba,Bergshoeff:2024ipq}, henceforth abbreviated $p$NC. In this notation, 0NC refers to the original, particle Newton-Cartan formalism \cite{Cartan:1923zea,Cartan:1924yea,Hartong:2022lsy,Bergshoeff:2022eog}, and 1NC refers to the string Newton-Cartan (SNC) framework \cite{Andringa:2012uz,Harmark:2017rpg,Bergshoeff:2018yvt,Harmark:2018cdl,Gomis:2019zyu,Gallegos:2019icg,Harmark:2019upf,Bergshoeff:2021bmc,Bidussi:2021ujm,Oling:2022fft} recalled in Section~\ref{sncsubsec}. More precisely, one should bear in mind that for a given value of $p$ there can be more than one NR brane theory, each calling for a somewhat different generalization of the Newton-Cartan formalism. The name 1NC, for instance, could refer to the framework relevant for NRF1 or for NRD1 theory, which involve different scalings. For a more complete specification, it would be necessary to allude to `X$p$NC' geometry (where X again is a placeholder for the brane type), but in this paper we will omit the X, to avoid clutter.

The key elements of $p$NC are again the longitudinal and transverse vielbeine
$\tau_{\mu}^{A}(x)$ and $E_{\mu}^{A'}(x)$, which allow us to generalize (\ref{nrdplimit}) by replacing $dx^A\to\tau_{\mu}^{A}dx^{\mu}$ and $dx^{A'}\to E_{\mu}^{A'}dx^{\mu}$. The statement that is U-dual to (\ref{nrf1bkg}) is that, \emph{if} a parameter $\omega\to\infty$ can be found such that an ordinary IIA/B supergravity background obeys the scaling \cite{Ebert:2021mfu,Blair:2023noj,Gomis:2023eav,Blair:2024aqz}
\begin{equation}
\begin{aligned}
\hat{G}_{\mu\nu}&=\omega\,\tau_{\mu\nu}
+\omega^{-1}E_{\mu\nu}~,
\\
\hat{C}_{(p+1)}&=\omega^{\frac{p+1}{2}}\, e^{-\Phi}\tau^0\wedge\cdots\wedge\tau^p + C_{(p+1)}~,
\\
\exp(\hat{\Phi})&=\omega^{\frac{p-3}{2}}\Phi~,
\\
\hat{C}_{(q+1)}&=C_{(q+1)}\quad\mbox{if}\; q\neq p~,
\\
\hat{B}_{(2)}&=B_{(2)}~,
\end{aligned}
\label{nrdpbkg}    
\end{equation}
then this limit defines NRD$p$ theory on the $p$NC geometry characterized by the NSNS fields 
$\tau_{\mu\nu},E_{\mu\nu},B_{(2)},\Phi$ and the RR fields $C_{(p+1)},C_{(q+1)}$.

\subsection{Black RR $p$-brane in NRD$p$: connection with holography}
\label{holographysubsec}

{}From the findings of \cite{Danielsson:2000mu,Guijosa:2023qym,Avila:2023aey}, we known that, generically, collections of objects in NRD$p$ gravitate, producing effects that can be described geometrically in terms of black brane backgrounds that schematically consist of two regions: an inner tubular bubble where physics is \emph{relativistic}, and an outer, asymptotically flat $p$NC geometry. Exiting the relativistic bubble corresponds to exiting the given collection of objects.  

Since D$p$-branes with positive longitudinal wrapping are the fundamental objects of NRD$p$, the most natural collection of objects to consider is a stack of $N$ static coincident longitudinal D$p$s. As recalled in the Introduction, Refs.~\cite{Danielsson:2000gi,Gomis:2000bd} showed that the NRD$p$ limit (\ref{nrdplimit}) is precisely the endpoint of Seiberg's derivation \cite{Seiberg:1997ad} of the Matrix theory description for DLCQ M theory on a transverse $p$-torus, and for this reason, the worldvolume description of a stack of $N$ D$p$s in NRD$p$ theory is given precisely by $U(N)$ maximally supersymmetric Yang-Mills (MSYM) in $p+1$ dimensions (i.e., the dimensional reduction of 10-dimensional $\cN=1$ SYM to the indicated number of dimensions). 

One can then inquire about the geometric description of this D$p$ stack as a RR black $p$-brane. This question is U-dual to the derivation of the black string in NRF1 theory, carried out in \cite{Avila:2023aey} and recalled above, in Section~\ref{blackstringsubsec}. In complete parallel with our discussion there, the simplest route is to start with the RR black $p$-brane of relativistic Type II string theory \cite{Horowitz:1991cd}, and apply the 
NRD$p$ limit (\ref{nrdplimit}). The explicit calculation was only carried out very recently in \cite{Blair:2024aqz} (with a different motivation), and led to the remarkable discovery that the resulting black $p$-brane,\footnote{See footnote~\ref{Lfoot}.}
\begin{equation}
\begin{aligned}
\widehat{ds}^2&=\left(\frac{r}{L}\right)^{\frac{7-p}{2}} dx^A dx^B \eta_{AB}
+\left(\frac{L}{r}\right)^{\frac{7-p}{2}}dx^{A'} dx^{B'} {\delta}_{A'B'}~,\\
\hat{g}_s^{-1}\hat{C}_{(p+1)}&=
{g}_s^{-1}\left(\frac{r}{L}\right)^{7-p} dt\wedge dx_1\wedge \ldots\wedge dx_p
\\
e^{\hat{\Phi}}&=\left(\frac{r}{L}\right)^{\frac{(7-p)(p-3)}{4}}g_s~,
\\
L^{7-p}&\equiv (4\pi)^{\frac{5-p}{2}}\Gamma\left(\frac{\scriptstyle 7-p}{\scriptstyle 2}\right) Ng_s l_s^{7-p}~,
\end{aligned}
\label{nrblackbrane}
\end{equation}
is the exact same AdS-type background that has been at play for a quarter century in the D$p$-brane implementation of holography \cite{Maldacena:1997re,Itzhaki:1998dd}! We reiterate that we have purposefully reported here this pivotal finding of \cite{Blair:2024aqz} already incorporating the viewpoint of \cite{Guijosa:2023qym,Avila:2023aey}, which is what allows us to further recognize that  (\ref{nrblackbrane}) is in fact the RR black $p$-brane that encodes the physics of a stack of D$p$-branes in NRD$p$ theory. 

To better assimilate this surprise, it is convenient to review its derivation in \cite{Blair:2024aqz} for the special case of a stack of D3-branes in NRD3 theory. Famously, the extremal black 3-brane in relativistic IIB string theory is \cite{Horowitz:1991cd}
\begin{equation}
\begin{aligned}
\widehat{ds}^2&=H(\hat{r})^{-1/2}d\hat{x}^A d\hat{x}^B \eta_{AB}
+H(\hat{r})^{1/2}d\hat{x}^{A'} d\hat{x}^{B'} \delta_{A'B'}~,
\\
\hat{C}_{(4)}&=
H(\hat{r})^{-1}d\hat{t}\wedge d\hat{x}_1\wedge d\hat{x}_2\wedge d\hat{x}_3 + \mbox{self-dual}~,
\\
H(\hat{r})&=1+\frac{\hat{L}^4}{\hat{r}^4}~,
\\
\hat{r}^2&\equiv \hat{x}^{A'} \hat{x}^{B'} \delta_{A'B'}~,
\\
\hat{L}^4&\equiv 4\pi\hat{N}\hat{g}_s\hat{l}_s^4~. 
\end{aligned}
\label{ordinaryblackthreebrane}
\end{equation}
Compared to the standard convention for the RR field in this solution, we have purposefully shifted $\hat{C}_{0123}$ by a constant,  so that, instead of vanishing at infinity, it asymptotes to $1$, as required by (\ref{prelimitpbranebkg}).
If we now simply apply to this background the NRD3 limit (\ref{nrd3scaling}), we see that $\hat{L}=L$ and $\hat{r}^4=\omega^{-2} r^{4}$, so the harmonic function loses the 1 and becomes $H=\omega^2 L^4/r^4$, and we obtain\footnote{See footnote~\ref{Lfoot}.} 
\begin{equation}
\begin{aligned}
\widehat{ds}^2&=\frac{r^2}{L^2}dx^A dx^B \eta_{AB}
+\frac{L^2}{r^2}dx^{A'} dx^{B'} \delta_{A'B'}~,
\\
\hat{C}_{(4)}&=
\frac{r^4}{L^4}dt\wedge dx_1\wedge dx_2\wedge dx_3 + \mbox{self-dual}~,
\\
L^4&\equiv 4\pi N g_s l_s^4
\end{aligned}
\label{adsbkg}
\end{equation}
which we recognize as the one and only
 AdS$_5\times$S$^5$. We thus verify that, as elucidated in \cite{Blair:2024aqz}, {\bf the NRD$p$ limit of \cite{Danielsson:2000gi,Gomis:2000bd} is \emph{exactly} the same as Maldacena's near-horizon limit \cite{Maldacena:1997re,Itzhaki:1998dd}}! 

Coupling this result with the insights of \cite{Danielsson:2000mu,Guijosa:2023qym,Avila:2023aey} reviewed in the preceding section, we come to understand that the profusely-studied AdS$_5\times$S$^5$ background is nothing more and nothing less than the black 3-brane within NRD3 theory. And, knowing that (\ref{adsbkg}) is the image under T$_{23}\,\circ$\,S duality of the black string of NRF1, Eq.~(\ref{blackstring}), everything we said in Subsection~\ref{blackstringsubsec} applies analogously to AdS$_5\times$S$^5$.
We will elaborate on these properties in the following section.

\section{Nonrelativity in Relativistic Holography}
\label{nrinrsec}

At the end of the previous section, we have found that, by bringing together the developments in \cite{Danielsson:2000mu,Guijosa:2023qym,Avila:2023aey} and \cite{Blair:2020ops,Blair:2023noj,Gomis:2023eav,Blair:2024aqz}, 
holographic duality for D$p$-branes \cite{Maldacena:1997re,Itzhaki:1998dd} is  revealed to be exactly the same statement as  open-string/closed-string duality, i.e., D-brane/black-brane duality, in the nonrelativistic D$p$-brane theory of \cite{Danielsson:2000gi,Gomis:2000bd}. In more detail, the starting point for the classic derivation of AdS/CFT is the statement 
\cite{Polchinski:1995mt,Strominger:1996sh,Gubser:1996wt,Callan:1996dv,Gubser:1996de,Maldacena:1996ix,Klebanov:1997kc,Gubser:1997yh,Gubser:1997se,Kruczenski:2007jg}
that 
\begin{equation}
\fbox{
\begin{minipage}{6cm}
\begin{center}
IIA/B string theory on \\
stack of $N$ D$p$-branes \\ 
in flat 10-dim Minkowski
\end{center}
\end{minipage} 
\huge
$=$ 
\normalsize
\begin{minipage}{6cm}
\begin{center}
IIA/B string theory on\\
asymptotically flat \\
RR black $p$-brane
\end{center}
\end{minipage}}~,
\label{polchinski}
\end{equation}
and Maldacena's low-energy/near-horizon limit transmutes this equivalence into\footnote{Here we write AdS$_{p+2}\times$S$^{8-p}$ 
as a metonym for  the conformally-related \cite{Skenderis:1998dq,Behrndt:1999mk} near-horizon geometry of the RR black $p$-brane \cite{Itzhaki:1998dd}, Eq.~(\ref{nrblackbrane}). And for conciseness, in the gravity side of this box we do not make reference to the various alternative string or M theory descriptions that are needed in various energy regimes when 
$p\neq 3$ \cite{Itzhaki:1998dd}.}  
\begin{equation}
\fbox{
\begin{minipage}{6cm}
\begin{center}
$SU(N)$ MSYM on \\
$(p+1)$-dim Minkowski 
\end{center}
\end{minipage} 
\huge
$=$ 
\normalsize
\begin{minipage}{6cm}
\begin{center}
IIA/B string theory on\\
AdS$_{p+2}\times$S$^{8-p}$
\end{center}
\end{minipage}}~.
\label{maldacena}
\end{equation}
Based on recent advances, particularly \cite{Avila:2023aey} and \cite{Blair:2024aqz}, in the present paper we are emphasizing that the identity between the near-horizon and NRD$p$ limits implies that {\bf the AdS/CFT\footnote{For simplicity, we employ the AdS/CFT denomination even for $p\neq 3$, where the setup is in truth non-AdS/non-CFT.} correspondence (\ref{maldacena}) can in fact be reinterpreted as the familiar statement of open-string/closed-string duality (\ref{polchinski}) in NRD$p$ theory}, namely,
\begin{equation}
\fbox{
\begin{minipage}{6cm}
\begin{center}
NRD$p$ theory on \\
stack of $N$ D$p$-branes \\ 
in flat 10-dim $p$NC spacetime
\end{center}
\end{minipage} 
\huge
$=$ 
\normalsize
\begin{minipage}{6cm}
\begin{center}
NRD$p$ theory on\\
asymptotically flat $p$NC \\
RR black $p$-brane
\end{center}
\end{minipage}}~.
\label{polchinskip}
\end{equation}
 As follows from our discussion in the preceding sections --- and as we will elaborate below --- both sides of the equivalence (\ref{polchinskip}) exhibit a mix of relativistic and nonrelativistic features. This structure unfolds in such a way that the present reinterpretation does not entail any contradiction with the well-known relativistic character of (\ref{maldacena}).

This very direct connection between standard holography and nonrelativistic brane theories holds equally for the cases based on M2-,  M5- or NS5-branes (described from the holographic perspective in \cite{Maldacena:1997re,Itzhaki:1998dd,Aharony:1998ub}, and from the NR brane perspective in \cite{Danielsson:2000gi}\cite{Gomis:2000bd}), as well as for cases associated with intersections of branes of various types, a rather thorough analysis of which was given recently in \cite{Blair:2024aqz}. 

In this section, we will present additional observations that expand on this main idea. We expect most of our comments to apply for general $p$, but for concreteness, we will focus on the $p=3$ case,
\begin{equation}
\fbox{
\begin{minipage}{6cm}
\begin{center}
NRD3 theory on \\
stack of $N$ D3-branes \\ 
in flat 10-dim $3$NC spacetime
\end{center}
\end{minipage} 
\huge
$=$ 
\normalsize
\begin{minipage}{6cm}
\begin{center}
NRD3 theory on\\
asymptotically flat 3NC \\
RR black $3$-brane
\end{center}
\end{minipage}}~.
\label{polchinski3}
\end{equation}

Taking advantage of our newly-gained understanding that (\ref{polchinski3}) is exactly the same equivalence as the 
acclaimed
 $p=3$ instance of the AdS/CFT correspondence (\ref{maldacena}), in the remainder of the paper we will alternate between these two alternative interpretations, by making repeated use the following two facts:
\begin{enumerate}
\item The `open string' or `field theory' side (left-hand side) of (\ref{polchinski3}) is described precisely by\footnote{For the difference between $U(N)$ and $SU(N)$, see Section~\ref{cmsubsec}.} $U(N)$  maximally supersymmetric Yang-Mills (MSYM) theory on 4-dimensional Minkowski.
\item The `closed string' or `gravity' side (right-hand side) of (\ref{polchinski3}) involves precisely AdS$_5\times$S$^5$.  
\end{enumerate}

\subsection{\emph{Not} an awful waste of space}
\label{awfulsubsec}

Since AdS refocuses timelike geodesics, we are used to thinking of it as a box that prevents massive objects from escaping to infinity. It thus seems like the infinite volume of space that is adjacent to the conformal boundary is  unexcitable, essentially wasted. It is the realm of \emph{non}-normalizable modes, which require infinite energy to activate, and take us to a different theory. 

The coupled insights of \cite{Avila:2023aey} and \cite{Blair:2024aqz} refine this intuition. Knowing that AdS$_5\times$S$^5$ \emph{is in fact an asymptotically flat 3NC geometry}, we deduce that objects carrying positive D3 charge should in fact be able to escape into the flat region.\footnote{The reason this does not contradict the refocusing property of AdS is of course that charged objects do not follow geodesics.} In other words, that infinite volume of space out there can in fact be colonized.
 
\begin{figure}[thb]
\begin{center}
 \begin{picture}(75,85)
   \includegraphics[width=7.0cm]{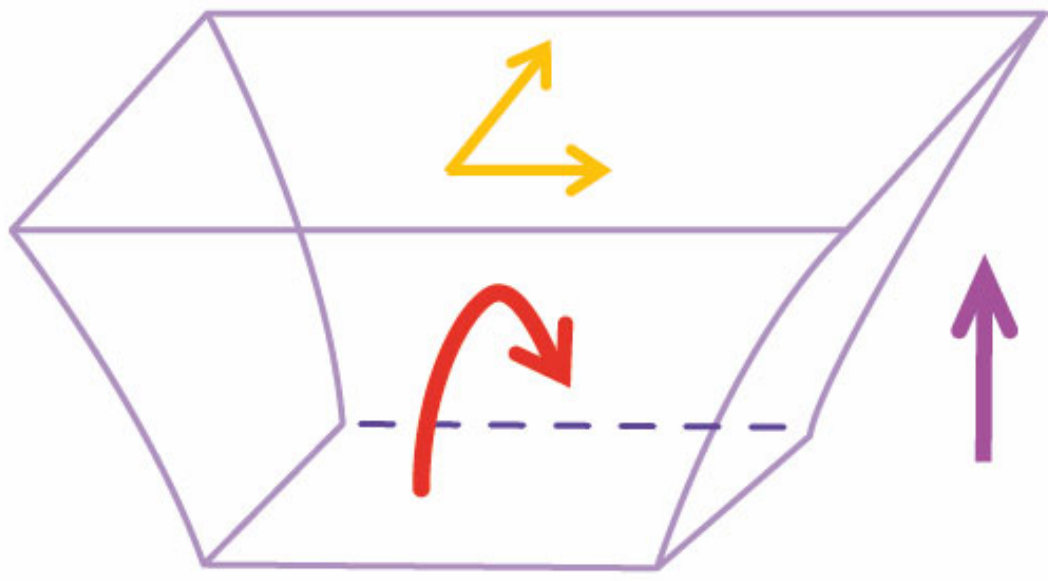}
   \put(-5,67){$r$}
   \put(-33,76){$\,\vec{x}$}
  \put(-15,14){$2\pi R$}
 \end{picture}
\begin{picture}(150,40) 
  \includegraphics[width=7.0cm]{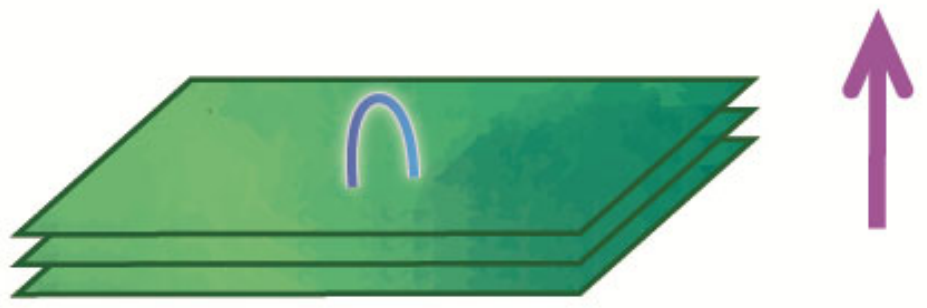}
  \hspace*{0.8cm}
  \includegraphics[width=7.0cm]{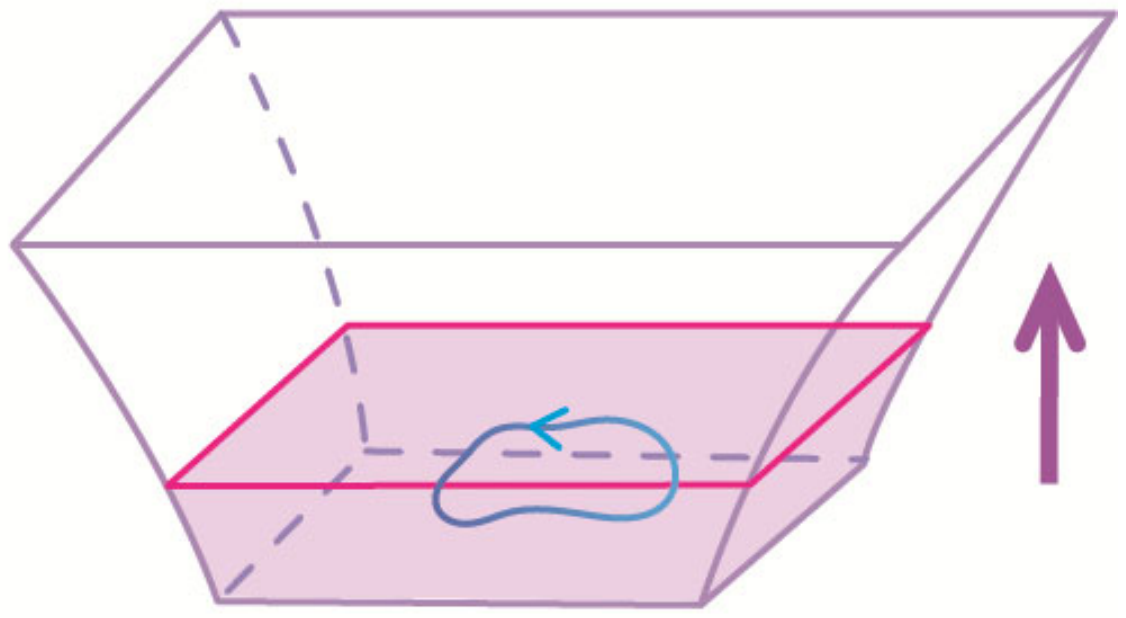}
  \end{picture}
 \vspace*{-4.5cm}
 \end{center}
\vspace*{-0.5cm}
  \caption{Representations of anti-de Sitter (AdS) space.
  Top: Standard schematic depiction of AdS in Poincar\'e coordinates. The red arrow serves as a reminder of the usual understanding of AdS as a gravitational box that traps massive objects. 
  Bottom: {}AdS$_5\times$S$^5$ as the RR black 3-brane (right) of nonrelativistic D3-brane (NRD3) theory, which provides an alternative description of a stack of coincident D3-branes in 10-dimensional flat 3-Newton-Cartan (3NC) spacetime (right). The black brane consists schematically of two qualitatively different regions: an inner relativistic bubble, dual to the stack, and an outer asymptotically flat 3NC region. {}From this perspective we understand that the bubble serves as a box that naturally traps its intrinsic excitations, which are dual to excitations of the D-brane stack. E.g., a closed string in the bubble cannot escape, because it corresponds to a bound state of open strings living within the D-brane stack.}
  \label{fig:ads} 
  \end{figure}

The `box' portion of AdS is none other than the relativistic tubular bubble generated by a stack of D3-branes in NRD3, so objects that are trapped by it are precisely the ones whose existence is only made possible by the bubble's emergence. They cannot escape to the flat region because \emph{they are not properly bulk excitations of NRD3}.\footnote{Notice that, with respect to standard AdS/CFT usage, here we have been led to a near-complete reversal of the notion of `bulk', by which we mean the 3NC region outside of the D3 stack (schematically, $r\geq L$).} They can only slide along the D3 stack, or its gravitational equivalent. See Figure~\ref{fig:ads}.

The most rightful inhabitants of any location on a (fully or asymptotically) 3NC geometry are evidently the D3-branes themselves. Within D3NR theory, D3s are not solitonic objects; they are instead the lightest excitations. 
\emph{A priori} we should not think of them in the usual manner as objects defined by the property that their excitations take the form of open strings. Out in the 3NC region, such open strings would have infinite energy,\footnote{Unless they are purely transverse--- see Section~\ref{f1d1subsec},} as we know directly from the effect of the NRD3 limit, and as we can also intuit from the possibility of realizing F1s as thin D3 tubes that carry a worldvolume electric field \cite{Callan:1997kz,Gibbons:1997xz}. {}From this second perspective, a portion of the open string excitation would necessarily be oriented as a \emph{negatively-}wrapped D3 (i.e., an anti-D3), which is forbidden.  
Instead,  the degrees of freedom of D3s in NRD3 are only their embedding fields and their internal gauge field excitations (plus superpartners). As we recalled in Section~\ref{introsec}, their dynamics are  precisely described by $(3+1)$-dimensional $U(N)$ MSYM. 
This theory of course does include the non-Abelian degrees of freedom that from the parent string theory perspective are due to open strings with endpoints of any one of the $N$ D3s in question. This well-known fact reflects the special way in which D3s within NRD3 collaborate to produce joint excitations.

\subsection{MSYM is truly 10-dimensional and nonrelativistic}
\label{trulysubsec}

We normally think of MSYM as a relativistic theory defined on a strictly $(3+1)$-dimensional spacetime. This is natural from the perspective of the D3-brane stack in the parent string theory, because the low-energy Maldacena limit that reduces the worldvolume action to MSYM form drastically shrinks the proper size of the directions perpendicular to the D3 stack, seemingly collapsing them down to zilch. 

Now that we know \cite{Blair:2024aqz} that the $p=3$ version of the Maldacena limit is precisely the same as the limit that transmutes standard Type IIB string theory into the NRD3 theory, we are afforded a different perspective. As stated in and around (\ref{polchinski3}), starting with the D3s of the parent theory and applying the limit, what we obtain is simply a stack of D3s within the flat 10-dimensional 3NC geometry. In other words, \emph{MSYM truly lives on a 10-dimensional manifold, equipped with a 3NC foliation that distinguishes between the four longitudinal directions $x^{A}$ and the six transverse directions $x^{A'}$, with the corresponding vielbeine being flat}, $\tau^A_{\mu}=\delta^A_{\mu},\, E^{A'}_{\mu}=\delta^{A'}_{\mu}$. With respect to this intrinsic notion of transverse proper distance, directions $x_4,\ldots,x_9$ are still infinitely large. 

The MSYM (equivalently, D3-in-NRD3) action has the well-known form
\begin{equation}
S_{\mbox{\tiny MSYM}}=-\int d^4 x\,
\tr{\left(
\frac{1}{2g^2_{\mbox{\tiny YM}}}F_{AB}F^{AB}
+D_A \Phi^{A'}D^A\Phi_{A'}
-\frac{g^2_{\mbox{\tiny YM}}}{2}[\Phi^{A'},\Phi^{B'}][\Phi_{A'},\Phi_{B'}]+\ldots\right)},
\label{msym}
\end{equation}
where $g^2_{\mbox{\tiny YM}}\equiv 2\pi g_s$, and the dots allude to the terms with fermions that provide the $\cN=4$ supersymmetric completion.  
The authors of \cite{Blair:2024aqz} showed how $S_{\mbox{\tiny MSYM}}$ follows from taking the NRD3 limit of the non-Abelian Dirac-Born-Infeld (DBI) action worked out in \cite{Myers:1999ps}.
The extension of (\ref{msym}) to more general 3NC vielbeine together with NSNS and RR background fields was also found in \cite{Blair:2024aqz}.

Our assertion that (\ref{msym})
describes in truth a ten-dimensional theory might seem severely misplaced, because it evidently involves an integral over a spacetime that is only 4-dimensional. In this familiar presentation of the action, there is a marked distinction between the longitudinal directions $x^A$, which describe spacetime itself, and the transverse directions $x^{A'}$, which appear only through the scalar matrix fields $X^{A'}\equiv (2\pi l_s^2) \Phi^{A'}$. 
This difference, however, stems merely from the choice of static gauge, a choice that is standard even in the relativistic context, and is particularly well suited to the system after the NRD3 limit, due to the ensuing positive-longitudinal-wrapping requirement  \cite{Danielsson:2000gi,Gomis:2000bd}. 

The explicit worldvolume-covariant form of the MSYM action on flat 3NC space is not currently known. To derive it through the same procedure as in \cite{Blair:2024aqz}, one would need to start from the covariant version of the Myers action \cite{Myers:1999ps}, which in itself is not known to date in manageable form--- see
\cite{DeBoer:2001uk,Brecher:2004qi,Brecher:2005sj,Ferrari:2013pi}, and especially \cite{Howe:2006rv}. 

This is just a technical limitation: even if the specifics are not known, it is clear that there ought to exist a covariant non-Abelian action with ten embedding fields $X^{\mu}(\sigma)$, which reduces to (\ref{msym}) upon choosing the static gauge. Of course, it will not involve all embedding fields on exactly the same footing, because, much like the static-gauge NRD$p$ worldvolume action in background fields derived recently in \cite{Blair:2024aqz}, it must reflect the relativistic nature of the longitudinal $X^{A}(\sigma)$ and the nonrelativistic character of the transverse $X^{A'}(\sigma)$. 
But even so, in that presentation it would be manifest that, even though the worldvolume is 4-dimensional, the \emph{target space} is fully 10-dimensional, just like for D3s in relativistic string theory.
What is special about the NRD3 theory is that, due to the effect of the limit, (\ref{msym}) or its covariant counterpart capture the \emph{complete} dynamics of the theory in the sector with $N$ D3-branes.  This is somewhat like the worldsheet formulation of relativistic string theory, or of F1s in NRF1 theory, but with the advantage that the non-Abelian action (\ref{msym}) in fact provides a second-quantized description of D3-branes in NRD3.  

One aspect of NRD3 that \emph{is} manifest in (\ref{msym}) is the nonrelativistic nature of the transverse dynamics. Indeed, the fact that the kinetic term is simply $\propto (\p_t\vec{X}_{\perp})^2$, instead of the usual DBI form  $\propto\sqrt{1-(\p_t\vec{X}_{\perp})^2}$ \cite{Leigh:1989jq,Myers:1999ps}, clearly indicates that there is no limiting speed for transverse motion. In the standard interpretation of the scalar fields $X^{A'}$ as variables in an abstract \emph{internal} space, we expect not to find an association with a relativistic lightcone. The NC limit is thus recognized as the natural way to turn a spatial direction of a Lorentzian spacetime into an internal space direction.  This points to a generalization of the standard Kaluza-Klein story, where any field theory defined on a $(p+1)$-dimensional spacetime and having a $(d-p-1)$-dimensional internal space should now be describable in terms of some variant of a $d$-dimensional $p$NC geometry.

\subsection{D3 probe: relativistic-to-nonrelativistic escape}
\label{d3subsec}

As we have emphasized, on \emph{both} sides of the AdS/CFT correspondence the theory is 10-dimensional, and includes the flat 3NC region that is external to the D3s. The simplest way to explore the full geometry is using as a probe one of the D3s themselves, say the first one, and allowing it to move along one of the six transverse Cartesian directions, say $x^9$. As usual, this is achieved by giving an expectation value to the specific scalar field component $\vev{X^9_{11}}$.
  
\begin{figure}[thb]
\begin{center}
\begin{picture}(150,85) 
  \includegraphics[width=7.0cm]{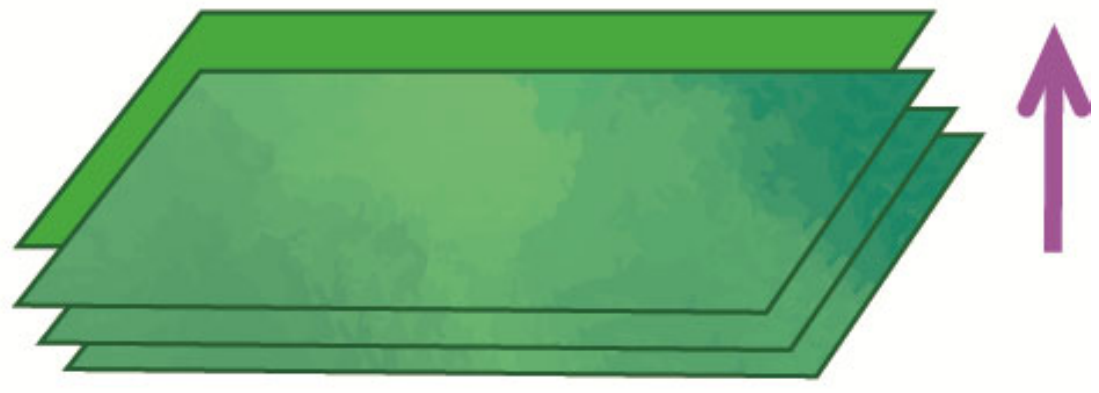}
  \hspace*{0.8cm}
  \includegraphics[width=7.0cm]{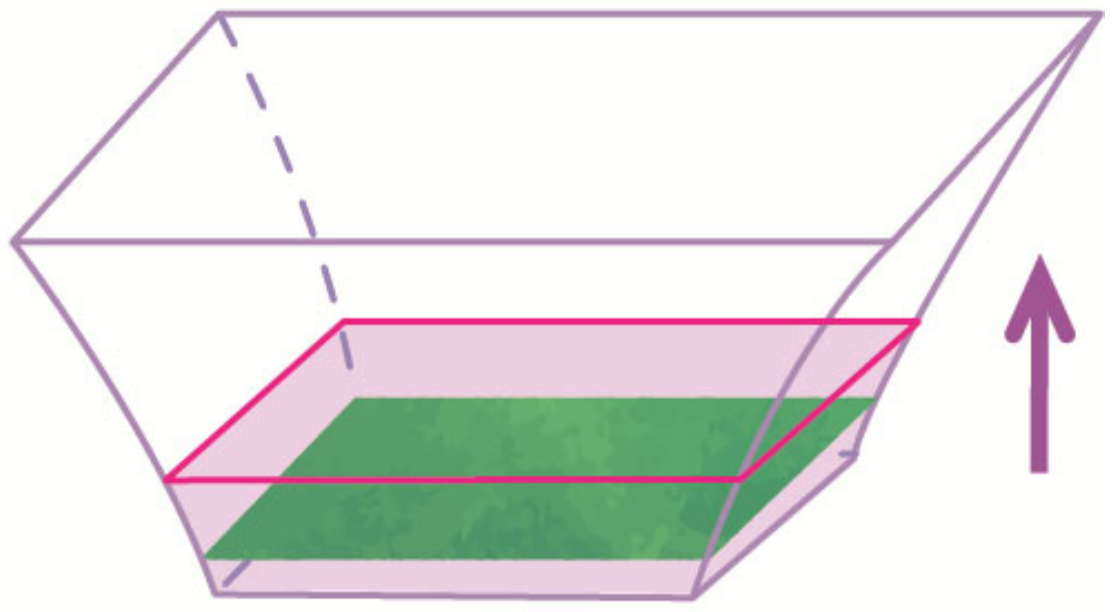}
 \end{picture}
 \begin{picture}(150,45)
   \includegraphics[width=7.0cm]{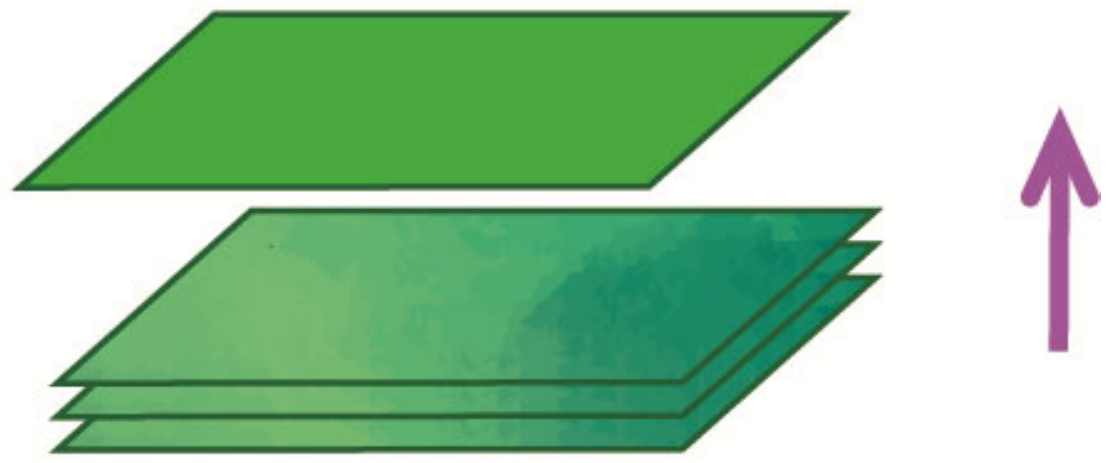}
    \hspace*{0.8cm}
  \includegraphics[width=7.0cm]{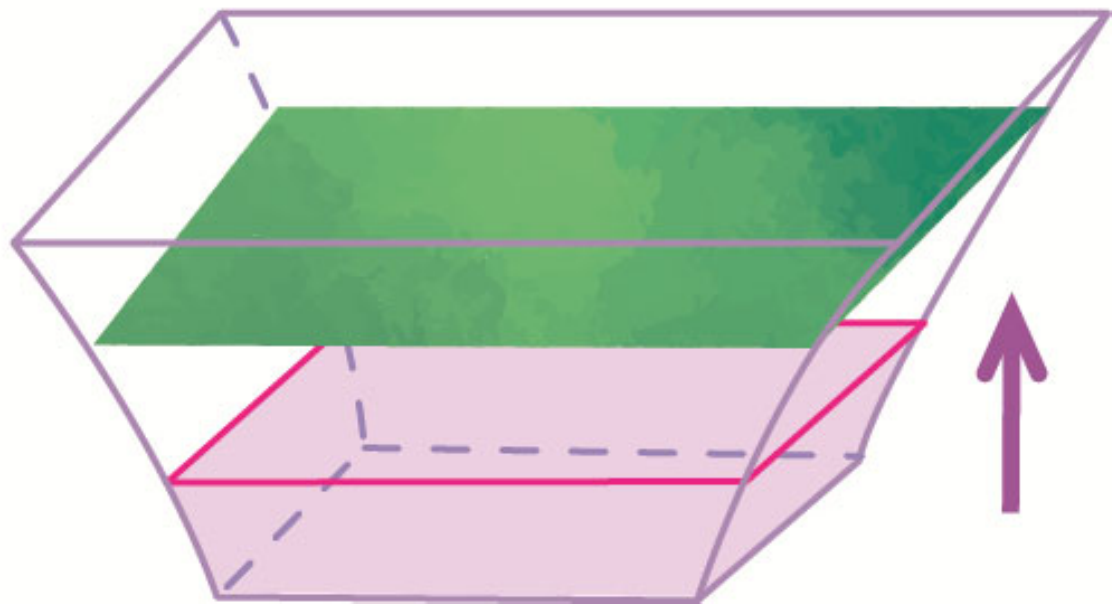}
 \end{picture}
 \end{center}
\vspace*{-5.2cm}
  \caption{Schematic depiction of the setup in nonrelativistic D3-brane (NRD3) theory where a lone D3-brane probes a stack of D3-branes embedded in a 10-dimensional flat 3Newton-Cartan (3NC) background. 
  Top left: The probe inside the D-brane stack. 
  Top right: The probe inside the relativistic bubble within AdS$_5\times$S$^5$ that is dual to the brane stack. The effect of the bubble (or equivalently, the stack) results in a relativistic action for the probe, Eq.~(\ref{d3probeads}).
  Bottom left: The probe in the flat 3NC region surrounding the D-brane stack.
  Bottom right: The probe in the asymptotically flat 3NC region outside of the relativistic bubble in AdS$_5\times$S$^5$. The probe's action (\ref{d3probenr}) correctly encodes motion in the flat \emph{nonrelativistic} geometry.}
  \label{fig:d3probe} 
  \end{figure}

The effective action for such a probe D3 is obtained in the field theory side by integrating out the complementary $U(N-1)$ degrees of freedom, and in the gravity side, by evaluating the familiar relativistic D3-brane action on the AdS background \cite{Maldacena:1997re},
\begin{equation}
    S_{\mbox{\tiny D3}}=T_{\mbox{\tiny D3}} \int d^4 x \left(-\sqrt{-\det \left(g_{AB} 
    +2\pi l_s^2 F_{AB}\right)} + c_{0124}\right),
\label{d3action}    
\end{equation}
with $g$ the induced metric, i.e., $g_{AB}\equiv\p_{A}X^{\mu}\p_{B}X^{\nu}G_{\mu\nu}$ in terms of the embedding fields  $(x^A,X^{A'}(x^A))$, and likewise for $c_{(4)}$. The brane tension is T$_{\mbox{\tiny D3}}\equiv 1/(2\pi)^{3}\gs\ls^{4}$. 
It is appropriate that (\ref{d3action}) is a \emph{relativistic} brane action, because we do expect the probe D3 to have that character within the bubble of Lorentzian spacetime generated (when $N g_s\gg 1$) by the stack of $N-1\gg 1$ D3s. See the top row of images in Fig.~\ref{fig:d3probe}.

If we turn on a constant value for the worldvolume electric or magnetic field strength $F$, the probe will describe a D3-F1 or D3-D1 bound state \cite{Douglas:1995bn}.
We will consider that possibility in Section~\ref{f1d1subsec} below. For now, we set $F=0$. Substituting (\ref{adsbkg}) into (\ref{d3action}) and letting R$^2\equiv X^{A'}X^{B'}\delta_{A'B'}$, 
we obtain
\begin{equation}
    S_{\mbox{\tiny D3}}=T_{\mbox{\tiny D3}} \int d^4 x \left(-\frac{\mbox{R}^2}{L^4}\sqrt{\mbox{R}^4-L^4(\p_{A}X^{A'}\p_{B}X^{B'}\eta^{AB}\delta_{A'B'})} 
    + \frac{\mbox{R}^4}{L^4}\right)~.
\label{d3probeads}    
\end{equation}
At any R, if we turn off the derivative terms, then we have a completely unexcited probe, and the action vanishes as it must, due to the BPS character of the configuration.

More interesting is the fact that when we move the probe out to $x^9\gg L$ (and consequently, $\mathrm{R}\gg L$), the action asymptotes to the finite functional
\begin{equation}
    S_{\mbox{\tiny D3}}=T_{\mbox{\tiny D3}}\int d^4 x \left(\frac{1}{2}(\p_{A}X^{A'}\p_{B}X^{B'}\eta^{AB}\delta_{A'B'})\right)~.
\label{d3probenr}    
\end{equation}
The quadratic character of this action  shows that, in the gravity side, the transverse dynamics is \emph{asymptotically nonrelativistic}. This result clearly demonstrates that, even though in AdS/CFT we are used to thinking of the entire AdS spacetime as corresponding to the interior of the (post-near-horizon-limit) D3-brane stack, our probe D3-brane is in fact able to completely exit the D3 stack and move arbitrarily far away into the flat 3NC region.  See the bottom row of images in Fig.~\ref{fig:d3probe}. This finding is U-dual to the fact described in Section~\ref{blackstringsubsec} that, in the black string background of NRF1 theory, a probe F1 can completely depart from the F1 stack, at zero energy cost and acquiring a quadratic, NR action in the process \cite{Avila:2023aey}.

\subsection{F1/D1 probe}
\label{f1d1subsec}

A different type of probe is a long F1, extended along one of the Cartesian directions. If the string is longitudinal, the form of the metric (\ref{adsbkg}) makes it clear that the energy cost for transporting a longitudinal string out of the bubble and into the flat 3NC region is infinite. This is the expected feature that an isolated longitudinal F1 can only exist within the relativistic bubble 
that gives a geometric description of the D3s, where it is dual to chromoelectric flux in MSYM.

A purely transverse F1, on the other hand, can extend arbitrarily far out into the flat region, as long as it terminates somewhere--- e.g., on the probe D3 of the preceding subsection. It is then dual to a very heavy W boson, which transforms as a quark with respect to the $U(N-1)$ group. This is of course the basis for the holographic computation of Wilson loops \cite{Maldacena:1998im,Rey:1998ik}, which has led to many fascinating lessons \cite{Sonnenschein:1999if,Semenoff:2002kk,Chernicoff:2011xv,Zarembo:2016bbk}.

The fact that longitudinal F1s do not survive the NRD3 limit, while transverse F1s do, can be intuited from the Born-Infeld (BI) string \cite{Callan:1997kz,Gibbons:1997xz} description of a fundamental string as a tubular D3-brane carrying electric flux. In the case of a longitudinal BI string, the tubular brane would necessarily have a portion with negative longitudinal wrapping (i.e., oriented like an anti-D3-brane), which is forbidden. A transverse BI string, on the other hand, can be described as a spike on a positively-oriented longitudinal D3-brane, and is consequently allowed. 

Parallel remarks apply for a probe long D1-brane, which is the holographic dual of chromomagnetic flux, and leads to the AdS/CFT description of 't~Hooft loops \cite{Minahan:1998xb}.

{}From the definition of NRD3 theory, we expect that the only way in which longitudinal F1s or D1s can move out to the asymptotically flat 3NC region is if they carry positive D3 charge. Due to the dimensionality, we phrase this the other way around, speaking of a D3 that carries F1 or D1 charge. This entails forming a D3-F1 or D3-D1 bound state, which as recalled above, is simply a D3-brane with a constant value of the electric or magnetic field strength. 
Using (\ref{adsbkg}) in (\ref{d3action}), we find that for this type of probe the asymptotic action indeed yields a finite result,
\begin{equation}
    S_{\mbox{\tiny D3}}=T_{\mbox{\tiny D3}}\int d^4 x \left(\frac{1}{2}(\p_{A}X^{A'}\p_{B}X^{B'}\eta^{AB}\delta_{A'B'}) 
    + F_{AB}F^{AB}\right)~.
\label{d3probenrf}    
\end{equation}
Again, the quadratic nature of this action clearly indicates that we are describing a D3-F1 or D3-D1 bound state on flat 10-dimensional 3NC space, within NRD3 theory. 

The worldsheet action for a fundamental string in the NRD$p$ limit was worked out in \cite{Gomis:2023eav}.

\subsection{Relation with short and long strings in AdS$_3$}
\label{ads3subsec}

The probe results in Secs.~\ref{d3subsec} and \ref{f1d1subsec} are analogous to the scenario of NRF1 theory examined in \cite{Danielsson:2000mu} and recalled in Section~\ref{dbranesubsec}, where a probe F1 was shown to be able to exit a D$p$-F1 bound state, with a finite energy cost that matches the change in the bound state energy. In parallel with our discussion here, \cite{Danielsson:2000mu} presented this result as evidence that the black brane background in question includes an asymptotic flat region, external to the D$p$-F1 stack.
   
As stated in that reference, the existence of long strings that can move out to the boundary and short strings that can only exist in an inner region is reminiscent of the long and short strings on AdS$_3$ with RR or NSNS flux, analyzed in \cite{Maldacena:1998uz,Seiberg:1999xz} and relevant to the AdS$_3$/CFT$_2$ correspondence \cite{Maldacena:1997re,Giveon:1998ns,deBoer:1998gyt,Maldacena:2000hw,Eberhardt:2019ywk}.

Thanks to the recent discovery of \cite{Blair:2024aqz}, the reason for this similarity is now evident: the instances of holographic duality in question also admit a direct interpretation as black branes in nonrelativistic brane theories.  E.g., it was shown explicitly in \cite{Blair:2024aqz} that the familiar AdS$_3\times$S$^3\times$T$^4$ background with RR flux, 
\begin{equation}
\begin{aligned}
\widehat{ds}^2&=\frac{r^2}{L_1 L_5}\left(-dt^2+dx_1^2\right)
+\frac{L_1}{L_5}\left(dx_2^2+\ldots+dx_5^2\right)
+\frac{L_1 L_5}{r^2} \left(dx_6^2+\ldots+dx_9^2\right)~,
\\
\hat{C}_{(2)}&=
\frac{r^2}{L_1^2}\,dt\wedge dx_1~,
\qquad
\hat{C}_{(6)}=
\frac{r^2}{L_5^2}\,dt\wedge dx_1\wedge \ldots\wedge dx_5~,
\\
e^{\hat{\Phi}}&=\frac{L_1}{L_5}g_s~,
\qquad\qquad\quad\;\,
r^2\equiv x_6^2+\ldots+x_9^2~, 
\\
L_1^2&\equiv \frac{(2\pi)^4}{V_4} N_1 g_s l_s^{6}~,
\qquad\,
L_5^2\equiv N_5 g_s l_s^{2}~,
\end{aligned}
\label{ads3}
\end{equation}
follows from a \emph{double} application of the NRD$p$ limit (\ref{nrdplimit}), with $p=1$ \emph{and} $p=5$. In the language of \cite{Danielsson:2000gi,Danielsson:2000mu,Guijosa:2023qym,Avila:2023aey}, this identifies (\ref{ads3}) as the black 5-brane that carries electric and magnetic 2-form RR charge in nonrelativistic D1-D5 theory (NRD1D5). The asymptotics of the black brane define what we naturally call a flat $(1,5)$NC asymptotic region, characterized by three separate vielbeine for the three different sets of directions: the longitudinal-longitudinal $x^a$ with $a=0,1$, the transverse-longitudinal $x^I$ with $I=2,\ldots,5$, and the transverse-transverse $x^i$ with $i=6,\ldots,9$. The most rightful inhabitants of that region are objects carrying both D1 and D5 charge. Since these are threshold bound states, objects carrying either one of these charges are also allowed. The simplest example is the long D-string considered in \cite{Maldacena:1998uz,Seiberg:1999xz}. 

Corresponding statements hold for the NSNS background S-dual to (\ref{ads3}), which describes a black 5-brane in nonrelativistic F1-NS5 theory (NRF1NS5), and has an asymptotically flat $(1,5)$NC region that is welcoming to the long fundamental strings considered in \cite{Seiberg:1999xz,Maldacena:2000hw} and many later works. Evidently, one can generate an infinite family of exactly analogous setups by making use of the full $SL(2,Z)$ S-duality group. 
 
In all such setups, the contrast between long strings, which can be placed anywhere, and short strings, which are trapped by the relativistic bubble that exists in the inner region, is explained by the fact that the very existence of the short strings is only made possible by the bubble. In other words, short strings are dual to excitations intrinsic to the stack of 5-branes$+$1-branes.

\subsection{Asymptotically \emph{non}-flat Newton-Cartan geometries}
\label{nonflatsubsec}

It is important to stress tha there is no \emph{a priori} requirement that the $p$NC geometry in these dualities be flat. In the present paper we are focusing on that case for simplicity, but other scenarios can be easily devised. To name just one class of examples, going back to the parent, relativistic Type IIB string theory, we can follow    \cite{Klebanov:1998hh} and place the stack of D3-branes not on 10-dimensional Minkowski $M_{10}$, but at the conical singularity of $M_4\times Y_6$, where the second factor is a Ricci-flat space with a singularity at $r=0$. In this case, the near-horizon/NRD3 limit is known to yield a RR black 3-brane of the form AdS$_5\times X^5$, where the second factor is an Einstein manifold of positive curvature. The asymptotic region of this ten-dimensional background is a 3NC geometry that is \emph{not} flat. One particular instance is when $Y_6$ is the conifold: the factor $X^5$ that replaces S$^5$ is then known to be $T^{1,1}=(SU(2)\times SU(2))/U(1)$, and the D3-brane worldvolume theory is an $\cN=1$ superconformal $U(N)\times U(N)$ gauge theory with four chiral bifundamental fields \cite{Klebanov:1998hh}. 

\subsection{AdS$_5\times$S$^5$ is indeed a black brane in NRD3 theory}
\label{indeedsubsec}

Having understood that the AdS$_5\times\mathrm{S}^5$ geometry (\ref{adsbkg}) is precisely U-dual to the black string (\ref{blackstring}) in NRF1 theory, we know that all of our remarks in Section~\ref{blackstringsubsec} apply here as well. 
Pivotal among these is the insistence that the uberknown fact that AdS$_5\times\mathrm{S}^5$ solves the equations of motion of relativistic Type IIB supergravity should \emph{not} be taken to indicate that we are somehow back in the \emph{parent} relativistic Type IIB string theory. The crucial point, again, is that the theory is defined not just by the equations of motion, but also by the boundary conditions.   

To go from the original Lorentzian black 3-brane (\ref{ordinaryblackthreebrane}) to its near-horizon region (\ref{adsbkg}), we applied the NRD3 limit (\ref{nrd3scaling}). Consistent with this is the property, discovered in \cite{Blair:2024aqz}, that upon identifying $\omega\equiv r^2/L^2$ at $r\gg L$, (\ref{adsbkg}) precisely agrees with (\ref{nrd3scaling}). Undoubtedly, then, the asymptopia of AdS$_5\times\mathrm{S}^5$ is flat 3NC, as befits states in \emph{nonrelativistic} D3-brane theory, and in sharp contrast with the Minkowskian asymptotics of the original black 3-brane  (\ref{ordinaryblackthreebrane}). In Sections~\ref{d3subsec} and \ref{f1d1subsec} we have verified that outside the relativistic bubble of AdS$_5\times\mathrm{S}^5$ (more specifically, at $r\gg L$) conditions are welcoming only for objects rightfully in the spectrum of NRD3 theory. 
 Due to the direct connection with (\ref{blackstring}), we know that the relativistic character of the local physics on (\ref{adsbkg}) is due to the fact that the D3-brane stack sources a term that is U-dual to the $\lambda\bar{\lambda}$ deformation discussed in Section~\ref{blackstringsubsec}, with a coefficient that is appreciable inside the inner bubble (which corresponds to being inside the stack), but switches off smoothly as one moves into the outer region (away from the stack).

Even without any reference to NC geometry and NRD3 theory, it is well understood that Type IIB string theory on (asymptotically) Minkowski$_{10}$ \emph{is not the same theory} as Type IIB string theory on (asymptotically) AdS$_5\times\mathrm{S}^5$. A sharp indicator of this is the fact that to go back from the near-horizon background (\ref{adsbkg}) to the full Lorentzian 3-brane geometry (\ref{ordinaryblackthreebrane}) one needs to turn on a \emph{non}-normalizable mode on AdS, which corresponds to deforming MSYM by the addition of an irrelevant operator $\mathcal{O}_8\sim \mathrm{Tr}(F^4)$ \cite{Gubser:1998iu,Intriligator:1999ai,Danielsson:2000ze,Amador:2003ju}.

Just like for the black string solution of NRF1 theory, worked out in \cite{Avila:2023aey} and reviewed in Section~\ref{blackstringsubsec}, we can ask whether there is any way to start with the parent black 3-brane (\ref{ordinaryblackthreebrane}) and take a limit that yields a pure 3NC geometry, with no relativistic bubble. The answer is again that for this we must implement the NRD3 limit (\ref{nrd3scaling}) in a way that completely depletes the gravitating source, formally scaling $\hat{N}=\omega^{-\alpha}N\to 0$, with $\alpha\ge 2$. For 
$\alpha>2$, the resulting 3NC geometry is flat, while for $\alpha=2$, it is modulated by the surviving harmonic function $H(r)=1+L^4/r^4$. As before, the physical conclusion is that pure 3NC backgrounds correspond to the absence of D3-branes.      

\subsection{Center-of-mass motion}
\label{cmsubsec}

As is well-known, upon rewriting the MSYM gauge group as $U(N)\simeq U(1)\times SU(N)/\bZ_N$, the $U(1)$  part describes the center-of-mass motion of the stack, while the $SU(N)$ part describes the relative degrees of freedom, and has a single zero-energy bound state with a definite transverse size \cite{Witten:1995im,Sethi:1997pa,Lin:2014wka}. Ever since \cite{Maldacena:1997re},  the AdS/CFT duality is usually taken to involve only the $SU(N)$ part of MSYM, because the $U(1)$ part is free, unlike the fields in the interior of the AdS geometry. 

The $U(1)$ degrees of freedom of the gauge theory are known to be dual to the massless vector multiplet that assembles states of various towers of Kaluza-Klein (KK) modes in the spectrum of Type IIB supergravity on AdS$_5\times$S$^5$ \cite{Gunaydin:1984fk,Kim:1985ez}. 
In particular, the six scalars that describe the position of the D3s along the transverse directions are the $k=1$ modes within one of the two KK towers of modes arising from the trace of the graviton on the 5-sphere and the 4-form RR photon also with 5-sphere indices. Due to gauge invariances, these `singleton' or `doubleton' modes have no propagating degrees of freedom, and live at the boundary of AdS \cite{Kim:1985ez}. With an appropriate choice of boundary conditions, they are related to the $B$-field modes studied in \cite{Witten:1998wy}, and match with the center-of-mass degrees of freedom of the D3s \cite{Aharony:1999ti,Bilal:1999ph,Mansfield:2003gs}. Their residence at the AdS boundary is consistent with the fact that, before the limit, they indicate the location of the threebrane throat within the surrounding (relativistic) flat spacetime. Up to now, the decision on whether or not to include them in AdS/CFT has seemed optional.

Knowing that AdS$_5\times$S$^5$ is nothing but the black threebrane geometry produced by a stack of $N$ coincident D3s in NRD3 theory, which consequently includes an outer 3NC flat region, it is in fact more natural to keep the $U(1)$ modes, as an encoding of the center-of-mass position, within the underlying flat 3NC geometry, of the bubble generated by the D3 stack. {}From the NRD3 perspective, then, the duality involves MSYM with gauge group $U(N)$. 

Indeed, through U-duality we know that the MSYM action (\ref{msym}) is a non-Abelian and static-gauge counterpart of the Gomis-Ooguri action for the fundamental string in NRF1 theory, Eq.~(\ref{go}), which at the quantum level leads to the NR spectrum (\ref{closednrf1}). For D3-branes in NRD3 theory, the distinctive $\vec{p}^{\,2}_{\perp}$ contribution to the energy arises precisely from the $U(1)$ center-of-mass component of the $U(N)$ gauge theory.  

Note that this issue of the position of the center of mass is confounded by the large isometry group of AdS$_5\times$S$^5$, which is in turn a reflection of the conformal nature of MSYM in $3+1$ dimensions. Before the near-horizon/NRD3 limit, there is a clearly identifiable origin for the throat; but after the limit, any two points in AdS$_5$ can be related through an isometry, and therefore appear to be equivalent. The situation is different in the non-conformal black $p$-brane geometries dual to D$p$s with $p\neq 3$ \cite{Itzhaki:1998dd,Skenderis:1998dq}, and also in the geometries dual to renormalization-group flows, e.g., \cite{Girardello:1998pd,Freedman:1999gp,Girardello:1999bd,Polchinski:2000uf}.

In the next subsection we will see that multi-centered brane geometries give us a more direct handle on the location of the corresponding brane stacks. 

\subsection{Disassembling anti-de Sitter space}
\label{disassemblingsubsec}

In Section~\ref{d3subsec} we separated a single D3 from the stack and treated it as a probe. We can alternatively separate the stack into various groups of D3s at different transverse positions, exploring more fully the Coulomb branch of MSYM. In the gravity side, this is described by the near-horizon region of the multi-centered black 3-brane solution
\cite{Kraus:1998hv,Klebanov:1999tb,Skenderis:2006di}, which in the present work we understand to be simply the multi-centered black 3-brane solution within NRD3 theory. 

\begin{figure}[thb]
\begin{picture}(150,70) 
  \centering
  \hspace*{0.5cm}
  \includegraphics[width=7.0cm]{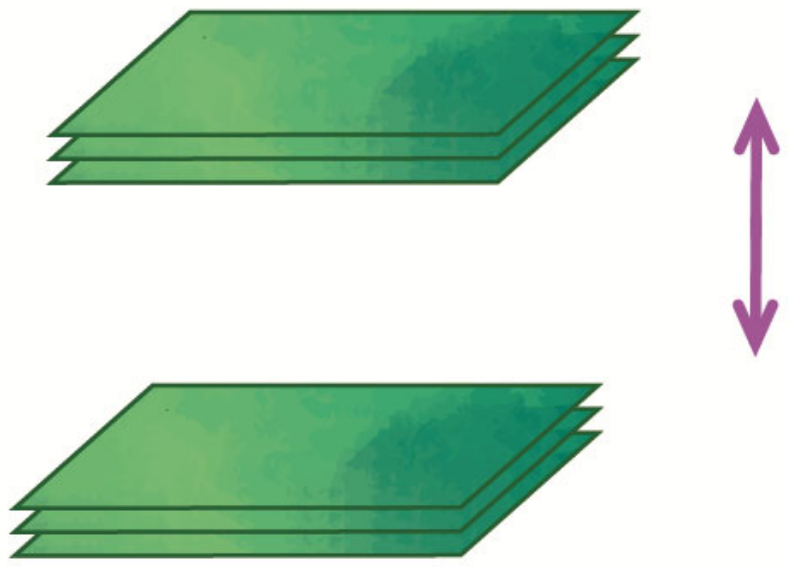}
  \hspace*{-0.5cm}
  \includegraphics[width=7.0cm]{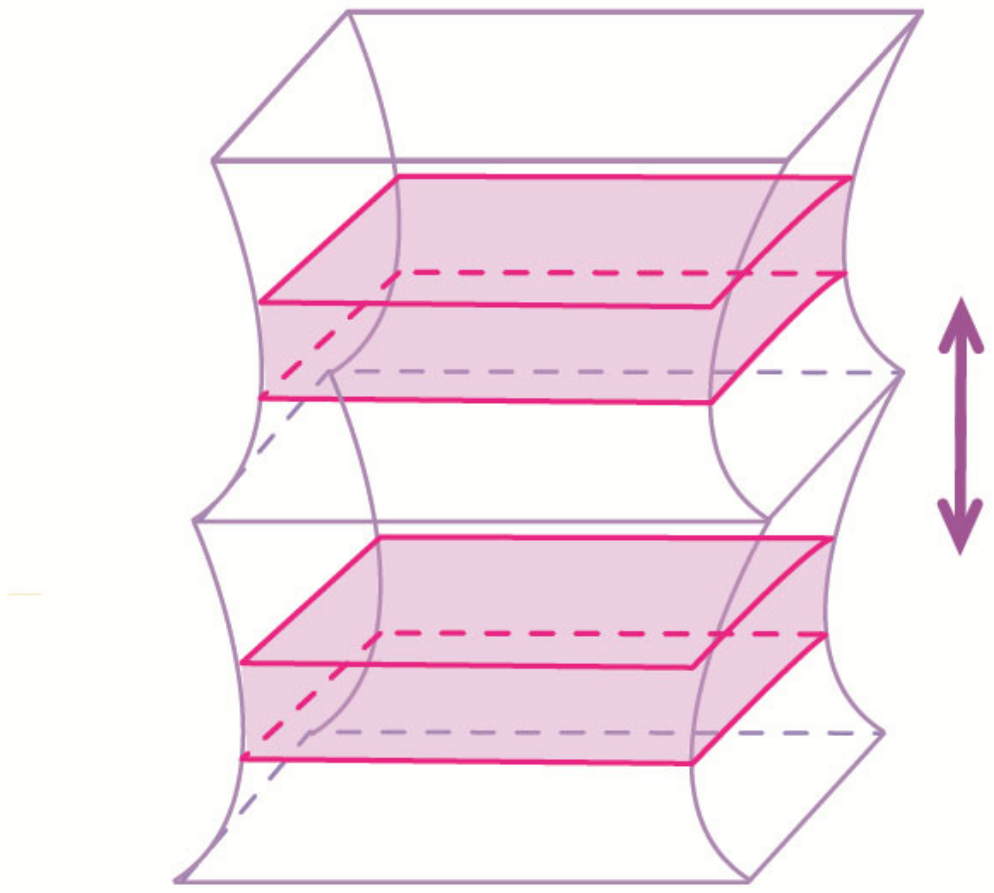}
 \put(-11,52){$r$}
 \end{picture}
\vspace*{-2.5cm}
  \caption{Pulling the D-branes apart. Left: Two stacks of $N/2$ coincident D3-branes each, a distance $\ell$ apart. Right: The corresponding two-center black 3-brane is an asymptotically AdS$_5\times$S$^5$ spacetime that contains two separate relativistic bubbles, each dual to one of the D-brane stacks. For large inter-stack/inter-bubble separation, the spacetime in the middle becomes flat and nonrelativistic (3-Newton-Cartan).
  }
  \label{fig:2bubbles} 
  \end{figure}

For concreteness, we will confine our attention here to the simplest case of separating into just two stacks, with $N/2$ D3s each, a distance $2l$ apart along the $x^9$ direction. The relevant NRD3 background is then
\begin{align}
\widehat{ds}^2&=\left[\frac{L^4/2}{(\rho^2+(x_9-l)^2)^2}+\frac{L^4/2}{(\rho^2+(x_9+l)^2)^2}\right]^{-1/2}dx^A dx^B \eta_{AB}
\nonumber\\
&\quad+\left[\frac{L^4/2}{(\rho^2+(x_9-l)^2)^2}+\frac{L^4/2}{(\rho^2+(x_9+l)^2)^2}\right]^{1/2}dx^{A'} dx^{B'} {\delta}_{A'B'}~,
\label{adscoulombbkg}\\
\hat{C}_{01234}&=
\left[\frac{L^4/2}{(\rho^2+(x_9-l)^2)^2}+\frac{L^4/2}{(\rho^2+(x_9+l)^2)^2}\right]^{-1}~,
\nonumber
\end{align}
where $\rho^2\equiv x_4^2+\ldots+x_8^2$. Schematically, this background consists of two relativistic bubbles separated by and surrounded with 10-dimensional flat 3NC space. See Fig.~\ref{fig:2bubbles}.

\vspace*{1.2cm}

Now we can put the inter-stack distance $l$ to good use. If we stand in the neighborhood of $x^9=0$, halfway between the two stacks, and pick any arbitrarily large size (UV cutoff) $\rho_{\mbox{\tiny UV}}$ along $x^9$ and $\rho$, then upon taking $l\gg \rho_{\mbox{\tiny UV}}$ in (\ref{adscoulombbkg}), the background is scaled according to 
\begin{align}
\widehat{ds}^2&=\left(\frac{L^4}{l^4}\right)^{-1/2}dx^A dx^B \eta_{AB}
+\left(\frac{L^4}{l^4}\right)^{1/2}dx^{A'} dx^{B'} {\delta}_{A'B'}~,
\label{midpointbkg}\\
\hat{C}_{01234}&=
\left(\frac{L^4}{l^4}\right)^{-1}~.
\nonumber
\end{align}
As $l\to\infty$, within the gravity description we would normally have regarded this background as degenerate and unreliable in the supergravity approximation. But within our present perspective, we notice immediately that (\ref{midpointbkg}) is simply the scaling  (\ref{nrd3scaling}) that defines the 10-dimensional flat 3NC geometry, with 
$\omega=l^2/L^2$. We know that \emph{Type IIB string theory is well-defined in this limit, and becomes NRD3 theory}. This again shows that the flat 3NC geometry is what remains when we have no D3s present. 

Importantly, this analysis reveals a second reason why our natural inclination to regard multi-center solutions like (\ref{adscoulombbkg}) as \emph{Lorentzian} backgrounds does not provide us with a complete account of the situation. Aside from the non-Lorentzian asymptotics emphasized in Section~\ref{indeedsubsec}, in the present subsection we have verified that for large inter-stack separations the respective relativistic bubbles no longer overlap, and \emph{the underlying non-Lorentzian (3NC) geometry is laid bare even in the central regions of the solution}.  

\subsection{Scattering D3s}
\label{d3scatsubsec}

In the AdS/CFT context, we are used to thinking of the pure AdS$_5\times$S$^5$ geometry, dual to the symmetric vacuum of $U(N)$ MSYM, as \emph{the} natural ground state around which we perturb. The connection between Coulomb-branch vacua and multi-centered 3-brane geometries has long been known \cite{Kraus:1998hv,Klebanov:1999tb,Skenderis:2006di}, but somehow the important role of the 3NC geometry as the underlying stage for the D3 dynamics had been hitherto overlooked. 

The discussion in Section~\ref{disassemblingsubsec} makes it clear that the curved region surrounding each center in a Coulomb-branch background is precisely the relativistic bubble dual to the corresponding stack, and that these bubbles exist on and can move through the (otherwise) flat 3NC geometry. 
The most basic scattering amplitudes of the theory are then understood to involve scattering of these bubbles against one another. And indeed, this is precisely what was computed in the classic Matrix theory calculations \cite{Bachas:1995kx,Lifschytz:1996iq,Douglas:1996yp,Becker:1997wh,Becker:1997xw,Dine:1997sz,Becker:1997cp,Buchbinder:2001ui,Taylor:2001vb}.
More specifically, this type of D3 scattering is U-dual to the scattering of gravitons in DLCQ M-Theory on a transverse 3-torus. It is also U-dual to the scattering of wound closed strings in NRF1 \cite{Danielsson:2000gi,Gomis:2000bd}.

\subsection{Scattering Newtonian gravitons}
\label{gravitonscatsubsec}

Having recognized that the most basic scattering process in NRD3 theory is D3 scattering, it is natural to wonder about the interpretation in this language of the renowned GKPW recipe \cite{Gubser:1998bc,Witten:1998qj} for correlators in AdS/CFT. It is well-understood \cite{Aharony:1999ti}, and explicit in the motivation of \cite{Gubser:1998bc}, that such correlators descend from what prior to the near-horizon limit would have been amplitudes for scattering closed strings off the D-branes.
Indeed, there is a one-to-one match \cite{Witten:1998qj} between the spectrum of chiral primary operators in MSYM and the modes of Type IIB supergravity (massless modes of Type IIB string theory) on AdS$_5\times$S$^5$, expanded as KK modes on the S$^5$ \cite{Gunaydin:1984fk,Kim:1985ez}. And, importantly, this pairing is additionally related to the way in which the non-Abelian D3-brane action couples to background supergravity modes \cite{Gubser:1998bc,Das:1998ei,Aharony:1999ti}. 

So, before taking the Maldacena/near-horizon/NRD3 limit, one is computing scattering amplitudes of closed strings, off the black 3-brane in the closed string description, and off the stack of D3s in the open string description.  Following the effect of the limit, we conclude then that the GKPW recipe computes scattering amplitudes of \emph{Newtonian gravitons} (and their superpartners) in NRD3 theory. As explained in \cite{Danielsson:2000mu}, in the field theory or D-brane side these modes  have vanishing transverse momentum when on shell,\footnote{For this reason, they are of measure zero in the space of asymptotic states, but here we are led to setting up the scattering process in such a way as to precisely pick up their contributions.} which implies that they are completely delocalized along such directions, and will bathe with equal strength all of the D3-branes embedded in the 3NC background, exciting them via the insertion of gauge-invariant operators. On the other hand, they can carry arbitrary longitudinal momentum, which is why they can lead to insertion of operators that are fully localized along the D3-brane worldvolume. 

The scattering process considered in this subsection is U-dual to having a stack of wound strings in NRF1 theory and using Newtonian gravitons to excite waves on them, which can then scatter off one another and  later reemit gravitons. 

Altogether then, the situation is more closely analogous than previously realized to scattering in the full relativistic string theory: 
the preceding subsection describes the analog of direct closed string scattering, whereas the present one involves scattering of closed strings off of D-branes.

\subsection{Entanglement entropy}
\label{entanglesubsec}

The entanglement entropy of a spatial region $\mathcal{A}$ in MSYM is given by the celebrated Ryu-Takayanagi formula \cite{Ryu:2006bv,Ryu:2006ef,Hubeny:2007xt,Lewkowycz:2013nqa,Dong:2016hjy}, originally written down purely within the AdS$_5$ factor
 (for reviews, see, e.g., \cite{Rangamani:2016dms,Nishioka:2018khk}). For our purposes here, we are interested in its natural extension to the full ten-dimensional AdS$_5\times$S$^5$ geometry 
\cite{Mollabashi:2014qfa,Karch:2014pma,Taylor:2015kda},
\begin{equation}
\mathcal{S}_{\mathcal{A}}=\frac{\mbox{Area}(\Gamma_{\mathcal{A}})}{4G^{\mbox{\tiny (10)}}_{\mbox{\scriptsize  N}}}~,
\label{rt10}
\end{equation}
where $\Gamma_{\mathcal{A}}$ denotes the minimal-area codimension-2 surface anchored on the entanglement surface $\p \mathcal{A}$. 
In the standard interpretation, the choice of $\p \mathcal{A}$ reflects a bipartitioning of both the spatial and the $SU(4)$ degrees of freedom of MSYM.  

Since $S_\mathcal{A}$ is UV divergent, its computation requires the introduction of a UV cutoff surface near the AdS boundary. The leading-order term when removing the cutoff gives the familiar area-law divergence, whose coefficient is scheme-dependent. Beyond it, one can find subleading divergences, and eventually finite and/or logarithmically-divergent contributions that encode physical information about the theory. One can alternatively define a renormalized version of the entanglement entropy that directly isolates these contributions \cite{Taylor:2016aoi,Anastasiou:2018rla,Anastasiou:2019ldc}. 

An interesting feature is that the surfaces $\Gamma_\mathcal{A}$ are always anchored on a UV cutoff surface near the conformal boundary of AdS, which is to say, away from the D3s, in the asymptotically flat 3NC region. It is therefore natural to inquire into RT calculations directly within the purely flat 3NC geometry that we get if we remove all of the D3s. 

To orient ourselves, consider first the calculation of \cite{Mollabashi:2014qfa} in the 2-stack geometry (\ref{adscoulombbkg}), where instead of bipartitioning the MSYM spatially, the authors bipartitioned the internal $SU(4)$. Specifically, they determined the entanglement entropy between the upper and lower halves of the S$^5$, placing the entangling surface $\p \mathcal{A}$ on the S$^4$ at the equator, i.e., polar angle $\Theta_1=\pi/2$. In the specific $SU(N/2)\times SU(N/2)$ setting discussed in Section~\ref{disassemblingsubsec}, this quantifies the entanglement between the two stacks. In the coordinates  of (\ref{adscoulombbkg}), $\p \mathcal{A}$ lies at $x^9=0$, on the UV cutoff surface $\rho=\rho_{\mbox{\tiny UV}}$, and extends additionally along $x^1,x^2,x^3$. By symmetry, the minimal-area surface is precisely $x^9=0$, halfway between the stacks. Using (\ref{adscoulombbkg}) in (\ref{rt10}), one finds 
\begin{equation}
\mathcal{S}\propto V_3\int_0^{\rho_{\mbox{\tiny UV}}} \frac{\rho^4\,d\rho }{\rho^2+l^2}
=V_3\left[\frac{\rho^3_{\mbox{\tiny UV}}}{3}-l^2\rho_{\mbox{\tiny UV}}+l^3\arctan\left(\frac{\rho_{\mbox{\tiny UV}}}{l}\right)\right]~,
\label{stackentanglement}
\end{equation}
where $V_3$ denotes the volume along the spatial directions of the field theory. The appearance of $V_3$ indicates that the entropy indeed follows the \emph{volume} law expected from the fact that we are quantifying entanglement between two sets of degrees of freedom that inhabit each and every point of space. 
For $l\ll\rho_{\mbox{\tiny UV}}$, (\ref{stackentanglement}) simplifies to  
$\mathcal{S}\propto V_3\,\rho^3_{\mbox{\tiny UV}}$.

Of interest to us is the fact that in the limit $l/\rho_{\mbox{\tiny UV}}\to\infty$, where we move both stacks arbitrarily far away, and we know from Section~\ref{disassemblingsubsec} that we are left with a minimal surface in the 10-dimensional flat 3NC geometry (\ref{midpointbkg}), the entropy (\ref{stackentanglement}) vanishes: $\mathcal{S}\propto \rho^5_{\mbox{\tiny UV}}/l^2\to 0$. As explained in \cite{Mollabashi:2014qfa}, the precise scaling of this outcome makes good sense from the point of view of the gauge theory. The complementary perspective for which we wish to advocate here is that this result can also be interpreted as a computation of entanglement entropy \emph{in the flat 3NC geometry}. Again, our basic point is that the 3NC background is what remains when there are no D3s present, so we indeed expect entanglement to be fully depleted, because of the complete absence of degrees of freedom. 

To reinforce this perspective, let us now study the system without D3s more directly. Consider the pure 3NC background (\ref{nrd3scaling}), placing our cutoff surface at $|x^9|=\rho_{\mbox{\tiny UV}}, \rho=\rho_{\mbox{\tiny UV}}$,\footnote{As in the preceding paragraphs, $\rho^2\equiv x_4^2+\ldots+x_8^2$.} and think of bipartitioning now the longitudinal space into halves, $x^1<0$ and $x^1>0$, by placing the entangling surface $\p \mathcal{A}$ at $x^1=0,x^9=\rho_{\mbox{\tiny UV}}$.\footnote{The overall position of our setup along $x^9$ is of course immaterial, due to transverse translational invariance.}
By symmetry, the corresponding RT surface is then simply $x^1=0$ extending vertically across the full range $x^9\in[-\rho_{\mbox{\tiny UV}},\rho_{\mbox{\tiny UV}}]$. It spans two longitudinal and six transverse directions in the metric (\ref{nrd3scaling}), so its area, and consequently the associated entanglement entropy, scales like $\mathcal{S}\propto \sqrt{\omega^2\omega^{-6}}=\omega^{-2}\to 0$. 

In the same spirit, we can consider a standard strip bipartitioning (as in \cite{Ryu:2006bv,Ryu:2006ef}), with the system $\mathcal{A}$ now chosen to be the region $|x^9|=\rho_{\mbox{\tiny UV}}, -a<x^1<a$, with $2a$ the width of the strip. By symmetry, the profile of the minimal surface can be expressed as $X_1(x_9)$. The area functional is thus
\begin{equation}
\mbox{Area}=2V_7\int_{x^9_{\mbox{\tiny min}}}^{\rho_{\mbox{\tiny UV}}}dx_9 \sqrt{\omega^2\omega^{-5}\left(\omega^{-1}+\omega X_1^{\prime\,2} \right)}~,
\label{entanglestrip}
\end{equation}
where $x^9=x^9_{\mbox{\tiny min}}$ is the putative turnaround depth for the surface. 
Extremization enforces the condition that the momentum conjugate to $X_1$ be conserved, and from this in turn it follows that $X_1^{\prime}=\mbox{constant}$. This is unsurprising, because (\ref{nrd3scaling}) is just flat space in rescaled coordinates, so planes should indeed be minimal surfaces. The only two choices that yield smooth (locally) minimal surfaces correctly anchored at $\p \mathcal{A}$ are then $X_1^{\prime}=0$, describing two purely vertical planes at $x^1=\pm a$, and $X_1^{\prime}=\infty$, describing a purely horizontal plane at $x^9=\rho_{\mbox{\tiny UV}}$. The smaller area is obtained in the vertical case, so through (\ref{rt10}) we find again that  $\mathcal{S}\propto \sqrt{\omega^2\omega^{-6}}=\omega^{-2}\to 0$. (The area of the horizontal surface vanishes as well, only more slowly.)

\begin{figure}[thb]
\begin{picture}(150,52) 
  \centering
  \includegraphics[width=7.0cm]{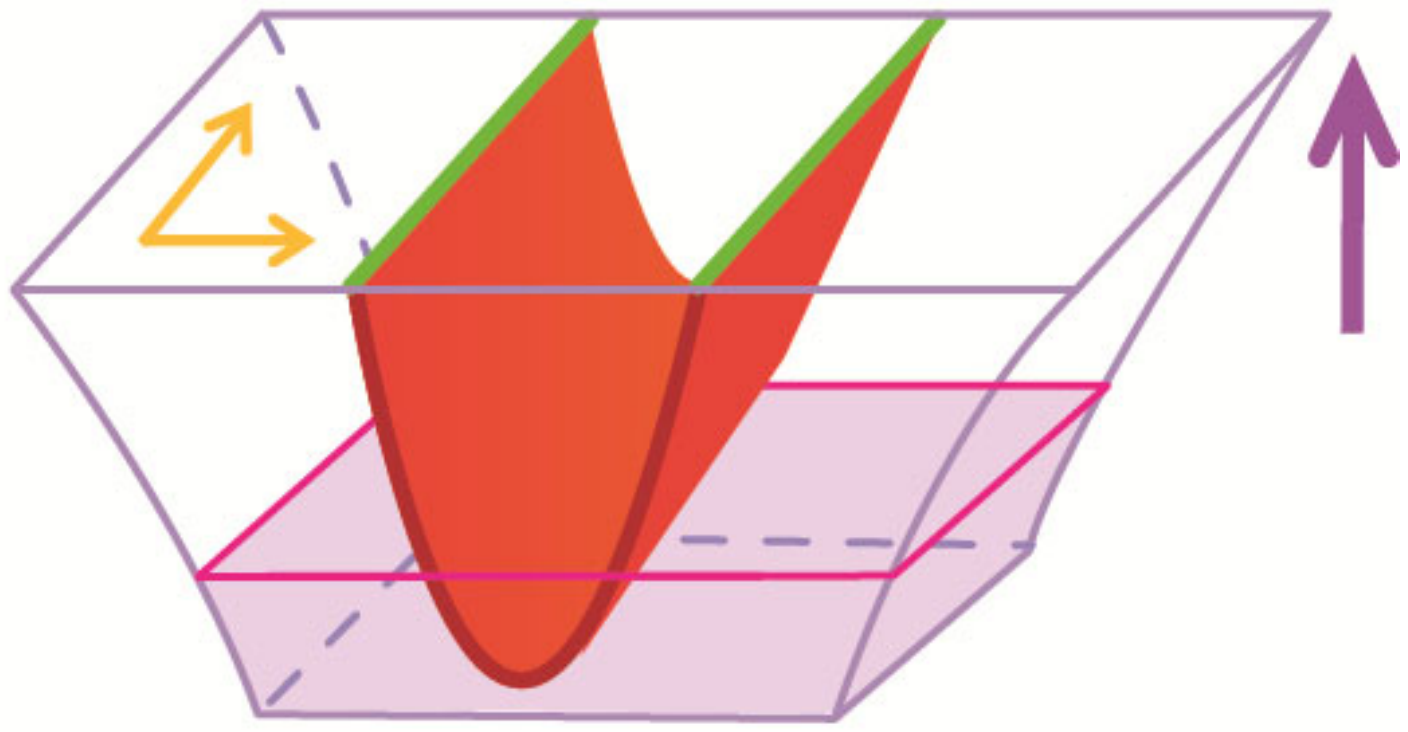}
  \hspace*{0.8cm}
  \includegraphics[width=7.0cm]{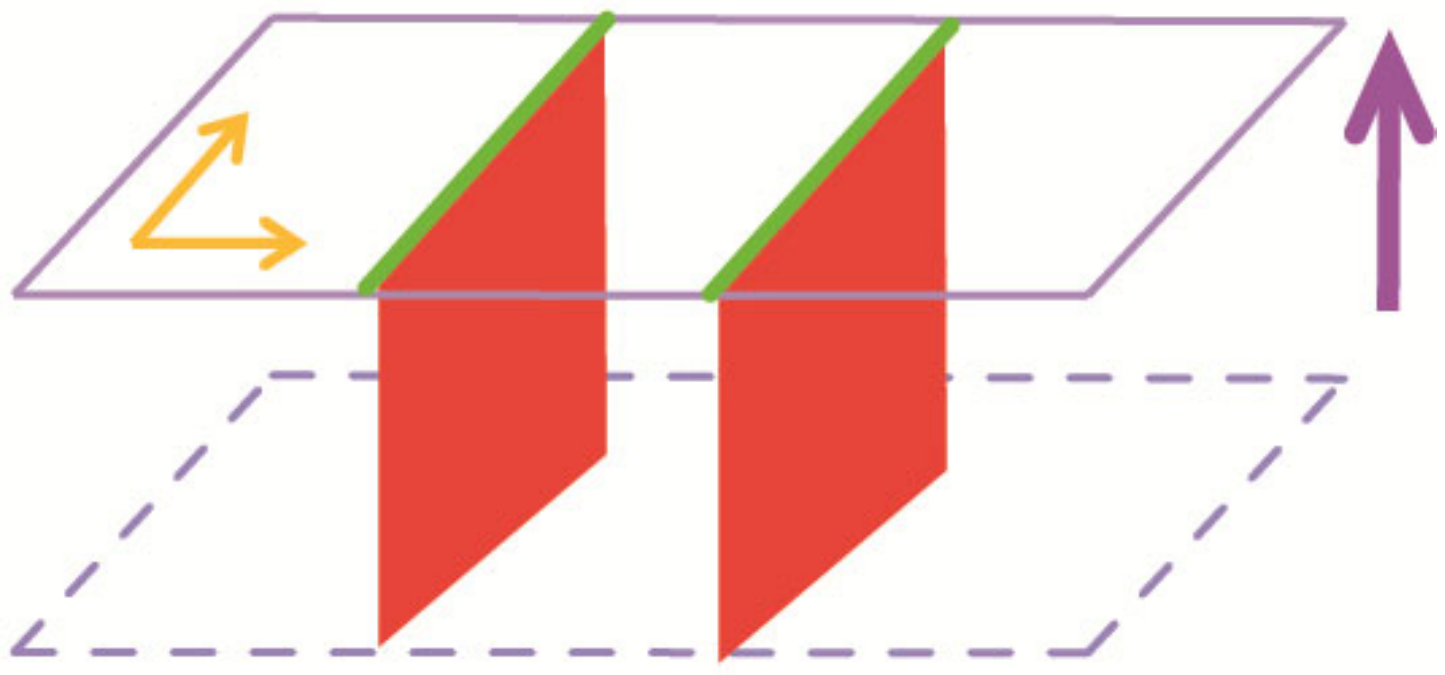}
  \put(-53,42){$\vec{x}$}
  \put(-2,44){$r$}
    \put(-132,40){$\vec{x}$}
    \put(-83,41){$r$}
  \end{picture}
\vspace*{-1.5cm}
  \caption{Minimal surfaces (in red) involved in the entanglement entropy calculation for a strip (bounded by the green lines). Left: For $g_sN\gg 1$, the calculation is carried out in the familiar AdS$_5\times$S$^5$ geometry, and the presence of the relativistic bubble (dual to the D3-brane stack) makes it favorable for the surface to extend longitudinally at a sufficiently deep (i.e., small $r$) location. Right: In the complete absence of D3-branes, the geometry is pure 3-Newton-Cartan (3NC), and the minimal surface is purely vertical, i.e., transverse, and its proper area~(\ref{entanglestrip}) vanishes, correctly reflecting the total absence of degrees of freedom.}
  \label{fig:entanglement} 
  \end{figure}

We see then that the RT formula correctly predicts zero entanglement entropy for the pure 3NC geometry (\ref{nrd3scaling}), which corresponds to having no D3s or any other objects present---i.e., it is $U(N)$ MSYM with $N=0$. On the gravitational side, the reason  for this general result is easy to intuit. The geometry of AdS$_5$ is only \emph{asymptotically} 3NC: as we move inwards in (\ref{adsbkg}), we have a warped version of (\ref{nrd3scaling}), which transitions smoothly from Newton-Cartan to Einsteinian geometry. Due to the warping, the outlook for would-be minimal surfaces is different at different radial depths. There is then a competition between the area gained by descending further along the transverse directions and the area decrease that this allows when extending across in the longitudinal directions. This leads to the familiar U-shaped or cap-shaped minimal surfaces. But in the pure 3NC geometry (\ref{nrd3scaling}), there is no such competition: it is always beneficial to extend purely along the transverse directions. With some qualifications depending on the choice of $\p \mathcal{A}$, this generally leads to strictly transverse RT surfaces, which have vanishing proper area. See Fig.~\ref{fig:entanglement}.

The fact that the 3NC geometry carries no entropy indicates that it is \emph{not} built up from entanglement of some pre-geometric degrees of freedom, unlike its relativistic counterparts \cite{Swingle:2009bg,VanRaamsdonk:2010pw,Maldacena:2013xja,Pastawski:2015qua,Hayden:2016cfa,Bao:2018pvs,VanRaamsdonk:2018zws}. Instead, $p$NC geometry provides the natural  
backdrop on which branes of various types can be embedded. It is on top of this Newtonian geometry that a relativistic, dynamical spacetime is built by the entanglement among the brane degrees of freedom. 

The central example of this paper has been the pure AdS$_5\times$S$^5$ geometry itself, which is a bubble of relativistic spacetime that emerges from the dynamics of $N$ D3s in NRD3 theory. 
As we emphasized, this is directly U-dual to the black string in nonrelativistic string theory, and to the fundamental black $p$-branes of all the other nonrelativistic brane theories, including the D$p$, NS5, M2 and M5 cases \cite{Maldacena:1997re,Itzhaki:1998dd,Aharony:1998ub}. 
And exactly the same picture is at play in systems of intersecting branes of various types, such as the AdS$_3$ system discussed in Section~\ref{ads3subsec}, and the many other examples examined recently in \cite{Fontanella:2024kyl,Lambert:2024uue,Lambert:2024yjk,Fontanella:2024rvn,Blair:2024aqz,Lambert:2024ncn}.  

The direct connection between standard holographic backgrounds such as (\ref{nrblackbrane}) and the black string (\ref{blackstring}) makes it clear that, in each case, the significant distortion that gives rise to a relativistic bubble is produced by a deformation dual to the $\lambda\bar{\lambda}$ term of the nonrelativistic string theory, discussed in Sections~\ref{introsec}, \ref{flatsubsec}, \ref{blackstringsubsec} and \ref{blackpbranesubsec}. For the NRD$p$ family of theories, the deformations in question are akin to the $T\bar{T}$ deformation \cite{Blair:2020ops}. Some aspects have been studied recently in \cite{Blair:2024aqz}. 

Interesting questions that are left for future work are the potential implications of the NR perspective for modern developments in relativistic holography, such as subregion duality \cite{Czech:2012bh,Wall:2012uf,Engelhardt:2014gca,Headrick:2014cta,Almheiri:2014lwa,Chen:2019gbt,Guijosa:2022jdo}, quantum error correction \cite{Almheiri:2014lwa,Harlow:2016vwg,Akers:2021fut}, tensor networks \cite{Swingle:2009bg,Pastawski:2015qua,Hayden:2016cfa,Bao:2018pvs}, entanglement islands and wormholes \cite{Penington:2019npb,Almheiri:2019psf,Almheiri:2019hni,Penington:2019kki,Almheiri:2019qdq,Geng:2020qvw,Chen:2020tes,Geng:2020fxl,Geng:2021hlu,Balasubramanian:2022gmo,Geng:2024xpj,Geng:2025rov} and operator algebras \cite{Leutheusser:2021qhd,Leutheusser:2021frk,Witten:2021unn,Chandrasekaran:2022cip,Chandrasekaran:2022eqq,Leutheusser:2022bgi,Bahiru:2022oas,Witten:2023xze,Kudler-Flam:2023qfl,Engelhardt:2023xer,Kudler-Flam:2024psh,Geng:2024dbl}

\section*{Note added}

While this paper was in preparation (and after its findings had been presented in \cite{guijosa:2025talk}), we received \cite{Harmark:2025ikv}, which builds on \cite{Guijosa:2023qym,Avila:2023aey} and \cite{Blair:2023noj,Blair:2024aqz} to obtain interesting results that enrich the understanding of nonrelativistic string theory summarized here in Section~\ref{stringsec}.

\section*{Acknowledgements}
We thank Ignacio Araya, Rodrigo Aros, Glenn Barnich and Miguel Campiglia for useful conversations, and Andr\'es Argandoña, Jes\'us Cruz, J.~Antonio Garc\'{\i}a and Mart\'{\i}n Kruczenski for valuable discussions. We are also very grateful to Andr\'es Argandoña and J.~Antonio Garc\'{\i}a for comments on the manuscript, and to Karen Ram\'{\i}rez for generating the figures.
Our work was partially supported by DGAPA-UNAM grant IN116823. 

\bibliography{holographynr}

\providecommand{\href}[2]{#2}\begingroup\raggedright\begin{thebibliography}{100}

\bibitem{Danielsson:2000gi}
U.~H. Danielsson, A.~Güijosa, and M.~Kruczenski, ``{IIA/B, wound and
  wrapped},'' {\em JHEP} {\bf 10} (2000) 020,
\href{http://www.arXiv.org/abs/hep-th/0009182}{{\tt hep-th/0009182}}.

\bibitem{Gomis:2000bd}
J.~Gomis and H.~Ooguri, ``{Nonrelativistic closed string theory},'' {\em J.
  Math. Phys.} {\bf 42} (2001) 3127--3151,
\href{http://www.arXiv.org/abs/hep-th/0009181}{{\tt hep-th/0009181}}.

\bibitem{Danielsson:2000mu}
U.~H. Danielsson, A.~Güijosa, and M.~Kruczenski, ``{Newtonian gravitons and
  D-brane collective coordinates in wound string theory},'' {\em JHEP} {\bf 03}
  (2001) 041,
\href{http://www.arXiv.org/abs/hep-th/0012183}{{\tt hep-th/0012183}}.

\bibitem{Banks:1996vh}
T.~Banks, W.~Fischler, S.~H. Shenker, and L.~Susskind, ``{M theory as a matrix
  model: A Conjecture},'' {\em Phys. Rev. D} {\bf 55} (1997) 5112--5128,
  \href{http://www.arXiv.org/abs/hep-th/9610043}{{\tt hep-th/9610043}}.

\bibitem{Susskind:1997cw}
L.~Susskind, ``{Another conjecture about M(atrix) theory},''
  \href{http://www.arXiv.org/abs/hep-th/9704080}{{\tt hep-th/9704080}}.

\bibitem{Maldacena:1997re}
J.~M. Maldacena, ``{The Large N limit of superconformal field theories and
  supergravity},'' {\em Adv. Theor. Math. Phys.} {\bf 2} (1998) 231--252,
  \href{http://www.arXiv.org/abs/hep-th/9711200}{{\tt hep-th/9711200}}.

\bibitem{Gubser:1998bc}
S.~S. Gubser, I.~R. Klebanov, and A.~M. Polyakov, ``{Gauge theory correlators
  from noncritical string theory},'' {\em Phys. Lett. B} {\bf 428} (1998)
  105--114, \href{http://www.arXiv.org/abs/hep-th/9802109}{{\tt
  hep-th/9802109}}.

\bibitem{Witten:1998qj}
E.~Witten, ``{Anti-de Sitter space and holography},'' {\em Adv. Theor. Math.
  Phys.} {\bf 2} (1998) 253--291,
  \href{http://www.arXiv.org/abs/hep-th/9802150}{{\tt hep-th/9802150}}.

\bibitem{Oling:2022fft}
G.~Oling and Z.~Yan, ``{Aspects of Nonrelativistic Strings},'' {\em Front. in
  Phys.} {\bf 10} (2022) 832271,
  \href{http://www.arXiv.org/abs/2202.12698}{{\tt 2202.12698}}.

\bibitem{Guijosa:2023qym}
A.~Güijosa and I.~C. Rosas-L\'opez, ``{Geometry from D-branes in
  nonrelativistic string theory},'' {\em Int. J. Mod. Phys. A} {\bf 39} (2024),
  no.~17n18, 2450031, \href{http://www.arXiv.org/abs/2312.03332}{{\tt
  2312.03332}}.

\bibitem{Avila:2023aey}
D.~\'Avila, A.~Güijosa, and R.~Olmedo, ``{Asymptotically nonrelativistic
  string backgrounds},'' {\em Int. J. Mod. Phys. A} {\bf 39} (2024), no.~15n16,
  2450047, \href{http://www.arXiv.org/abs/2312.13243}{{\tt 2312.13243}}.

\bibitem{Blair:2020ops}
C.~D.~A. Blair, ``{Non-relativistic duality and $T \bar T$ deformations},''
  {\em JHEP} {\bf 07} (2020) 069,
  \href{http://www.arXiv.org/abs/2002.12413}{{\tt 2002.12413}}.

\bibitem{Blair:2023noj}
C.~D.~A. Blair, J.~Lahnsteiner, N.~A. Obers, and Z.~Yan, ``{Unification of
  Decoupling Limits in String and M Theory},'' {\em Phys. Rev. Lett.} {\bf 132}
  (2024), no.~16, 161603, \href{http://www.arXiv.org/abs/2311.10564}{{\tt
  2311.10564}}.

\bibitem{Gomis:2023eav}
J.~Gomis and Z.~Yan, ``{Worldsheet Formalism for Decoupling Limits in String
  Theory},'' \href{http://www.arXiv.org/abs/2311.10565}{{\tt 2311.10565}}.

\bibitem{Blair:2024aqz}
C.~D.~A. Blair, J.~Lahnsteiner, N.~A. Obers, and Z.~Yan, ``{Matrix Theory
  Reloaded: A BPS Road to Holography},''
  \href{http://www.arXiv.org/abs/2410.03591}{{\tt 2410.03591}}.

\bibitem{Seiberg:2000ms}
N.~Seiberg, L.~Susskind, and N.~Toumbas, ``{Strings in background electric
  field, space / time noncommutativity and a new noncritical string theory},''
  {\em JHEP} {\bf 06} (2000) 021,
\href{http://www.arXiv.org/abs/hep-th/0005040}{{\tt hep-th/0005040}}.

\bibitem{Gopakumar:2000na}
R.~Gopakumar, J.~M. Maldacena, S.~Minwalla, and A.~Strominger, ``{S duality and
  noncommutative gauge theory},'' {\em JHEP} {\bf 06} (2000) 036,
  \href{http://www.arXiv.org/abs/hep-th/0005048}{{\tt hep-th/0005048}}.

\bibitem{Klebanov:2000pp}
I.~R. Klebanov and J.~M. Maldacena, ``{(1+1)-dimensional NCOS and its U(N)
  gauge theory dual},'' {\em Int. J. Mod. Phys.} {\bf A16} (2001) 922--935,
  \href{http://www.arXiv.org/abs/hep-th/0006085}{{\tt hep-th/0006085}}.
[Adv. Theor. Math. Phys.4,283(2000)].

\bibitem{Cartan:1923zea}
E.~Cartan, ``{Sur les vari\'et\'es \`a connexion affine et la th\'eorie de la
  relativit\'e g\'en\'eralis\'ee. (premi\`ere partie)},'' {\em Annales Sci.
  Ecole Norm. Sup.} {\bf 40} (1923) 325--412.

\bibitem{Cartan:1924yea}
E.~Cartan, ``{Sur les vari\'et\'es \`a connexion affine et la th\'eorie de la
  relativit\'e g\'en\'eralis\'ee. (premi\`ere partie) (Suite).},'' {\em Annales
  Sci. Ecole Norm. Sup.} {\bf 41} (1924) 1--25.

\bibitem{Hartong:2022lsy}
J.~Hartong, N.~A. Obers, and G.~Oling, ``{Review on Non-Relativistic
  Gravity},'' \href{http://www.arXiv.org/abs/2212.11309}{{\tt 2212.11309}}.

\bibitem{Bergshoeff:2022eog}
E.~Bergshoeff, J.~Figueroa-O'Farrill, and J.~Gomis, ``{A non-lorentzian
  primer},'' {\em SciPost Phys. Lect. Notes} {\bf 69} (2023) 1,
  \href{http://www.arXiv.org/abs/2206.12177}{{\tt 2206.12177}}.

\bibitem{Andringa:2012uz}
R.~Andringa, E.~Bergshoeff, J.~Gomis, and M.~de~Roo, ``{'Stringy' Newton-Cartan
  Gravity},'' {\em Class. Quant. Grav.} {\bf 29} (2012) 235020,
  \href{http://www.arXiv.org/abs/1206.5176}{{\tt 1206.5176}}.

\bibitem{Harmark:2017rpg}
T.~Harmark, J.~Hartong, and N.~A. Obers, ``{Nonrelativistic strings and limits
  of the AdS/CFT correspondence},'' {\em Phys. Rev. D} {\bf 96} (2017), no.~8,
  086019, \href{http://www.arXiv.org/abs/1705.03535}{{\tt 1705.03535}}.

\bibitem{Bergshoeff:2018yvt}
E.~Bergshoeff, J.~Gomis, and Z.~Yan, ``{Nonrelativistic String Theory and
  T-Duality},'' {\em JHEP} {\bf 11} (2018) 133,
\href{http://www.arXiv.org/abs/1806.06071}{{\tt 1806.06071}}.

\bibitem{Harmark:2018cdl}
T.~Harmark, J.~Hartong, L.~Menculini, N.~A. Obers, and Z.~Yan, ``{Strings with
  Non-Relativistic Conformal Symmetry and Limits of the AdS/CFT
  Correspondence},'' {\em JHEP} {\bf 11} (2018) 190,
  \href{http://www.arXiv.org/abs/1810.05560}{{\tt 1810.05560}}.

\bibitem{Gomis:2019zyu}
J.~Gomis, J.~Oh, and Z.~Yan, ``{Nonrelativistic String Theory in Background
  Fields},'' {\em JHEP} {\bf 10} (2019) 101,
\href{http://www.arXiv.org/abs/1905.07315}{{\tt 1905.07315}}.

\bibitem{Gallegos:2019icg}
A.~D. Gallegos, U.~G\"ursoy, and N.~Zinnato, ``{Torsional Newton Cartan gravity
  from non-relativistic strings},'' {\em JHEP} {\bf 09} (2020) 172,
  \href{http://www.arXiv.org/abs/1906.01607}{{\tt 1906.01607}}.

\bibitem{Harmark:2019upf}
T.~Harmark, J.~Hartong, L.~Menculini, N.~A. Obers, and G.~Oling, ``{Relating
  non-relativistic string theories},'' {\em JHEP} {\bf 11} (2019) 071,
  \href{http://www.arXiv.org/abs/1907.01663}{{\tt 1907.01663}}.

\bibitem{Bergshoeff:2021bmc}
E.~A. Bergshoeff, J.~Lahnsteiner, L.~Romano, J.~Rosseel, and C.~\c{S}im\c{s}ek,
  ``{A non-relativistic limit of NS-NS gravity},'' {\em JHEP} {\bf 06} (2021)
  021, \href{http://www.arXiv.org/abs/2102.06974}{{\tt 2102.06974}}.

\bibitem{Bidussi:2021ujm}
L.~Bidussi, T.~Harmark, J.~Hartong, N.~A. Obers, and G.~Oling, ``{Torsional
  string Newton-Cartan geometry for non-relativistic strings},'' {\em JHEP}
  {\bf 02} (2022) 116, \href{http://www.arXiv.org/abs/2107.00642}{{\tt
  2107.00642}}.

\bibitem{Dabholkar:1989jt}
A.~Dabholkar and J.~A. Harvey, ``{Nonrenormalization of the Superstring
  Tension},'' {\em Phys. Rev. Lett.} {\bf 63} (1989)
478.

\bibitem{Dabholkar:1990yf}
A.~Dabholkar, G.~W. Gibbons, J.~A. Harvey, and F.~Ruiz~Ruiz, ``{Superstrings
  and Solitons},'' {\em Nucl. Phys.} {\bf B340} (1990)
33--55.

\bibitem{Yan:2019xsf}
Z.~Yan and M.~Yu, ``{Background Field Method for Nonlinear Sigma Models in
  Nonrelativistic String Theory},'' {\em JHEP} {\bf 03} (2020) 181,
  \href{http://www.arXiv.org/abs/1912.03181}{{\tt 1912.03181}}.

\bibitem{Gomis:2020fui}
J.~Gomis, Z.~Yan, and M.~Yu, ``{Nonrelativistic Open String and Yang-Mills
  Theory},'' {\em JHEP} {\bf 03} (2021) 269,
  \href{http://www.arXiv.org/abs/2007.01886}{{\tt 2007.01886}}.

\bibitem{Yan:2021lbe}
Z.~Yan, ``{Torsional deformation of nonrelativistic string theory},'' {\em
  JHEP} {\bf 09} (2021) 035, \href{http://www.arXiv.org/abs/2106.10021}{{\tt
  2106.10021}}.

\bibitem{Smirnov:2016lqw}
F.~A. Smirnov and A.~B. Zamolodchikov, ``{On space of integrable quantum field
  theories},'' {\em Nucl. Phys. B} {\bf 915} (2017) 363--383,
  \href{http://www.arXiv.org/abs/1608.05499}{{\tt 1608.05499}}.

\bibitem{Cavaglia:2016oda}
A.~Cavagli\`a, S.~Negro, I.~M. Sz\'ecs\'enyi, and R.~Tateo, ``{$T
  \bar{T}$-deformed 2D Quantum Field Theories},'' {\em JHEP} {\bf 10} (2016)
  112, \href{http://www.arXiv.org/abs/1608.05534}{{\tt 1608.05534}}.

\bibitem{Jiang:2019epa}
Y.~Jiang, ``{A pedagogical review on solvable irrelevant deformations of 2D
  quantum field theory},'' {\em Commun. Theor. Phys.} {\bf 73} (2021), no.~5,
  057201, \href{http://www.arXiv.org/abs/1904.13376}{{\tt 1904.13376}}.

\bibitem{Seiberg:1997ad}
N.~Seiberg, ``{Why is the matrix model correct?},'' {\em Phys. Rev. Lett.} {\bf
  79} (1997) 3577--3580, \href{http://www.arXiv.org/abs/hep-th/9710009}{{\tt
  hep-th/9710009}}.

\bibitem{Sen:1997we}
A.~Sen, ``{D0-branes on T$^n$ and matrix theory},'' {\em Adv. Theor. Math.
  Phys.} {\bf 2} (1998) 51--59,
  \href{http://www.arXiv.org/abs/hep-th/9709220}{{\tt hep-th/9709220}}.

\bibitem{Hellerman:1997yu}
S.~Hellerman and J.~Polchinski, ``{Compactification in the lightlike limit},''
  {\em Phys. Rev. D} {\bf 59} (1999) 125002,
  \href{http://www.arXiv.org/abs/hep-th/9711037}{{\tt hep-th/9711037}}.

\bibitem{Guijosa:1998rq}
A.~Güijosa, ``{Is physics in the infinite momentum frame independent of the
  compactification radius?},'' {\em Nucl. Phys. B} {\bf 533} (1998) 406--426,
  \href{http://www.arXiv.org/abs/hep-th/9804034}{{\tt hep-th/9804034}}.

\bibitem{Bilal:1998ys}
A.~Bilal, ``{DLCQ of M theory as the lightlike limit},'' {\em Phys. Lett. B}
  {\bf 435} (1998) 312--318,
  \href{http://www.arXiv.org/abs/hep-th/9805070}{{\tt hep-th/9805070}}.

\bibitem{Dijkgraaf:1997vv}
R.~Dijkgraaf, E.~P. Verlinde, and H.~L. Verlinde, ``{Matrix string theory},''
  {\em Nucl. Phys. B} {\bf 500} (1997) 43--61,
  \href{http://www.arXiv.org/abs/hep-th/9703030}{{\tt hep-th/9703030}}.

\bibitem{Hyun:2000cw}
S.~Hyun, ``{U-duality between NCOS theory and matrix theory},'' {\em Nucl.
  Phys. B} {\bf 598} (2001) 276--290,
  \href{http://www.arXiv.org/abs/hep-th/0008213}{{\tt hep-th/0008213}}.

\bibitem{Witten:1995im}
E.~Witten, ``{Bound states of strings and p-branes},'' {\em Nucl. Phys. B} {\bf
  460} (1996) 335--350, \href{http://www.arXiv.org/abs/hep-th/9510135}{{\tt
  hep-th/9510135}}.

\bibitem{Gopakumar:2000ep}
R.~Gopakumar, S.~Minwalla, N.~Seiberg, and A.~Strominger, ``{(OM) theory in
  diverse dimensions},'' {\em JHEP} {\bf 08} (2000) 008,
\href{http://www.arXiv.org/abs/hep-th/0006062}{{\tt hep-th/0006062}}.

\bibitem{Garcia:2002fa}
J.~García, A.~Güijosa, and J.~Vergara, ``{A Membrane action for OM theory},''
  {\em Nucl. Phys. B} {\bf 630} (2002) 178--202,
  \href{http://www.arXiv.org/abs/hep-th/0201140}{{\tt hep-th/0201140}}.

\bibitem{Seiberg:1997zk}
N.~Seiberg, ``{New theories in six-dimensions and matrix description of M
  theory on T$^5$ and T$^5$/Z$_2$},'' {\em Phys. Lett. B} {\bf 408} (1997)
  98--104, \href{http://www.arXiv.org/abs/hep-th/9705221}{{\tt
  hep-th/9705221}}.

\bibitem{Losev:1997hx}
A.~Losev, G.~W. Moore, and S.~L. Shatashvili, ``{M \& m's},'' {\em Nucl. Phys.
  B} {\bf 522} (1998) 105--124,
  \href{http://www.arXiv.org/abs/hep-th/9707250}{{\tt hep-th/9707250}}.

\bibitem{Aharony:2019zsx}
O.~Aharony, M.~Evtikhiev, and A.~Feldman, ``{Little String Theories on Curved
  Manifolds},'' {\em JHEP} {\bf 10} (2019) 180,
  \href{http://www.arXiv.org/abs/1908.02642}{{\tt 1908.02642}}.

\bibitem{Sheikh-Jabbari:1997qke}
M.~M. Sheikh-Jabbari, ``{More on mixed boundary conditions and D-branes bound
  states},'' {\em Phys. Lett. B} {\bf 425} (1998) 48--54,
  \href{http://www.arXiv.org/abs/hep-th/9712199}{{\tt hep-th/9712199}}.

\bibitem{Sheikh-Jabbari:1998aur}
M.~M. Sheikh-Jabbari, ``{SuperYang-Mills theory on noncommutative torus from
  open strings interactions},'' {\em Phys. Lett. B} {\bf 450} (1999) 119--125,
  \href{http://www.arXiv.org/abs/hep-th/9810179}{{\tt hep-th/9810179}}.

\bibitem{Seiberg:1999vs}
N.~Seiberg and E.~Witten, ``{String theory and noncommutative geometry},'' {\em
  JHEP} {\bf 09} (1999) 032,
  \href{http://www.arXiv.org/abs/hep-th/9908142}{{\tt hep-th/9908142}}.

\bibitem{Bergshoeff:2022iss}
E.~A. Bergshoeff, K.~T. Grosvenor, J.~Lahnsteiner, Z.~Yan, and U.~Zorba,
  ``{Branched SL(2,\ensuremath{\mathbb{Z}}) duality},'' {\em JHEP} {\bf 10}
  (2022) 131, \href{http://www.arXiv.org/abs/2208.13815}{{\tt 2208.13815}}.

\bibitem{Fontanella:2024rvn}
A.~Fontanella and J.~M. Nieto~Garc\'\i{}a, ``{Constructing Non-Relativistic
  AdS$_5$/CFT$_4$ Holography},''
  \href{http://www.arXiv.org/abs/2403.02379}{{\tt 2403.02379}}.

\bibitem{Lambert:2024uue}
N.~Lambert and J.~Smith, ``{Non-relativistic M2-branes and the AdS/CFT
  correspondence},'' {\em JHEP} {\bf 06} (2024) 009,
  \href{http://www.arXiv.org/abs/2401.14955}{{\tt 2401.14955}}.

\bibitem{Lambert:2024yjk}
N.~Lambert and J.~Smith, ``{Non-relativistic intersecting branes, Newton-Cartan
  geometry and AdS/CFT},'' {\em JHEP} {\bf 07} (2024) 224,
  \href{http://www.arXiv.org/abs/2405.06552}{{\tt 2405.06552}}.

\bibitem{Fontanella:2024kyl}
A.~Fontanella and J.~M. Nieto~Garc\'\i{}a, ``{Nonrelativistic Holography from
  AdS5/CFT4},'' {\em Phys. Rev. Lett.} {\bf 133} (2024), no.~15, 151601,
  \href{http://www.arXiv.org/abs/2409.02267}{{\tt 2409.02267}}.

\bibitem{Lambert:2024ncn}
N.~Lambert and J.~Smith, ``{Reciprocal non-relativistic decoupling limits of
  String Theory and M-Theory},'' {\em JHEP} {\bf 12} (2024) 094,
  \href{http://www.arXiv.org/abs/2410.17074}{{\tt 2410.17074}}.

\bibitem{Cardona:2016ytk}
B.~Cardona, J.~Gomis, and J.~M. Pons, ``{Dynamics of Carroll Strings},'' {\em
  JHEP} {\bf 07} (2016) 050, \href{http://www.arXiv.org/abs/1605.05483}{{\tt
  1605.05483}}.

\bibitem{Bagchi:2023cfp}
A.~Bagchi, A.~Banerjee, J.~Hartong, E.~Have, K.~S. Kolekar, and M.~Mandlik,
  ``{Strings near black holes are Carrollian},'' {\em Phys. Rev. D} {\bf 110}
  (2024), no.~8, 086009, \href{http://www.arXiv.org/abs/2312.14240}{{\tt
  2312.14240}}.

\bibitem{Harksen:2024bnh}
M.~Harksen, D.~Hidalgo, W.~Sybesma, and L.~Thorlacius, ``{Carroll strings with
  an extended symmetry algebra},'' {\em JHEP} {\bf 05} (2024) 206,
  \href{http://www.arXiv.org/abs/2403.01984}{{\tt 2403.01984}}.

\bibitem{Bagchi:2024rje}
A.~Bagchi, A.~Banerjee, J.~Hartong, E.~Have, and K.~S. Kolekar, ``{Strings near
  black holes are Carrollian. Part II},'' {\em JHEP} {\bf 11} (2024) 024,
  \href{http://www.arXiv.org/abs/2407.12911}{{\tt 2407.12911}}.

\bibitem{Harmark:2014mpa}
T.~Harmark and M.~Orselli, ``{Spin Matrix Theory: A quantum mechanical model of
  the AdS/CFT correspondence},'' {\em JHEP} {\bf 11} (2014) 134,
  \href{http://www.arXiv.org/abs/1409.4417}{{\tt 1409.4417}}.

\bibitem{Lindstrom:1990qb}
U.~Lindstrom, B.~Sundborg, and G.~Theodoridis, ``{The Zero tension limit of the
  superstring},'' {\em Phys. Lett. B} {\bf 253} (1991) 319--323.

\bibitem{Lindstrom:1990ar}
U.~Lindstrom, B.~Sundborg, and G.~Theodoridis, ``{The Zero tension limit of the
  spinning string},'' {\em Phys. Lett. B} {\bf 258} (1991) 331--334.

\bibitem{Isberg:1992ia}
J.~Isberg, U.~Lindstrom, and B.~Sundborg, ``{Space-time symmetries of quantized
  tensionless strings},'' {\em Phys. Lett. B} {\bf 293} (1992) 321--326,
  \href{http://www.arXiv.org/abs/hep-th/9207005}{{\tt hep-th/9207005}}.

\bibitem{Bagchi:2013bga}
A.~Bagchi, ``{Tensionless Strings and Galilean Conformal Algebra},'' {\em JHEP}
  {\bf 05} (2013) 141, \href{http://www.arXiv.org/abs/1303.0291}{{\tt
  1303.0291}}.

\bibitem{Bagchi:2015nca}
A.~Bagchi, S.~Chakrabortty, and P.~Parekh, ``{Tensionless Strings from
  Worldsheet Symmetries},'' {\em JHEP} {\bf 01} (2016) 158,
  \href{http://www.arXiv.org/abs/1507.04361}{{\tt 1507.04361}}.

\bibitem{Bagchi:2016yyf}
A.~Bagchi, S.~Chakrabortty, and P.~Parekh, ``{Tensionless Superstrings: View
  from the Worldsheet},'' {\em JHEP} {\bf 10} (2016) 113,
  \href{http://www.arXiv.org/abs/1606.09628}{{\tt 1606.09628}}.

\bibitem{Gomis:2005pg}
J.~Gomis, J.~Gomis, and K.~Kamimura, ``{Non-relativistic superstrings: A New
  soluble sector of $AdS_5 \times S^5$},'' {\em JHEP} {\bf 12} (2005) 024,
  \href{http://www.arXiv.org/abs/hep-th/0507036}{{\tt hep-th/0507036}}.

\bibitem{Sakaguchi:2007ba}
M.~Sakaguchi and K.~Yoshida, ``{Holography of Non-relativistic String on AdS(5)
  x S**5},'' {\em JHEP} {\bf 02} (2008) 092,
  \href{http://www.arXiv.org/abs/0712.4112}{{\tt 0712.4112}}.

\bibitem{Harmark:2020vll}
T.~Harmark, J.~Hartong, N.~A. Obers, and G.~Oling, ``{Spin Matrix Theory String
  Backgrounds and Penrose Limits of AdS/CFT},'' {\em JHEP} {\bf 03} (2021) 129,
  \href{http://www.arXiv.org/abs/2011.02539}{{\tt 2011.02539}}.

\bibitem{Itzhaki:1998dd}
N.~Itzhaki, J.~M. Maldacena, J.~Sonnenschein, and S.~Yankielowicz,
  ``{Supergravity and the large N limit of theories with sixteen
  supercharges},'' {\em Phys. Rev. D} {\bf 58} (1998) 046004,
  \href{http://www.arXiv.org/abs/hep-th/9802042}{{\tt hep-th/9802042}}.

\bibitem{Balasubramanian:1997kd}
V.~Balasubramanian, R.~Gopakumar, and F.~Larsen, ``{Gauge theory, geometry and
  the large N limit},'' {\em Nucl. Phys. B} {\bf 526} (1998) 415--431,
  \href{http://www.arXiv.org/abs/hep-th/9712077}{{\tt hep-th/9712077}}.

\bibitem{Hyun:1997zt}
S.~Hyun, Y.~Kiem, and H.~Shin, ``{Infinite Lorentz boost along the M theory
  circle and nonasymptotically flat solutions in supergravities},'' {\em Phys.
  Rev. D} {\bf 57} (1998) 4856--4861,
  \href{http://www.arXiv.org/abs/hep-th/9712021}{{\tt hep-th/9712021}}.

\bibitem{Hyun:1998bi}
S.~Hyun, ``{The Background geometry of DLCQ supergravity},'' {\em Phys. Lett.
  B} {\bf 441} (1998) 116--122,
  \href{http://www.arXiv.org/abs/hep-th/9802026}{{\tt hep-th/9802026}}.

\bibitem{Hyun:1998iq}
S.~Hyun and Y.~Kiem, ``{Background geometry of DLCQ M theory on a p - torus and
  holography},'' {\em Phys. Rev. D} {\bf 59} (1999) 026003,
  \href{http://www.arXiv.org/abs/hep-th/9805136}{{\tt hep-th/9805136}}.

\bibitem{deAlwis:1998ki}
S.~P. de~Alwis, ``{On the supergravity gauge theory correspondence and the
  matrix model},'' {\em Phys. Rev. D} {\bf 59} (1999) 044029,
  \href{http://www.arXiv.org/abs/hep-th/9806178}{{\tt hep-th/9806178}}.

\bibitem{Polchinski:1999br}
J.~Polchinski, ``{M theory and the light cone},'' {\em Prog. Theor. Phys.
  Suppl.} {\bf 134} (1999) 158--170,
  \href{http://www.arXiv.org/abs/hep-th/9903165}{{\tt hep-th/9903165}}.

\bibitem{Kluson:2017abm}
J.~Kluson, ``{Note about Hamiltonian formalism for Newton\textendash{}Cartan
  string and p-brane},'' {\em Eur. Phys. J. C} {\bf 78} (2018), no.~6, 511,
  \href{http://www.arXiv.org/abs/1712.07430}{{\tt 1712.07430}}.

\bibitem{Pereniguez:2019eoq}
D.~Pere\~niguez, ``{$p$-brane Newton\textendash{}Cartan geometry},'' {\em J.
  Math. Phys.} {\bf 60} (2019), no.~11, 112501,
  \href{http://www.arXiv.org/abs/1908.04801}{{\tt 1908.04801}}.

\bibitem{Blair:2021waq}
C.~D.~A. Blair, D.~Gallegos, and N.~Zinnato, ``{A non-relativistic limit of
  M-theory and 11-dimensional membrane Newton-Cartan geometry},'' {\em JHEP}
  {\bf 10} (2021) 015, \href{http://www.arXiv.org/abs/2104.07579}{{\tt
  2104.07579}}.

\bibitem{Ebert:2021mfu}
S.~Ebert, H.-Y. Sun, and Z.~Yan, ``{Dual D-brane actions in nonrelativistic
  string theory},'' {\em JHEP} {\bf 04} (2022) 161,
  \href{http://www.arXiv.org/abs/2112.09316}{{\tt 2112.09316}}.

\bibitem{Novosad:2021tlq}
P.~Novosad, ``{Lagrangians for nonrelativistic gravity},'' {\em Phys. Rev. D}
  {\bf 105} (2022), no.~6, 064051,
  \href{http://www.arXiv.org/abs/2112.12648}{{\tt 2112.12648}}.

\bibitem{Bergshoeff:2023rkk}
E.~Bergshoeff, J.~Figueroa-O'Farrill, K.~van Helden, J.~Rosseel, I.~Rotko, and
  T.~ter Veldhuis, ``{$p$-brane Galilean and Carrollian geometries and
  gravities},'' {\em J. Phys. A} {\bf 57} (2024), no.~24, 245205,
  \href{http://www.arXiv.org/abs/2308.12852}{{\tt 2308.12852}}.

\bibitem{Ebert:2023hba}
S.~Ebert and Z.~Yan, ``{Anisotropic compactification of nonrelativistic
  M-theory},'' {\em JHEP} {\bf 11} (2023) 135,
  \href{http://www.arXiv.org/abs/2309.04912}{{\tt 2309.04912}}.

\bibitem{Bergshoeff:2024ipq}
E.~A. Bergshoeff, G.~Giorgi, and L.~Romano, ``{From Relativistic Gravity to the
  Poisson Equation},'' \href{http://www.arXiv.org/abs/2410.00692}{{\tt
  2410.00692}}.

\bibitem{Aharony:1998ub}
O.~Aharony, M.~Berkooz, D.~Kutasov, and N.~Seiberg, ``{Linear dilatons, NS
  five-branes and holography},'' {\em JHEP} {\bf 10} (1998) 004,
  \href{http://www.arXiv.org/abs/hep-th/9808149}{{\tt hep-th/9808149}}.

\bibitem{Klebanov:1998hh}
I.~R. Klebanov and E.~Witten, ``{Superconformal field theory on three-branes at
  a Calabi-Yau singularity},'' {\em Nucl. Phys. B} {\bf 536} (1998) 199--218,
  \href{http://www.arXiv.org/abs/hep-th/9807080}{{\tt hep-th/9807080}}.

\bibitem{Bachas:1995kx}
C.~Bachas, ``{D-brane dynamics},'' {\em Phys. Lett. B} {\bf 374} (1996) 37--42,
  \href{http://www.arXiv.org/abs/hep-th/9511043}{{\tt hep-th/9511043}}.

\bibitem{Lifschytz:1996iq}
G.~Lifschytz, ``{Comparing d-branes to black-branes},'' {\em Phys. Lett. B}
  {\bf 388} (1996) 720--726,
  \href{http://www.arXiv.org/abs/hep-th/9604156}{{\tt hep-th/9604156}}.

\bibitem{Douglas:1996yp}
M.~R. Douglas, D.~N. Kabat, P.~Pouliot, and S.~H. Shenker, ``{D-branes and
  short distances in string theory},'' {\em Nucl. Phys. B} {\bf 485} (1997)
  85--127, \href{http://www.arXiv.org/abs/hep-th/9608024}{{\tt
  hep-th/9608024}}.

\bibitem{Becker:1997wh}
K.~Becker and M.~Becker, ``{A Two loop test of M(atrix) theory},'' {\em Nucl.
  Phys. B} {\bf 506} (1997) 48--60,
  \href{http://www.arXiv.org/abs/hep-th/9705091}{{\tt hep-th/9705091}}.

\bibitem{Becker:1997xw}
K.~Becker, M.~Becker, J.~Polchinski, and A.~A. Tseytlin, ``{Higher order
  graviton scattering in M(atrix) theory},'' {\em Phys. Rev. D} {\bf 56} (1997)
  R3174--R3178, \href{http://www.arXiv.org/abs/hep-th/9706072}{{\tt
  hep-th/9706072}}.

\bibitem{Dine:1997sz}
M.~Dine and A.~Rajaraman, ``{Multigraviton scattering in the matrix model},''
  {\em Phys. Lett. B} {\bf 425} (1998) 77--85,
  \href{http://www.arXiv.org/abs/hep-th/9710174}{{\tt hep-th/9710174}}.

\bibitem{Becker:1997cp}
K.~Becker and M.~Becker, ``{On graviton scattering amplitudes in M theory},''
  {\em Phys. Rev. D} {\bf 57} (1998) 6464--6470,
  \href{http://www.arXiv.org/abs/hep-th/9712238}{{\tt hep-th/9712238}}.

\bibitem{Buchbinder:2001ui}
I.~L. Buchbinder, A.~Y. Petrov, and A.~A. Tseytlin, ``{Two loop N=4
  superYang-Mills effective action and interaction between D3-branes},'' {\em
  Nucl. Phys. B} {\bf 621} (2002) 179--207,
  \href{http://www.arXiv.org/abs/hep-th/0110173}{{\tt hep-th/0110173}}.

\bibitem{Taylor:2001vb}
W.~Taylor, ``{M(atrix) Theory: Matrix Quantum Mechanics as a Fundamental
  Theory},'' {\em Rev. Mod. Phys.} {\bf 73} (2001) 419--462,
  \href{http://www.arXiv.org/abs/hep-th/0101126}{{\tt hep-th/0101126}}.

\bibitem{Ryu:2006bv}
S.~Ryu and T.~Takayanagi, ``{Holographic derivation of entanglement entropy
  from AdS/CFT},'' {\em Phys. Rev. Lett.} {\bf 96} (2006) 181602,
  \href{http://www.arXiv.org/abs/hep-th/0603001}{{\tt hep-th/0603001}}.

\bibitem{Ryu:2006ef}
S.~Ryu and T.~Takayanagi, ``{Aspects of Holographic Entanglement Entropy},''
  {\em JHEP} {\bf 08} (2006) 045,
  \href{http://www.arXiv.org/abs/hep-th/0605073}{{\tt hep-th/0605073}}.

\bibitem{Gomis:2004pw}
J.~Gomis, K.~Kamimura, and P.~K. Townsend, ``{Non-relativistic superbranes},''
  {\em JHEP} {\bf 11} (2004) 051,
  \href{http://www.arXiv.org/abs/hep-th/0409219}{{\tt hep-th/0409219}}.

\bibitem{Batlle:2016iel}
C.~Batlle, J.~Gomis, and D.~Not, ``{Extended Galilean symmetries of
  non-relativistic strings},'' {\em JHEP} {\bf 02} (2017) 049,
  \href{http://www.arXiv.org/abs/1611.00026}{{\tt 1611.00026}}.

\bibitem{Seiberg:1999xz}
N.~Seiberg and E.~Witten, ``{The D1 / D5 system and singular CFT},'' {\em JHEP}
  {\bf 04} (1999) 017, \href{http://www.arXiv.org/abs/hep-th/9903224}{{\tt
  hep-th/9903224}}.

\bibitem{Guijosa:2001notes}
A.~Güijosa, ``Black branes in wound string theory.'' unpublished, 2001.

\bibitem{Bergshoeff:2019pij}
E.~A. Bergshoeff, J.~Gomis, J.~Rosseel, C.~Şimşek, and Z.~Yan, ``{String
  Theory and String Newton-Cartan Geometry},'' {\em J. Phys.} {\bf A53} (2020),
  no.~1, 014001,
\href{http://www.arXiv.org/abs/1907.10668}{{\tt 1907.10668}}.

\bibitem{Bergshoeff:2021tfn}
E.~A. Bergshoeff, J.~Lahnsteiner, L.~Romano, J.~Rosseel, and C.~Simsek,
  ``{Non-relativistic ten-dimensional minimal supergravity},'' {\em JHEP} {\bf
  12} (2021) 123, \href{http://www.arXiv.org/abs/2107.14636}{{\tt 2107.14636}}.

\bibitem{Schwarz:1995dk}
J.~H. Schwarz, ``{An SL(2,Z) multiplet of type IIB superstrings},'' {\em Phys.
  Lett. B} {\bf 360} (1995) 13--18,
  \href{http://www.arXiv.org/abs/hep-th/9508143}{{\tt hep-th/9508143}}.
  [Erratum: Phys.Lett.B 364, 252 (1995)].

\bibitem{Russo:1996if}
J.~G. Russo and A.~A. Tseytlin, ``{Waves, boosted branes and BPS states in m
  theory},'' {\em Nucl. Phys. B} {\bf 490} (1997) 121--144,
  \href{http://www.arXiv.org/abs/hep-th/9611047}{{\tt hep-th/9611047}}.

\bibitem{Costa:1996zd}
M.~S. Costa and G.~Papadopoulos, ``{Superstring dualities and p-brane bound
  states},'' {\em Nucl. Phys. B} {\bf 510} (1998) 217--231,
  \href{http://www.arXiv.org/abs/hep-th/9612204}{{\tt hep-th/9612204}}.

\bibitem{Lu:1999uca}
J.~X. Lu and S.~Roy, ``{Nonthreshold (f, Dp) bound states},'' {\em Nucl. Phys.
  B} {\bf 560} (1999) 181--206,
  \href{http://www.arXiv.org/abs/hep-th/9904129}{{\tt hep-th/9904129}}.

\bibitem{Harmark:2000wv}
T.~Harmark, ``{Supergravity and space-time noncommutative open string
  theory},'' {\em JHEP} {\bf 07} (2000) 043,
  \href{http://www.arXiv.org/abs/hep-th/0006023}{{\tt hep-th/0006023}}.

\bibitem{Horowitz:1991cd}
G.~T. Horowitz and A.~Strominger, ``{Black strings and P-branes},'' {\em Nucl.
  Phys. B} {\bf 360} (1991) 197--209.

\bibitem{Polchinski:1995mt}
J.~Polchinski, ``{Dirichlet Branes and Ramond-Ramond charges},'' {\em Phys.
  Rev. Lett.} {\bf 75} (1995) 4724--4727,
  \href{http://www.arXiv.org/abs/hep-th/9510017}{{\tt hep-th/9510017}}.

\bibitem{Douglas:1995bn}
M.~R. Douglas, ``{Branes within branes},'' {\em NATO Sci. Ser. C} {\bf 520}
  (1999) 267--275, \href{http://www.arXiv.org/abs/hep-th/9512077}{{\tt
  hep-th/9512077}}.

\bibitem{Dai:1989ua}
J.~Dai, R.~G. Leigh, and J.~Polchinski, ``{New Connections Between String
  Theories},'' {\em Mod. Phys. Lett. A} {\bf 4} (1989) 2073--2083.

\bibitem{Horava:1989ga}
P.~Horava, ``{Background Duality of Open String Models},'' {\em Phys. Lett. B}
  {\bf 231} (1989) 251--257.

\bibitem{Strominger:1996sh}
A.~Strominger and C.~Vafa, ``{Microscopic origin of the Bekenstein-Hawking
  entropy},'' {\em Phys. Lett. B} {\bf 379} (1996) 99--104,
  \href{http://www.arXiv.org/abs/hep-th/9601029}{{\tt hep-th/9601029}}.

\bibitem{Gubser:1996wt}
S.~S. Gubser, A.~Hashimoto, I.~R. Klebanov, and J.~M. Maldacena,
  ``{Gravitational lensing by p-branes},'' {\em Nucl. Phys. B} {\bf 472} (1996)
  231--248, \href{http://www.arXiv.org/abs/hep-th/9601057}{{\tt
  hep-th/9601057}}.

\bibitem{Callan:1996dv}
C.~G. Callan and J.~M. Maldacena, ``{D-brane approach to black hole quantum
  mechanics},'' {\em Nucl. Phys. B} {\bf 472} (1996) 591--610,
  \href{http://www.arXiv.org/abs/hep-th/9602043}{{\tt hep-th/9602043}}.

\bibitem{Gubser:1996de}
S.~S. Gubser, I.~R. Klebanov, and A.~W. Peet, ``{Entropy and temperature of
  black 3-branes},'' {\em Phys. Rev. D} {\bf 54} (1996) 3915--3919,
  \href{http://www.arXiv.org/abs/hep-th/9602135}{{\tt hep-th/9602135}}.

\bibitem{Maldacena:1996ix}
J.~M. Maldacena and A.~Strominger, ``{Black hole grey body factors and d-brane
  spectroscopy},'' {\em Phys. Rev. D} {\bf 55} (1997) 861--870,
  \href{http://www.arXiv.org/abs/hep-th/9609026}{{\tt hep-th/9609026}}.

\bibitem{Klebanov:1997kc}
I.~R. Klebanov, ``{World volume approach to absorption by nondilatonic
  branes},'' {\em Nucl. Phys. B} {\bf 496} (1997) 231--242,
  \href{http://www.arXiv.org/abs/hep-th/9702076}{{\tt hep-th/9702076}}.

\bibitem{Gubser:1997yh}
S.~S. Gubser, I.~R. Klebanov, and A.~A. Tseytlin, ``{String theory and
  classical absorption by three-branes},'' {\em Nucl. Phys. B} {\bf 499} (1997)
  217--240, \href{http://www.arXiv.org/abs/hep-th/9703040}{{\tt
  hep-th/9703040}}.

\bibitem{Gubser:1997se}
S.~S. Gubser and I.~R. Klebanov, ``{Absorption by branes and Schwinger terms in
  the world volume theory},'' {\em Phys. Lett. B} {\bf 413} (1997) 41--48,
  \href{http://www.arXiv.org/abs/hep-th/9708005}{{\tt hep-th/9708005}}.

\bibitem{Kruczenski:2007jg}
M.~Kruczenski, ``{Summing planar diagrams},'' {\em JHEP} {\bf 10} (2008) 075,
  \href{http://www.arXiv.org/abs/hep-th/0703218}{{\tt hep-th/0703218}}.

\bibitem{Skenderis:1998dq}
K.~Skenderis, ``{Field theory limit of branes and gauged supergravities},''
  {\em Fortsch. Phys.} {\bf 48} (2000) 205--208,
  \href{http://www.arXiv.org/abs/hep-th/9903003}{{\tt hep-th/9903003}}.

\bibitem{Behrndt:1999mk}
K.~Behrndt, E.~Bergshoeff, R.~Halbersma, and J.~P. van~der Schaar, ``{On domain
  wall / QFT dualities in various dimensions},'' {\em Class. Quant. Grav.} {\bf
  16} (1999) 3517--3552, \href{http://www.arXiv.org/abs/hep-th/9907006}{{\tt
  hep-th/9907006}}.

\bibitem{Callan:1997kz}
C.~G. Callan and J.~M. Maldacena, ``{Brane death and dynamics from the
  Born-Infeld action},'' {\em Nucl. Phys. B} {\bf 513} (1998) 198--212,
  \href{http://www.arXiv.org/abs/hep-th/9708147}{{\tt hep-th/9708147}}.

\bibitem{Gibbons:1997xz}
G.~W. Gibbons, ``{Born-Infeld particles and Dirichlet p-branes},'' {\em Nucl.
  Phys. B} {\bf 514} (1998) 603--639,
  \href{http://www.arXiv.org/abs/hep-th/9709027}{{\tt hep-th/9709027}}.

\bibitem{Myers:1999ps}
R.~C. Myers, ``{Dielectric branes},'' {\em JHEP} {\bf 12} (1999) 022,
  \href{http://www.arXiv.org/abs/hep-th/9910053}{{\tt hep-th/9910053}}.

\bibitem{DeBoer:2001uk}
J.~De~Boer and K.~Schalm, ``{General covariance of the nonAbelian DBI
  action},'' {\em JHEP} {\bf 02} (2003) 041,
  \href{http://www.arXiv.org/abs/hep-th/0108161}{{\tt hep-th/0108161}}.

\bibitem{Brecher:2004qi}
D.~Brecher, K.~Furuuchi, H.~Ling, and M.~Van~Raamsdonk, ``{Generally covariant
  actions for multiple D-branes},'' {\em JHEP} {\bf 06} (2004) 020,
  \href{http://www.arXiv.org/abs/hep-th/0403289}{{\tt hep-th/0403289}}.

\bibitem{Brecher:2005sj}
D.~Brecher, P.~Koerber, H.~Ling, and M.~Van~Raamsdonk, ``{Poincare invariance
  in multiple D-brane actions},'' {\em JHEP} {\bf 01} (2006) 151,
  \href{http://www.arXiv.org/abs/hep-th/0509026}{{\tt hep-th/0509026}}.

\bibitem{Ferrari:2013pi}
F.~Ferrari, ``{On Matrix Geometry and Effective Actions},'' {\em Nucl. Phys. B}
  {\bf 871} (2013) 181--221, \href{http://www.arXiv.org/abs/1301.3722}{{\tt
  1301.3722}}.

\bibitem{Howe:2006rv}
P.~S. Howe, U.~Lindstrom, and L.~Wulff, ``{On the covariance of the
  Dirac-Born-Infeld-Myers action},'' {\em JHEP} {\bf 02} (2007) 070,
  \href{http://www.arXiv.org/abs/hep-th/0607156}{{\tt hep-th/0607156}}.

\bibitem{Leigh:1989jq}
R.~G. Leigh, ``{Dirac-Born-Infeld Action from Dirichlet Sigma Model},'' {\em
  Mod. Phys. Lett. A} {\bf 4} (1989) 2767.

\bibitem{Maldacena:1998im}
J.~M. Maldacena, ``{Wilson loops in large N field theories},'' {\em Phys. Rev.
  Lett.} {\bf 80} (1998) 4859--4862,
  \href{http://www.arXiv.org/abs/hep-th/9803002}{{\tt hep-th/9803002}}.

\bibitem{Rey:1998ik}
S.-J. Rey and J.-T. Yee, ``{Macroscopic strings as heavy quarks in large N
  gauge theory and anti-de Sitter supergravity},'' {\em Eur. Phys. J. C} {\bf
  22} (2001) 379--394, \href{http://www.arXiv.org/abs/hep-th/9803001}{{\tt
  hep-th/9803001}}.

\bibitem{Sonnenschein:1999if}
J.~Sonnenschein, ``{What does the string / gauge correspondence teach us about
  Wilson loops?},'' in {\em {Advanced School on Supersymmetry in the Theories
  of Fields, Strings and Branes}}, pp.~219--269.
\newblock 7, 1999.
\newblock \href{http://www.arXiv.org/abs/hep-th/0003032}{{\tt hep-th/0003032}}.

\bibitem{Semenoff:2002kk}
G.~W. Semenoff and K.~Zarembo, ``{Wilson loops in SYM theory: From weak to
  strong coupling},'' {\em Nucl. Phys. B Proc. Suppl.} {\bf 108} (2002)
  106--112, \href{http://www.arXiv.org/abs/hep-th/0202156}{{\tt
  hep-th/0202156}}.

\bibitem{Chernicoff:2011xv}
M.~Chernicoff, J.~A. García, A.~Güijosa, and J.~F. Pedraza, ``{Holographic
  Lessons for Quark Dynamics},'' {\em J. Phys. G} {\bf 39} (2012) 054002,
  \href{http://www.arXiv.org/abs/1111.0872}{{\tt 1111.0872}}.

\bibitem{Zarembo:2016bbk}
K.~Zarembo, ``{Localization and AdS/CFT Correspondence},'' {\em J. Phys. A}
  {\bf 50} (2017), no.~44, 443011,
  \href{http://www.arXiv.org/abs/1608.02963}{{\tt 1608.02963}}.

\bibitem{Minahan:1998xb}
J.~A. Minahan, ``{Quark - monopole potentials in large N superYang-Mills},''
  {\em Adv. Theor. Math. Phys.} {\bf 2} (1998) 559--569,
  \href{http://www.arXiv.org/abs/hep-th/9803111}{{\tt hep-th/9803111}}.

\bibitem{Maldacena:1998uz}
J.~M. Maldacena, J.~Michelson, and A.~Strominger, ``{Anti-de Sitter
  fragmentation},'' {\em JHEP} {\bf 02} (1999) 011,
  \href{http://www.arXiv.org/abs/hep-th/9812073}{{\tt hep-th/9812073}}.

\bibitem{Giveon:1998ns}
A.~Giveon, D.~Kutasov, and N.~Seiberg, ``{Comments on string theory on
  AdS(3)},'' {\em Adv. Theor. Math. Phys.} {\bf 2} (1998) 733--782,
  \href{http://www.arXiv.org/abs/hep-th/9806194}{{\tt hep-th/9806194}}.

\bibitem{deBoer:1998gyt}
J.~de~Boer, H.~Ooguri, H.~Robins, and J.~Tannenhauser, ``{String theory on
  AdS$_3$},'' {\em JHEP} {\bf 12} (1998) 026,
  \href{http://www.arXiv.org/abs/hep-th/9812046}{{\tt hep-th/9812046}}.

\bibitem{Maldacena:2000hw}
J.~M. Maldacena and H.~Ooguri, ``{Strings in AdS$_3$ and SL(2,R) WZW model 1:
  The Spectrum},'' {\em J. Math. Phys.} {\bf 42} (2001) 2929--2960,
  \href{http://www.arXiv.org/abs/hep-th/0001053}{{\tt hep-th/0001053}}.

\bibitem{Eberhardt:2019ywk}
L.~Eberhardt, M.~R. Gaberdiel, and R.~Gopakumar, ``{Deriving the
  AdS$_{3}$/CFT$_{2}$ correspondence},'' {\em JHEP} {\bf 02} (2020) 136,
  \href{http://www.arXiv.org/abs/1911.00378}{{\tt 1911.00378}}.

\bibitem{Gubser:1998iu}
S.~S. Gubser and A.~Hashimoto, ``{Exact absorption probabilities for the
  D3-brane},'' {\em Commun. Math. Phys.} {\bf 203} (1999) 325--340,
  \href{http://www.arXiv.org/abs/hep-th/9805140}{{\tt hep-th/9805140}}.

\bibitem{Intriligator:1999ai}
K.~A. Intriligator, ``{Maximally supersymmetric RG flows and AdS duality},''
  {\em Nucl. Phys. B} {\bf 580} (2000) 99--120,
  \href{http://www.arXiv.org/abs/hep-th/9909082}{{\tt hep-th/9909082}}.

\bibitem{Danielsson:2000ze}
U.~H. Danielsson, A.~Güijosa, M.~Kruczenski, and B.~Sundborg, ``{D3-brane
  holography},'' {\em JHEP} {\bf 05} (2000) 028,
  \href{http://www.arXiv.org/abs/hep-th/0004187}{{\tt hep-th/0004187}}.

\bibitem{Amador:2003ju}
X.~Amador, E.~Cáceres, H.~García-Compeán, and A.~Güijosa, ``{Conifold
  holography},'' {\em JHEP} {\bf 06} (2003) 049,
  \href{http://www.arXiv.org/abs/hep-th/0305257}{{\tt hep-th/0305257}}.

\bibitem{Sethi:1997pa}
S.~Sethi and M.~Stern, ``{D-brane bound states redux},'' {\em Commun. Math.
  Phys.} {\bf 194} (1998) 675--705,
  \href{http://www.arXiv.org/abs/hep-th/9705046}{{\tt hep-th/9705046}}.

\bibitem{Lin:2014wka}
Y.-H. Lin and X.~Yin, ``{On the Ground State Wave Function of Matrix Theory},''
  {\em JHEP} {\bf 11} (2015) 027,
  \href{http://www.arXiv.org/abs/1402.0055}{{\tt 1402.0055}}.

\bibitem{Gunaydin:1984fk}
M.~Gunaydin and N.~Marcus, ``{The Spectrum of the S$^5$ Compactification of the
  Chiral $\mathcal{N}=2$, $D=10$ Supergravity and the Unitary Supermultiplets
  of U(2, 2/4)},'' {\em Class. Quant. Grav.} {\bf 2} (1985) L11.

\bibitem{Kim:1985ez}
H.~J. Kim, L.~J. Romans, and P.~van Nieuwenhuizen, ``{The Mass Spectrum of
  Chiral $\mathcal{N}=2$ $D=10$ Supergravity on S$^5$},'' {\em Phys. Rev. D}
  {\bf 32} (1985) 389.

\bibitem{Witten:1998wy}
E.~Witten, ``{AdS / CFT correspondence and topological field theory},'' {\em
  JHEP} {\bf 12} (1998) 012,
  \href{http://www.arXiv.org/abs/hep-th/9812012}{{\tt hep-th/9812012}}.

\bibitem{Aharony:1999ti}
O.~Aharony, S.~S. Gubser, J.~M. Maldacena, H.~Ooguri, and Y.~Oz, ``{Large N
  field theories, string theory and gravity},'' {\em Phys. Rept.} {\bf 323}
  (2000) 183--386, \href{http://www.arXiv.org/abs/hep-th/9905111}{{\tt
  hep-th/9905111}}.

\bibitem{Bilal:1999ph}
A.~Bilal and C.-S. Chu, ``{A Note on the chiral anomaly in the AdS / CFT
  correspondence and 1 / N**2 correction},'' {\em Nucl. Phys. B} {\bf 562}
  (1999) 181--190, \href{http://www.arXiv.org/abs/hep-th/9907106}{{\tt
  hep-th/9907106}}.

\bibitem{Mansfield:2003gs}
P.~Mansfield, D.~Nolland, and T.~Ueno, ``{The Boundary Weyl anomaly in the N=4
  SYM / type IIB supergravity correspondence},'' {\em JHEP} {\bf 01} (2004)
  013, \href{http://www.arXiv.org/abs/hep-th/0311021}{{\tt hep-th/0311021}}.

\bibitem{Girardello:1998pd}
L.~Girardello, M.~Petrini, M.~Porrati, and A.~Zaffaroni, ``{Novel local CFT and
  exact results on perturbations of N=4 superYang Mills from AdS dynamics},''
  {\em JHEP} {\bf 12} (1998) 022,
  \href{http://www.arXiv.org/abs/hep-th/9810126}{{\tt hep-th/9810126}}.

\bibitem{Freedman:1999gp}
D.~Z. Freedman, S.~S. Gubser, K.~Pilch, and N.~P. Warner, ``{Renormalization
  group flows from holography supersymmetry and a c theorem},'' {\em Adv.
  Theor. Math. Phys.} {\bf 3} (1999) 363--417,
  \href{http://www.arXiv.org/abs/hep-th/9904017}{{\tt hep-th/9904017}}.

\bibitem{Girardello:1999bd}
L.~Girardello, M.~Petrini, M.~Porrati, and A.~Zaffaroni, ``{The Supergravity
  dual of N=1 superYang-Mills theory},'' {\em Nucl. Phys. B} {\bf 569} (2000)
  451--469, \href{http://www.arXiv.org/abs/hep-th/9909047}{{\tt
  hep-th/9909047}}.

\bibitem{Polchinski:2000uf}
J.~Polchinski and M.~J. Strassler, ``{The String dual of a confining
  four-dimensional gauge theory},''
  \href{http://www.arXiv.org/abs/hep-th/0003136}{{\tt hep-th/0003136}}.

\bibitem{Kraus:1998hv}
P.~Kraus, F.~Larsen, and S.~P. Trivedi, ``{The Coulomb branch of gauge theory
  from rotating branes},'' {\em JHEP} {\bf 03} (1999) 003,
  \href{http://www.arXiv.org/abs/hep-th/9811120}{{\tt hep-th/9811120}}.

\bibitem{Klebanov:1999tb}
I.~R. Klebanov and E.~Witten, ``{AdS / CFT correspondence and symmetry
  breaking},'' {\em Nucl. Phys. B} {\bf 556} (1999) 89--114,
  \href{http://www.arXiv.org/abs/hep-th/9905104}{{\tt hep-th/9905104}}.

\bibitem{Skenderis:2006di}
K.~Skenderis and M.~Taylor, ``{Holographic Coulomb branch vevs},'' {\em JHEP}
  {\bf 08} (2006) 001, \href{http://www.arXiv.org/abs/hep-th/0604169}{{\tt
  hep-th/0604169}}.

\bibitem{Das:1998ei}
S.~R. Das and S.~P. Trivedi, ``{Three-brane action and the correspondence
  between N=4 Yang-Mills theory and anti-De Sitter space},'' {\em Phys. Lett.
  B} {\bf 445} (1998) 142--149,
  \href{http://www.arXiv.org/abs/hep-th/9804149}{{\tt hep-th/9804149}}.

\bibitem{Hubeny:2007xt}
V.~E. Hubeny, M.~Rangamani, and T.~Takayanagi, ``{A Covariant holographic
  entanglement entropy proposal},'' {\em JHEP} {\bf 07} (2007) 062,
  \href{http://www.arXiv.org/abs/0705.0016}{{\tt 0705.0016}}.

\bibitem{Lewkowycz:2013nqa}
A.~Lewkowycz and J.~Maldacena, ``{Generalized gravitational entropy},'' {\em
  JHEP} {\bf 08} (2013) 090, \href{http://www.arXiv.org/abs/1304.4926}{{\tt
  1304.4926}}.

\bibitem{Dong:2016hjy}
X.~Dong, A.~Lewkowycz, and M.~Rangamani, ``{Deriving covariant holographic
  entanglement},'' {\em JHEP} {\bf 11} (2016) 028,
  \href{http://www.arXiv.org/abs/1607.07506}{{\tt 1607.07506}}.

\bibitem{Rangamani:2016dms}
M.~Rangamani and T.~Takayanagi, {\em {Holographic Entanglement Entropy}},
  vol.~931.
\newblock Springer, 2017.

\bibitem{Nishioka:2018khk}
T.~Nishioka, ``{Entanglement entropy: holography and renormalization group},''
  {\em Rev. Mod. Phys.} {\bf 90} (2018), no.~3, 035007,
  \href{http://www.arXiv.org/abs/1801.10352}{{\tt 1801.10352}}.

\bibitem{Mollabashi:2014qfa}
A.~Mollabashi, N.~Shiba, and T.~Takayanagi, ``{Entanglement between Two
  Interacting CFTs and Generalized Holographic Entanglement Entropy},'' {\em
  JHEP} {\bf 04} (2014) 185, \href{http://www.arXiv.org/abs/1403.1393}{{\tt
  1403.1393}}.

\bibitem{Karch:2014pma}
A.~Karch and C.~F. Uhlemann, ``{Holographic entanglement entropy and the
  internal space},'' {\em Phys. Rev. D} {\bf 91} (2015), no.~8, 086005,
  \href{http://www.arXiv.org/abs/1501.00003}{{\tt 1501.00003}}.

\bibitem{Taylor:2015kda}
M.~Taylor, ``{Generalized entanglement entropy},'' {\em JHEP} {\bf 07} (2016)
  040, \href{http://www.arXiv.org/abs/1507.06410}{{\tt 1507.06410}}.

\bibitem{Taylor:2016aoi}
M.~Taylor and W.~Woodhead, ``{Renormalized entanglement entropy},'' {\em JHEP}
  {\bf 08} (2016) 165, \href{http://www.arXiv.org/abs/1604.06808}{{\tt
  1604.06808}}.

\bibitem{Anastasiou:2018rla}
G.~Anastasiou, I.~J. Araya, and R.~Olea, ``{Topological terms, AdS$_{2n}$
  gravity and renormalized Entanglement Entropy of holographic CFTs},'' {\em
  Phys. Rev. D} {\bf 97} (2018), no.~10, 106015,
  \href{http://www.arXiv.org/abs/1803.04990}{{\tt 1803.04990}}.

\bibitem{Anastasiou:2019ldc}
G.~Anastasiou, I.~J. Araya, A.~Güijosa, and R.~Olea, ``{Renormalized AdS
  gravity and holographic entanglement entropy of even-dimensional CFTs},''
  {\em JHEP} {\bf 10} (2019) 221,
  \href{http://www.arXiv.org/abs/1908.11447}{{\tt 1908.11447}}.

\bibitem{Swingle:2009bg}
B.~Swingle, ``{Entanglement Renormalization and Holography},'' {\em Phys. Rev.
  D} {\bf 86} (2012) 065007, \href{http://www.arXiv.org/abs/0905.1317}{{\tt
  0905.1317}}.

\bibitem{VanRaamsdonk:2010pw}
M.~Van~Raamsdonk, ``{Building up spacetime with quantum entanglement},'' {\em
  Gen. Rel. Grav.} {\bf 42} (2010) 2323--2329,
  \href{http://www.arXiv.org/abs/1005.3035}{{\tt 1005.3035}}.

\bibitem{Maldacena:2013xja}
J.~Maldacena and L.~Susskind, ``{Cool horizons for entangled black holes},''
  {\em Fortsch. Phys.} {\bf 61} (2013) 781--811,
  \href{http://www.arXiv.org/abs/1306.0533}{{\tt 1306.0533}}.

\bibitem{Pastawski:2015qua}
F.~Pastawski, B.~Yoshida, D.~Harlow, and J.~Preskill, ``{Holographic quantum
  error-correcting codes: Toy models for the bulk/boundary correspondence},''
  {\em JHEP} {\bf 06} (2015) 149,
  \href{http://www.arXiv.org/abs/1503.06237}{{\tt 1503.06237}}.

\bibitem{Hayden:2016cfa}
P.~Hayden, S.~Nezami, X.-L. Qi, N.~Thomas, M.~Walter, and Z.~Yang,
  ``{Holographic duality from random tensor networks},'' {\em JHEP} {\bf 11}
  (2016) 009, \href{http://www.arXiv.org/abs/1601.01694}{{\tt 1601.01694}}.

\bibitem{Bao:2018pvs}
N.~Bao, G.~Penington, J.~Sorce, and A.~C. Wall, ``{Beyond Toy Models:
  Distilling Tensor Networks in Full AdS/CFT},'' {\em JHEP} {\bf 11} (2019)
  069, \href{http://www.arXiv.org/abs/1812.01171}{{\tt 1812.01171}}.

\bibitem{VanRaamsdonk:2018zws}
M.~Van~Raamsdonk, ``{Building up spacetime with quantum entanglement II: It
  from BC-bit},'' \href{http://www.arXiv.org/abs/1809.01197}{{\tt 1809.01197}}.

\bibitem{Czech:2012bh}
B.~Czech, J.~L. Karczmarek, F.~Nogueira, and M.~Van~Raamsdonk, ``{The Gravity
  Dual of a Density Matrix},'' {\em Class. Quant. Grav.} {\bf 29} (2012)
  155009, \href{http://www.arXiv.org/abs/1204.1330}{{\tt 1204.1330}}.

\bibitem{Wall:2012uf}
A.~C. Wall, ``{Maximin Surfaces, and the Strong Subadditivity of the Covariant
  Holographic Entanglement Entropy},'' {\em Class. Quant. Grav.} {\bf 31}
  (2014), no.~22, 225007, \href{http://www.arXiv.org/abs/1211.3494}{{\tt
  1211.3494}}.

\bibitem{Engelhardt:2014gca}
N.~Engelhardt and A.~C. Wall, ``{Quantum Extremal Surfaces: Holographic
  Entanglement Entropy beyond the Classical Regime},'' {\em JHEP} {\bf 01}
  (2015) 073, \href{http://www.arXiv.org/abs/1408.3203}{{\tt 1408.3203}}.

\bibitem{Headrick:2014cta}
M.~Headrick, V.~E. Hubeny, A.~Lawrence, and M.~Rangamani, ``{Causality \&
  holographic entanglement entropy},'' {\em JHEP} {\bf 12} (2014) 162,
  \href{http://www.arXiv.org/abs/1408.6300}{{\tt 1408.6300}}.

\bibitem{Almheiri:2014lwa}
A.~Almheiri, X.~Dong, and D.~Harlow, ``{Bulk Locality and Quantum Error
  Correction in AdS/CFT},'' {\em JHEP} {\bf 04} (2015) 163,
  \href{http://www.arXiv.org/abs/1411.7041}{{\tt 1411.7041}}.

\bibitem{Chen:2019gbt}
C.-F. Chen, G.~Penington, and G.~Salton, ``{Entanglement Wedge Reconstruction
  using the Petz Map},'' {\em JHEP} {\bf 01} (2020) 168,
  \href{http://www.arXiv.org/abs/1902.02844}{{\tt 1902.02844}}.

\bibitem{Guijosa:2022jdo}
A.~Güijosa, Y.~D. Olivas, and J.~F. Pedraza, ``{Holographic coarse-graining:
  correlators from the entanglement wedge and other reduced geometries},'' {\em
  JHEP} {\bf 08} (2022) 118, \href{http://www.arXiv.org/abs/2201.01786}{{\tt
  2201.01786}}.

\bibitem{Harlow:2016vwg}
D.~Harlow, ``{The Ryu\textendash{}Takayanagi Formula from Quantum Error
  Correction},'' {\em Commun. Math. Phys.} {\bf 354} (2017), no.~3, 865--912,
  \href{http://www.arXiv.org/abs/1607.03901}{{\tt 1607.03901}}.

\bibitem{Akers:2021fut}
C.~Akers and G.~Penington, ``{Quantum minimal surfaces from quantum error
  correction},'' {\em SciPost Phys.} {\bf 12} (2022), no.~5, 157,
  \href{http://www.arXiv.org/abs/2109.14618}{{\tt 2109.14618}}.

\bibitem{Penington:2019npb}
G.~Penington, ``{Entanglement Wedge Reconstruction and the Information
  Paradox},'' {\em JHEP} {\bf 09} (2020) 002,
  \href{http://www.arXiv.org/abs/1905.08255}{{\tt 1905.08255}}.

\bibitem{Almheiri:2019psf}
A.~Almheiri, N.~Engelhardt, D.~Marolf, and H.~Maxfield, ``{The entropy of bulk
  quantum fields and the entanglement wedge of an evaporating black hole},''
  {\em JHEP} {\bf 12} (2019) 063,
  \href{http://www.arXiv.org/abs/1905.08762}{{\tt 1905.08762}}.

\bibitem{Almheiri:2019hni}
A.~Almheiri, R.~Mahajan, J.~Maldacena, and Y.~Zhao, ``{The Page curve of
  Hawking radiation from semiclassical geometry},'' {\em JHEP} {\bf 03} (2020)
  149, \href{http://www.arXiv.org/abs/1908.10996}{{\tt 1908.10996}}.

\bibitem{Penington:2019kki}
G.~Penington, S.~H. Shenker, D.~Stanford, and Z.~Yang, ``{Replica wormholes and
  the black hole interior},'' {\em JHEP} {\bf 03} (2022) 205,
  \href{http://www.arXiv.org/abs/1911.11977}{{\tt 1911.11977}}.

\bibitem{Almheiri:2019qdq}
A.~Almheiri, T.~Hartman, J.~Maldacena, E.~Shaghoulian, and A.~Tajdini,
  ``{Replica Wormholes and the Entropy of Hawking Radiation},'' {\em JHEP} {\bf
  05} (2020) 013, \href{http://www.arXiv.org/abs/1911.12333}{{\tt 1911.12333}}.

\bibitem{Geng:2020qvw}
H.~Geng and A.~Karch, ``{Massive islands},'' {\em JHEP} {\bf 09} (2020) 121,
  \href{http://www.arXiv.org/abs/2006.02438}{{\tt 2006.02438}}.

\bibitem{Chen:2020tes}
Y.~Chen, V.~Gorbenko, and J.~Maldacena, ``{Bra-ket wormholes in gravitationally
  prepared states},'' {\em JHEP} {\bf 02} (2021) 009,
  \href{http://www.arXiv.org/abs/2007.16091}{{\tt 2007.16091}}.

\bibitem{Geng:2020fxl}
H.~Geng, A.~Karch, C.~Perez-Pardavila, S.~Raju, L.~Randall, M.~Riojas, and
  S.~Shashi, ``{Information Transfer with a Gravitating Bath},'' {\em SciPost
  Phys.} {\bf 10} (2021), no.~5, 103,
  \href{http://www.arXiv.org/abs/2012.04671}{{\tt 2012.04671}}.

\bibitem{Geng:2021hlu}
H.~Geng, A.~Karch, C.~Perez-Pardavila, S.~Raju, L.~Randall, M.~Riojas, and
  S.~Shashi, ``{Inconsistency of islands in theories with long-range
  gravity},'' {\em JHEP} {\bf 01} (2022) 182,
  \href{http://www.arXiv.org/abs/2107.03390}{{\tt 2107.03390}}.

\bibitem{Balasubramanian:2022gmo}
V.~Balasubramanian, A.~Lawrence, J.~M. Magan, and M.~Sasieta, ``{Microscopic
  Origin of the Entropy of Black Holes in General Relativity},'' {\em Phys.
  Rev. X} {\bf 14} (2024), no.~1, 011024,
  \href{http://www.arXiv.org/abs/2212.02447}{{\tt 2212.02447}}.

\bibitem{Geng:2024xpj}
H.~Geng, ``{Replica wormholes and entanglement islands in the Karch-Randall
  braneworld},'' {\em JHEP} {\bf 01} (2025) 063,
  \href{http://www.arXiv.org/abs/2405.14872}{{\tt 2405.14872}}.

\bibitem{Geng:2025rov}
H.~Geng, ``{The Mechanism behind the Information Encoding for Islands},''
  \href{http://www.arXiv.org/abs/2502.08703}{{\tt 2502.08703}}.

\bibitem{Leutheusser:2021qhd}
S.~Leutheusser and H.~Liu, ``{Causal connectability between quantum systems and
  the black hole interior in holographic duality},'' {\em Phys. Rev. D} {\bf
  108} (2023), no.~8, 086019, \href{http://www.arXiv.org/abs/2110.05497}{{\tt
  2110.05497}}.

\bibitem{Leutheusser:2021frk}
S.~A.~W. Leutheusser and H.~Liu, ``{Emergent Times in Holographic Duality},''
  {\em Phys. Rev. D} {\bf 108} (2023), no.~8, 086020,
  \href{http://www.arXiv.org/abs/2112.12156}{{\tt 2112.12156}}.

\bibitem{Witten:2021unn}
E.~Witten, ``{Gravity and the crossed product},'' {\em JHEP} {\bf 10} (2022)
  008, \href{http://www.arXiv.org/abs/2112.12828}{{\tt 2112.12828}}.

\bibitem{Chandrasekaran:2022cip}
V.~Chandrasekaran, R.~Longo, G.~Penington, and E.~Witten, ``{An algebra of
  observables for de Sitter space},'' {\em JHEP} {\bf 02} (2023) 082,
  \href{http://www.arXiv.org/abs/2206.10780}{{\tt 2206.10780}}.

\bibitem{Chandrasekaran:2022eqq}
V.~Chandrasekaran, G.~Penington, and E.~Witten, ``{Large N algebras and
  generalized entropy},'' {\em JHEP} {\bf 04} (2023) 009,
  \href{http://www.arXiv.org/abs/2209.10454}{{\tt 2209.10454}}.

\bibitem{Leutheusser:2022bgi}
S.~Leutheusser and H.~Liu, ``{Subregion-subalgebra duality: emergence of space
  and time in holography},'' \href{http://www.arXiv.org/abs/2212.13266}{{\tt
  2212.13266}}.

\bibitem{Bahiru:2022oas}
E.~Bahiru, A.~Belin, K.~Papadodimas, G.~Sarosi, and N.~Vardian,
  ``{State-dressed local operators in the AdS/CFT correspondence},'' {\em Phys.
  Rev. D} {\bf 108} (2023), no.~8, 086035,
  \href{http://www.arXiv.org/abs/2209.06845}{{\tt 2209.06845}}.

\bibitem{Witten:2023xze}
E.~Witten, ``{A background-independent algebra in quantum gravity},'' {\em
  JHEP} {\bf 03} (2024) 077, \href{http://www.arXiv.org/abs/2308.03663}{{\tt
  2308.03663}}.

\bibitem{Kudler-Flam:2023qfl}
J.~Kudler-Flam, S.~Leutheusser, and G.~Satishchandran, ``{Generalized black
  hole entropy is von Neumann entropy},'' {\em Phys. Rev. D} {\bf 111} (2025),
  no.~2, 025013, \href{http://www.arXiv.org/abs/2309.15897}{{\tt 2309.15897}}.

\bibitem{Engelhardt:2023xer}
N.~Engelhardt and H.~Liu, ``{Algebraic ER=EPR and complexity transfer},'' {\em
  JHEP} {\bf 07} (2024) 013, \href{http://www.arXiv.org/abs/2311.04281}{{\tt
  2311.04281}}.

\bibitem{Kudler-Flam:2024psh}
J.~Kudler-Flam, S.~Leutheusser, and G.~Satishchandran, ``{Algebraic
  Observational Cosmology},'' \href{http://www.arXiv.org/abs/2406.01669}{{\tt
  2406.01669}}.

\bibitem{Geng:2024dbl}
H.~Geng, ``{Quantum Rods and Clock in a Gravitational Universe},''
  \href{http://www.arXiv.org/abs/2412.03636}{{\tt 2412.03636}}.

\bibitem{guijosa:2025talk}
A.~Güijosa, ``On the underlying nonrelativistic nature of relativistic
  holography.'' Talk presented at the HolographyCL Farewell Meeting, January
  16, 2025.
\newblock https://holography.cl/activities/events/farewell.

\bibitem{Harmark:2025ikv}
T.~Harmark, J.~Lahnsteiner, and N.~A. Obers, ``{Gravitational solitons and
  non-relativistic string theory},''
  \href{http://www.arXiv.org/abs/2501.10178}{{\tt 2501.10178}}.

\end{thebibliography}\endgroup

\bibliographystyle{./utphys}

\end{document}